\documentclass[structabstract]{aa} 
\include{definitions}
\usepackage{graphicx,epstopdf}
\usepackage{epsfig}
\usepackage{txfonts}
\usepackage[utf8]{inputenc}
\usepackage{dcolumn}
\usepackage{natbib}
\usepackage{enumerate}
\usepackage{wrapfig}
\usepackage{rotating}
\usepackage{color}
\usepackage{grffile}
\usepackage{array,float}
\usepackage{multirow}
\usepackage{lscape}
\usepackage{xcolor}
\usepackage{hyperref} 
\usepackage{pdflscape}
\usepackage{comment}
\usepackage{afterpage}
\usepackage{capt-of}

\usepackage{etoolbox}
\makeatletter
\patchcmd\@combinedblfloats{\box\@outputbox}{\unvbox\@outputbox}{}{\errmessage{\noexpand patch failed}}
\makeatother

\usepackage{adjustbox}
\usepackage{array}
\newcolumntype{R}[2]{%
    >{\adjustbox{angle=#1,lap=\width-(#2)}\bgroup}%
    l%
    <{\egroup}%
}

\usepackage{pifont}
\newcommand{\cmark}{\ding{51}}%
\newcommand{\xmark}{\ding{55}}%

\newcolumntype{d}[1]{D{.}{\cdot}{#1}}

\newcolumntype{.}{D{.}{.}{-1}}

\newcommand{\mum}{$\mu$m}

\newcommand{\hi}{H~{\sc i}}
\newcommand{\hii}{H~{\sc ii}}
\newcommand{\uchii}{UC\,H~{\sc ii}}

\newcommand{\sex}{\texttt{SExtractor}}
\newcommand{\blobcat}{\texttt{BLOBCAT}}

\newcommand{\casa}{\texttt{CASA}}
\newcommand{\imfit}{\texttt{imfit}}

\newcommand{\blobs}{{1975 }}
\newcommand{\larges}{{\color{black}195 }}

\newcommand{\catsizetwo}{{1575 }}

\newcommand{\catsize}{{\color{black}1575 }}

\def\as {\ifmmode {\rlap.}$\,$''$\,$\! \else ${\rlap.}$\,$''$\,$\!$\fi}
\def\decsec {\ifmmode {\rlap.}$\,$^{\rm s}$\,$\! \else ${\rlap.}$\,$^{\rm
s}$\,$\!$\fi}\def\decs {\ifmmode {\rlap.}$\,$^{\rm s}$\,$\! \else
${\rlap.}$\,$^{\rm s}$\,$\!$\fi}

\begin{document}

\title{GLOSTAR --- Radio Source Catalog I: $28\degr < \ell < 36\degr$ and $|b| < 1$\degr \thanks{Full versions of Table\,\ref{tbl:glostar_cat} and Figures 7, 9 and 17 are only available in electronic form at the CDS via anonymous ftp to cdsarc.u-strasbg.fr (130.79.125.5) or via http://cdsweb.u-strasbg.fr/cgi-bin/qcat?J/A\&A/. }} 
\author{S.-N.\,X.\,Medina \inst{1}\thanks{E-mail: smedina@mpifr-bonn.mpg.de}, J.\,S.\,Urquhart\inst{2}, S.\,A.\,Dzib\inst{1}, A.\,Brunthaler\inst{1}, B.\,Cotton\inst{3}, K.\,M.\,Menten\inst{1}, F.\,Wyrowski\inst{1}, H.\, Beuther\inst{4}, S.\,J.\,Billington\inst{2}, C.\,Carrasco-Gonzalez\inst{5}, T.\,Csengeri\inst{1}, Y.\,Gong\inst{1}, P.\, Hofner\inst{6,7}, H.\,Nguyen\inst{1},   G.\,N.\,Ortiz-Le\'on\inst{1}, J.\,Ott\inst{7}, J.\,D.\,Pandian \inst{8}, N.\,Roy\inst{9}, E.\,Sarkar\inst{10}, Y.\,Wang\inst{4}, and B.\,Winkel\inst{1}.}


 \institute{Max-Planck-Institut f\"ur Radioastronomie (MPIfR), 
              Auf dem H\"ugel 69, 53121 Bonn, Germany. \and School of Physical Sciences, University of Kent, 
              Ingram Building, Canterbury, Kent CT2\,7NH, UK
          \and National Radio Astronomy Observatory (NRAO), 520 Edgemont Road, Charlottesville, VA 22903, USA. 
         \and Max Planck Institute for Astronomy, Koenigstuhl 17, 69117 Heidelberg, Germany.
         \and Instituto de Radioastronom\'{i}a y Astrof\'{i}sica (IRyA), Universidad Nacional Aut\'{o}noma de M\'{e}xico  Morelia, 58089, M\'{e}xico.
         \and Physics Department, New Mexico Tech, 801 Leroy Place, Socorro, NM 87801, USA.
         \and National Radio Astronomy Observatory, 1003 Lopezville Road, P.O. Box O, Socorro, NM 87801, USA 
         \and Department of Earth and Space Science, Indian Institute for Space Science and Tegnology, Trivandrum 695547, India.         
         \and Department of Physics, Indian Institute of Science, Bangalore 560012, India
          \and Sodankylä Geophysical Observatory, University of Oulu, 99600 Sodankylä, Finland. }

\date{Received / Accepted}
\authorrunning{S.-N.\,X.\,Medina et al.}

\abstract
{Radio continuum surveys of the Galactic plane are an excellent way to identify different source populations such as planetary nebulae, \hii\ regions, and radio stars and characterize their statistical properties. {The} GLOSTAR survey will study the star formation in the Galactic plane between $-2\degr < \ell < 85\degr$ and $|b| < 1\degr$ with unprecedented sensitivity in both, flux density ($\sim$40\,$\mu$Jy\,beam$^{-1}$) and range of angular scales ($\sim$1\as5 to the largest radio structures in the Galaxy).}
{In this paper we present the first results obtained from a radio continuum map of a 16 square degree sized region of the Galactic plane centered on $\ell = 32\degr$ and $b = 0\degr$ ($28\degr < \ell < 36\degr$ and $|b| < 1\degr$). This map has a resolution of 18\arcsec\ and a sensitivity of $\sim$60-150\,$\mu$Jy\,beam$^{-1}$.}
{We present data acquired in 40 hours of observations with the VLA in D-configuration. %
Two 1\,GHz wide sub-bands were observed simultaneously
and they were centred at 4.7 and 6.9\,GHz. These data were calibrated and imaged using the {\it Obit} software package. The source extraction has been performed using the \blobcat\ software package and verified through a combination of visual inspection and cross-matching with other radio and mid-infrared surveys. }
{The final catalog consists of \catsizetwo discrete radio sources and 27 large scale structures (including W43 and W44). By cross-matching with other catalogs and calculating the spectral indices ($S(\nu) \propto \nu^\alpha$), we have classified {231} continuum sources as \hii\ regions,  37 as ionization fronts, and 46 as planetary nebulae. The longitude and latitude distribution and negative spectral indices are all consistent with the vast majority of the unclassified sources being extragalactic background sources.}
{We present a catalog of  \catsizetwo radio continuum sources and discuss their physical properties, emission nature and relation with previously reported. These first GLOSTAR results have increased the number of reliable \hii\ regions in this part of the Galaxy by a factor of four.}


\maketitle

\section{\label{intro}Introduction}

One of the great challenges in modern astronomy is understanding the circumstances of the formation of high-mass stars ($>8$\,M$_{\sun}$). Massive stars dominate the energy budget of galaxies, regulate future star formation and drive their evolution (\citealt{kennicutt2005}). Over the past 10-15 years there has been a huge effort to map the Galactic mid-plane at infrared (e.g., GLIMPSE; \citealt{churchwell2009}, MIPSGAL; \citealt{carey2009}), (sub)millimeter (e.g., HiGAL; \citealt{molinari2010a}, ATLASGAL; \citealt{schuller2009}, BGPS; \citealt{aguirre2011}, \& SEDIGISM; \citealt{schuller2017}) and radio wavelengths (e.g., CORNISH; \citealt{hoare2012}, MAGPIS; \citealt{helfand2006}, THOR; \citealt{Beuther2016,bihr2015,wang2018}). These surveys have provided Galaxy-wide and unbiased samples of high-mass star forming regions (\citealt{urquhart2014_csc, urquhart2018_csc, elia2017}) that include all evolutionary stages and, for the first time, allow star formation to be studied in a global context.  

 The recent upgrade of the {\it Karl G. Jansky} Very Large Array (VLA) telescope of the NRAO\footnote{The National Radio Astronomy Observatory is operated by Associated Universities Inc. under cooperative agreement with the National Science Foundation.} has resulted in a significant increase in frequency coverage and sensitivity \citep{perley2011}. 
 This provides an excellent opportunity to conduct powerful and comprehensive radio-wavelength surveys of both, the ionized and the molecular, tracers of star formation in the Galactic plane that will complement previous surveys. The GLOSTAR (Global View of Star Formation in the Milky Way) survey\footnote{This VLA survey was part of the proposal for ERC Advanced Investigator Grant (247078) GLOSTAR that partially funded early stages of the research described here.} is one of these new radio-wavelength surveys. Combined with accurate distances determined by the BeSSeL trigonometric parallax survey (The Bar and Spiral Structure Legacy Survey; \citealt{reid2016}), {it} will provide a more complete view of star formation in the Milky Way.
 The GLOSTAR-VLA survey (Brunthaler et al. in prep.) uses the wideband (4-8\,GHz) C-band receivers of the VLA to conduct an unbiased survey to characterize star-forming regions in the Milky Way. This survey of the Galactic mid-plane allows {us} to detect tell-tale tracers of early phases of high-mass star formation: compact, ultra- and hyper-compact \hii\ regions, and  6.7\,GHz methanol (CH$_3$OH) masers, which trace some of the earliest evolutionary stages in the formation of high-mass stars (\citealt{minier2003}) and can be used to pinpoint the positions of very young stellar objects, many of them still deeply embedded in their natal material. The observations also cover emission from the 4.8\,GHz formaldehyde (H$_2$CO) and multiple Radio Recombination Lines (RRLs) all of which will be presented in future publications. GLOSTAR observations were made with the VLA B- and D-configurations. 

In this paper, we present the first source catalog covering the 16 square degrees of the Galactic mid-plane observed with the VLA in its most compact (D) configuration ($28\degr < \ell < 36\degr$ and  $|b| < 1\degr$). We provide {a} description of the observational and data reduction strategies used in Sect.\,\ref{sect:glostar_vla_survey}. In Sect.\,\ref{sect:catalogue_construction} we describe the source extraction and steps used to identify reliable sources and {dubious} sources that were removed from the final catalog. In the Sect.\,\ref{sect:source_verification} we compared the detected sources with other Galactic surveys to look for counterparts at other wavelengths, and therefore to verify our catalog. We discuss the physical properties and the resulting statistics of the sources in Sect.\,\ref{sect:catalogue_properties} and in Sect.\,\ref{sect:identification_emission_sources} we use the multi-wavelength data available to determine their nature {(e.g., \hii\ regions, planetary nebulae, extragalactic, etc.)}. In Sect.\,\ref{sect:properties_galactic_distribution_HII_regions} we focus on the properties and distribution of a sample of \hii\ regions identified from our source classification. Finally in Sect.\,\ref{conclusions} we give a summary of our results and  highlight our main findings.

\section{GLOSTAR-VLA survey observations}
\label{sect:glostar_vla_survey}

\subsection{Observation Strategy}
\label{subsec:observation Strategy}

The {\it Karl G. Jansky} Very Large Array (VLA) was used in D-configuration to observe the C-band (4$-$8\,GHz) continuum emission. The correlator setup consisted of two 1\,GHz wide sub-bands, centered at 4.7 and 6.9\,GHz. The primary beam FWHMs are $\sim9'$ and $\sim6'$ at 4.7 and 6.9\,GHz, respectively. Each sub-band was divided in eight spectral windows of 128~MHz, and each spectral window observed 64~channels of 2~MHz. Higher  spectral resolution  windows were used to cover the most prominent methanol maser emission line  at  6.7\,GHz, formaldehyde absorption at 4.8\,GHz, and seven  radio  recombination  lines. The results of these line observations will be reported in forthcoming papers. The chosen setup avoids strong and persistent Radio Frequency Interference (RFI) seen at 4.1 and 6.3\,GHz and allows a rough estimation of spectral indices.

The region of the GLOSTAR-VLA survey presented in this paper was observed over a total of eight epochs and a total of 40 hours of telescope time (project ID 13A-334). The observations were taken during the spring of 2013 (see Table\,\ref{tab:obs} for detailes). Each epoch consisted of observations of five hours. The phase calibrator was J1804$+$0101 and the amplitude calibrators 3C\,286 and 3C\,48 were observed periodically during each observing block. During each epoch, an area of $2\degr\times1\degr$ was covered by about 630 pointings. {The pointings were on a hexagonal grid with a spacing of $\theta_{hex}=3\as25$. This corresponds to $\theta_B/2$, where $\theta_B=6\as5$ is the FWHM width of the primary beam at the central frequency of the higher frequency continuum band (6.9 GHz). Each grid position was observed twice for 11 seconds which, after considering the slewing time, resulted in a total integration time of 15 seconds per position}. The theoretical noise level from these observations is $\sim$90\,$\mu$Jy\,beam$^{-1}$ per position and per sub-band. This noise is improved by a factor of two by combining the two observations of each position and both 1\,GHz sub-bands (theoretical noise $\sim$45\,$\mu$Jy\,beam$^{-1}$). 


\begin{table}
\centering
\caption{Observation epochs.}
\label{tab:obs}
\begin{tabular}{cc}\hline\hline
Region     &   Observation  date     \\
 ($\ell$)  & (yyyy.mm.dd) \\
\hline
28\degr - 29\degr& 2013.04.09 \\
29\degr - 30\degr& 2013.04.06 \\
30\degr - 31\degr& 2013.04.11 \\
31\degr - 32\degr& 2013.04.15 \\
32\degr - 33\degr& 2013.04.16 \\
33\degr - 34\degr& 2013.04.20 \\
34\degr - 35\degr& 2013.04.29 \\
35\degr - 36\degr& 2013.05.02 \\
\hline\hline
\end{tabular}
\end{table}

\subsection{Calibration, Data Reduction and Imaging of Continuum data}
\label{sect:Calibration_redution}

The data were calibrated using the {\it Obit} software packages \citep{cotton2008}, which was designed for handling radio astronomy data. {\it Obit} applications can be accessed through a Python interface (ObitTalk). {\it Obit} also inter-operates with Classic AIPS \citep{greisen2003} and has access to its tasks.

Observations made with the VLA are well standardized, as is the data calibration, and as such calibration pipelines are now available. These implement the data reduction and calibration using amplitude, phase, and bandpass calibrators (for an example consult the CASA pipeline webpage: https://science.nrao.edu/facilities/vla/data-processing/pipeline). Following the {methodology} of these pipelines, we have developed a pipeline in {\it Obit} to calibrate the GLOSTAR-VLA data with some differences. The details of these differences are described in what follows:

\begin{enumerate}

\item One of the first steps that most pipelines apply is to flag\footnote{``Flagging'' is a colloquial term used for marking faulty data to exclude them from further use.} the first record of every scan to consider the slew of the antennas. As our target scans last only a few seconds, {$\sim$11\,s,} this flagging covers an important fraction of our science data. Furthermore, the distance between the target pointings is only a few arcminutes so that the slewing time is {minimised}. Thus, our pipeline only performs the flagging of the first record on scans that observe calibrators. \\

\item The GLOSTAR observations use a mixed setup of continuum and spectral lines, which are not handled {well} by pipelines. Thus, the next step of our pipeline is to  distinguish between the different observational sets; i.e., for continuum, radio recombination line, formaldehyde and methanol data, and then load solely the continuum data. \\

\item A standard VLA calibration was applied in the {\it Obit} package which
alternates editing of data affected by instrumental problems or
interfering signals with various calibration steps.
External calibration includes VLA system temperature (T$_{\rm sys}$) calibration, group delay, bandpass,
amplitude and phase, and polarization calibration.  After a Hanning
smoothing of the data, an automated search for outliers in the data is
performed by a comparison against running medians in time and
frequency for each data product.  Calibrator data is checked for
adequate signal to noise ratio by an RMS/mean test.  After the 
initial pass at data
editing, the various parallel hand calibration steps are performed
applying further editing based on outliers in the calibration
solutions.  Following the first pass at calibration, a second set of
automated editing steps is done on the calibrated data. The calibration tables are then reset, the flagging table kept and
the calibration is repeated.  This assures that the calibration is
based on fully edited data.  After flagging cross hand outliers, the
polarization calibration is performed. Per pointing flagging is also included as part of the imaging. An initial, shallow CLEAN is done to determine the approximate flux density in the field. Data for that pointing is then clipped at 5 times the estimated pointing integrated flux density.\\

\item The {\it Obit} task MFImage was used for the continuum imaging (see \citealt{cotton2018} for details). First, we imaged the data after a shallow cleaning to identify strong sources that needed to be imaged in outlier fields in the final imaging. This program uses wide-band multifrequency synthesis by dividing the observed bandwidth into frequency bins, i.e. sections of the sub-band, with a fractional bandwidth of less than 5\%. The number of frequency bins for our observation is eight. These are narrow enough such that effects of variable spectral index and antenna pattern variations are minor performing a joint spectral deconvolution. \\

\item The CLEANing is driven by a weighted average of the residual images from these frequency bins. Each pointing is imaged separately with a $0.28\times0.28$ square degrees sized area, centered on the phase center, using a Briggs' weighting with robustness parameter 0 and using {its} native restoring beam and with a pixel size of $2\as5$. Outlier fields are used if a source with a integrated flux density of more than 10\,mJy is known from the NVSS or a previous imaging run from our own data out to a distance of 0.4 degrees. Multi-scale clean was used with three different sizes: a delta function, and tapers of 20 and 40 pixels (50\arcsec\ and 100\arcsec), respectively. Cleaning stopped when 5000 iterations were reached or the maximum in the residual image was below a level 0.25\,mJy.  Phase only, and amplitude plus phase self-calibration were applied for images with sufficiently bright peaks ($>$ 10\,mJy for phase only and $>$ 100\,mJy for amplitude plus phase self-calibrations). The shortest baselines of 30 meter allowed imaging of scales up to 4\arcmin.\\

\item After imaging the data from each pointing, the images were merged together in {\it Obit}. Since all pointings had a slightly different native beam size and orientation due do different $uv$-coverage, all pointings were smoothed in a first step to a circular synthesized beam of 18\arcsec\ in diameter. This beam size corresponds roughly to the major axis of the largest beam size of all the pointings. Primary beam corrections were applied during the merging of the images. After combining all the pointings at each frequency bin, a combined image at the reference frequency was generated.\\

\item The maximum sensitivity can be obtained by a weighted combination of the overlapping observed pointing images onto a set of images covering the surveyed region. All else being equal, weighting by the square of the antenna power pattern in a given direction gives the optimum sensitivity. The mosaic formation process for each image plane is given by the summation over overlapping pointing images: 

$$ M(x,y)\ =\ {{\sum_{i=1}^{n}A_{i}(x,y)\ I_{i}(x,y)}\over{\sum_{i=1}^{n}A^2_{i}(x,y)}}$$

where $A_{i}(x,y)$ is the antenna gain of pointing $i$ in the direction $(x,y)$ and $I_{i}(x,y)$ is the pointing $i$ pixel value interpolated to the direction $(x,y)$ and $M$ is the mosaic image. After combination, the spectrum in each pixel is redetermined. Note, the pointing images have already been corrected 
by the antenna gain pattern. A more detailed discussion of these image considerations is given by \citet{cotton2018}.\\

\end{enumerate}

\begin{sidewaysfigure*}
\centering

\includegraphics[width=0.98\textwidth, trim= 0 50 0 0, angle=0] {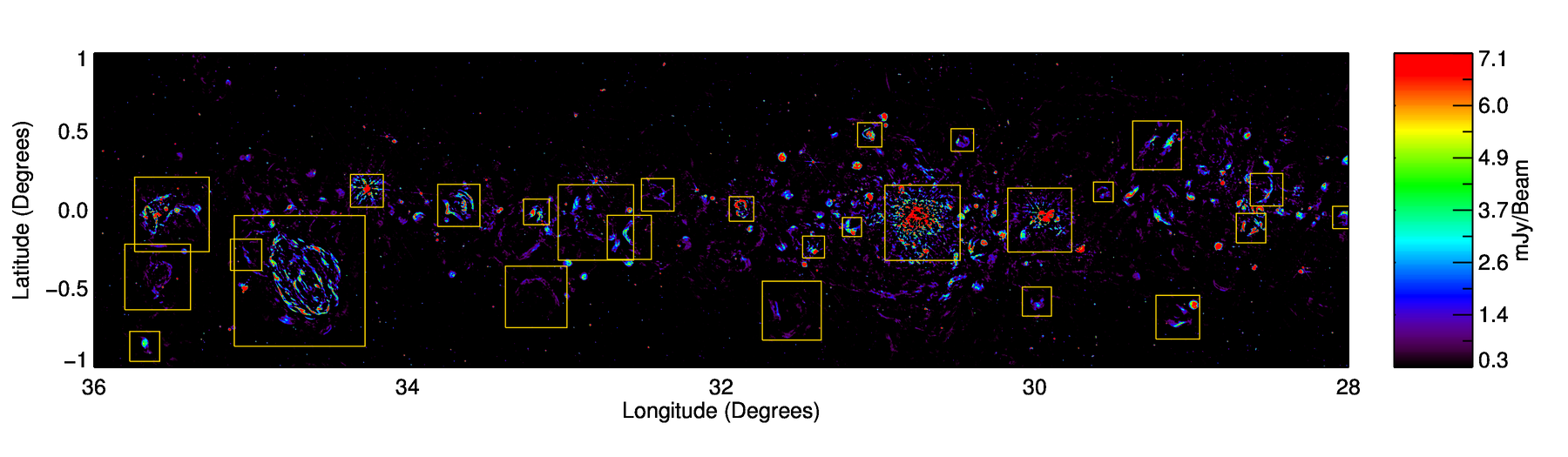}
\includegraphics[width=0.98\textwidth, trim= 0 0 0 50, angle=0] {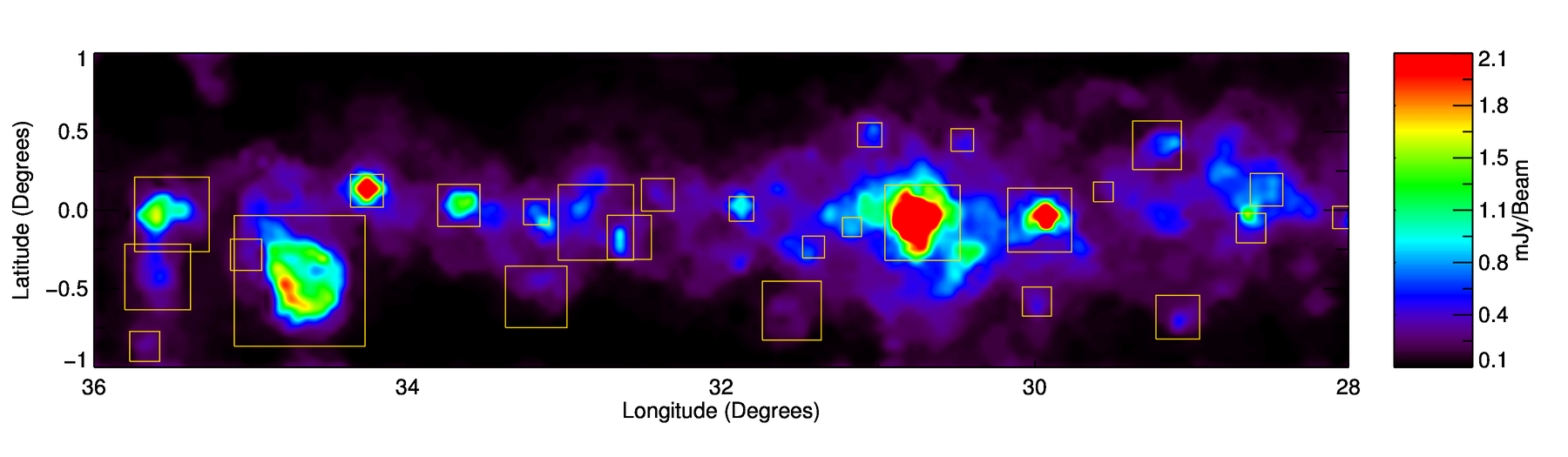} 

\caption{ Upper panel: GLOSTAR radio continuum map at 5.8~GHz of 16 sq degrees of the Galactic plane ($28\degr  < \ell < 36\degr$ and $|b| < 1\degr$). Lower panel: Full resolution noise background map determined by \sex\ using a mesh size of $80\times80$ pixels and threshold of 5$\sigma$. The yellow boxes correspond to the complexes that have been identified (see Sect.\,\ref{sect:large_scale_structures}  and Table\,\ref{tbl:LargeStruc} for more detailed and positions, respectively).   }
\label{fig.integrated_map}
\end{sidewaysfigure*}

\section{Analysis of the continuum map}
\label{sect:catalogue_construction}

\subsection{Radio continuum map}
\label{sect:RadioContiMap}

\begin{figure}
\centering
\includegraphics[width=0.49\textwidth, trim= 0 0 0 0, angle=0] {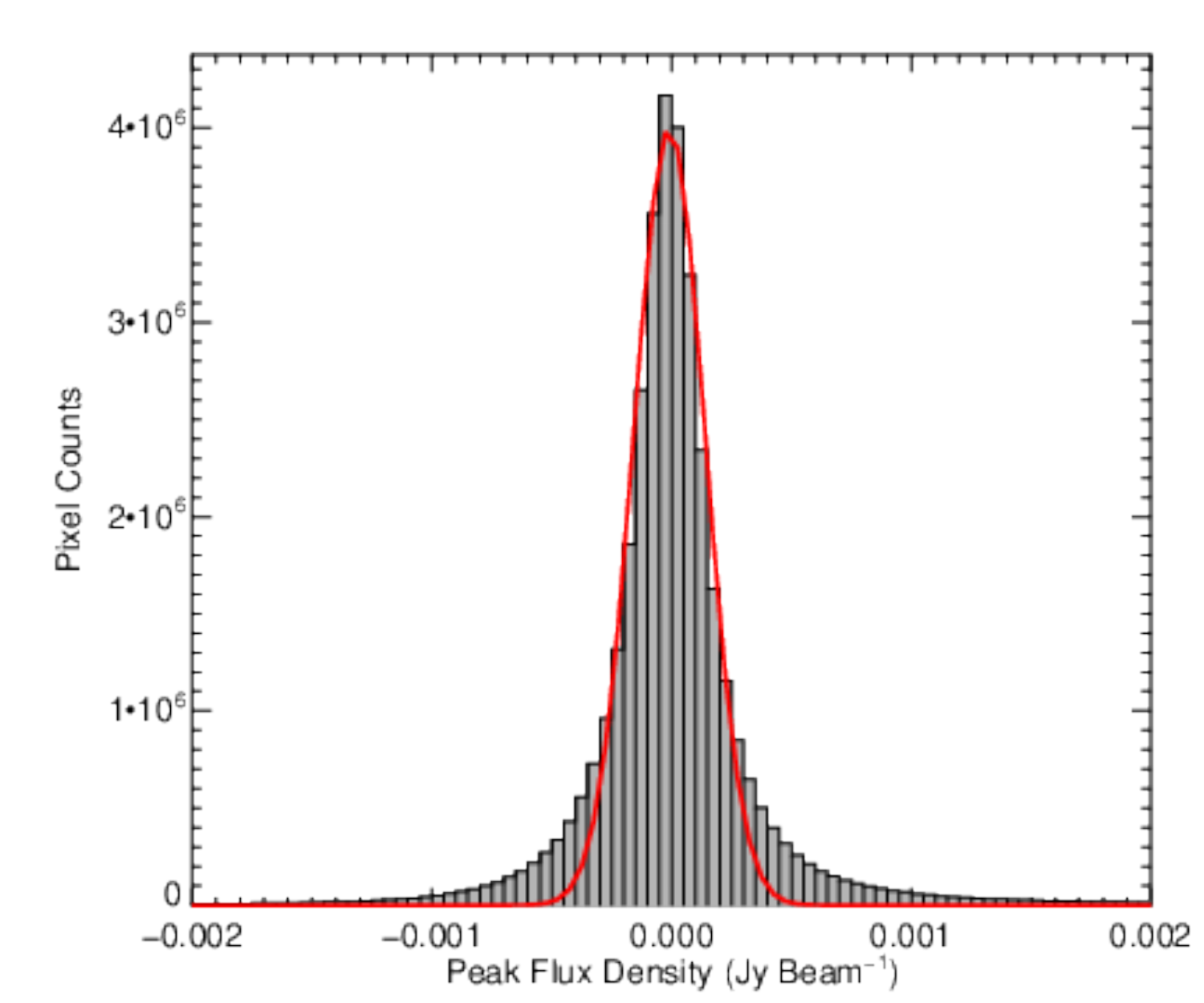}
\caption{Noise distribution of the map presented in Fig.\,\ref{fig.integrated_map}. We have restricted the range of pixel values to between $-$2 and 2\,mJy in order to estimate the noise in the map and determine its sensitivity. The red line shows the results of a Gaussian fit to the distribution, which gives the standard deviation of the noise as 150\,$\mu$Jy. The bin size is 50\,$\mu$Jy. }
\label{fig.noise_histogram}
\end{figure}

\begin{figure}
\centering
\includegraphics[width=0.49\textwidth, trim= 0 0 0 0, angle=0] {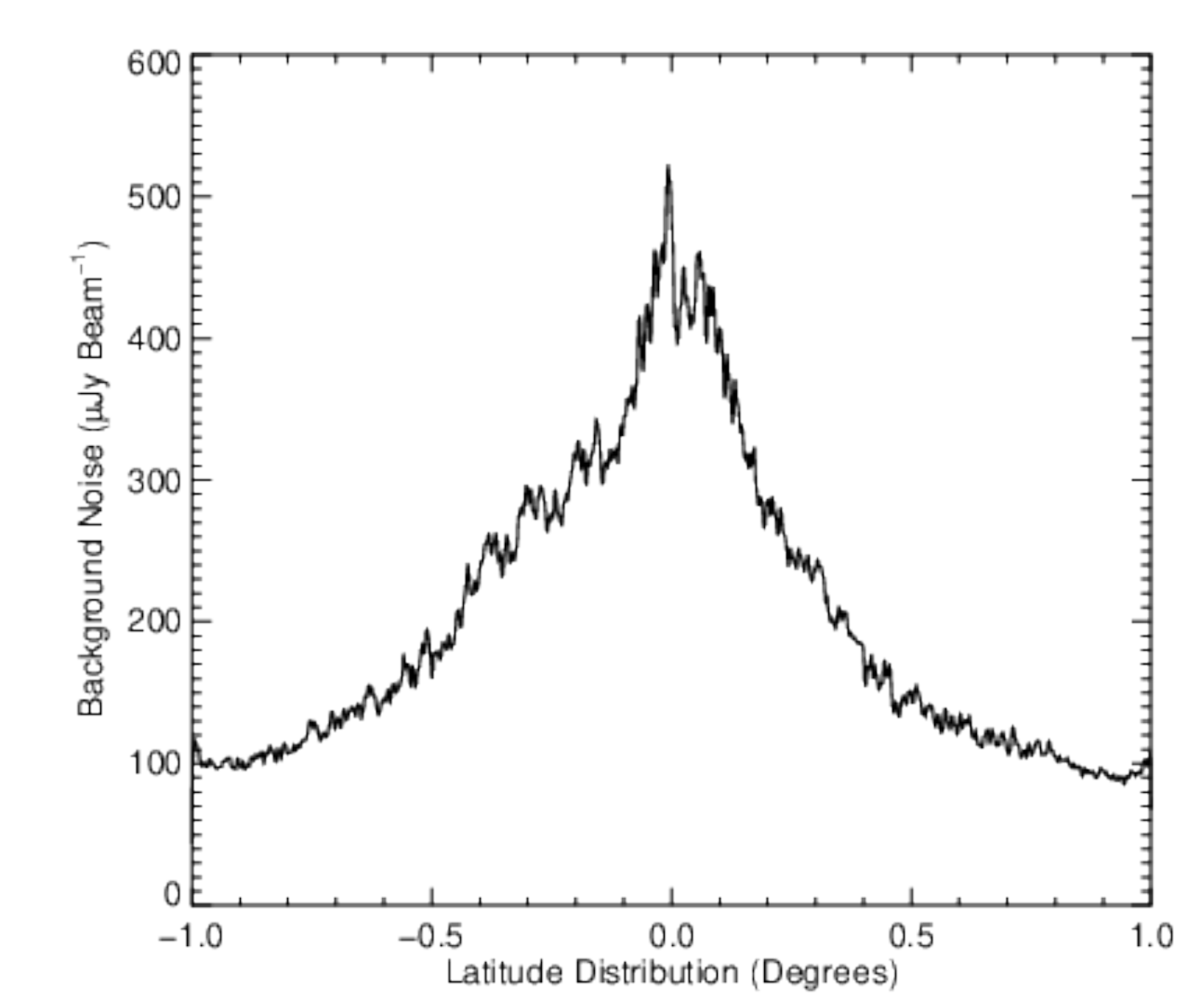}
\caption{Noise distribution of the map presented in Fig.\,\ref{fig.integrated_map} as a function of Galactic latitude. This has been produced in the same way as Fig.\,\ref{fig.noise_histogram} but for each increment in latitude.}
\label{fig.noise_histogram_fn_lat}
\end{figure}

A mosaic of the 16 square degrees that results from the 8 sub images 
integrated and discussed in this paper is presented in the upper 
panel of Fig.\,\ref{fig.integrated_map}. The effective frequency of the integrated image is 5.8\,GHz.  Although the observed region extends slightly beyond $|b| = 1$, the noise increases significantly in these regions due to poorer $uv$-coverage and differences in the beam sizes of the different sub-maps and therefore these edges have been removed. The final map has a spatial resolution of 18\arcsec. In the  lower panel of this figure, we present a map of the noise distribution  created using \sex\  \citep{bertin1996,holwerda2005}. A full description of the maps can be found in Sect.\,\ref{sect:source extraction}. In Fig.\,\ref{fig.noise_histogram} we present a histogram of the pixel values from which we estimate the sensitivity of the survey by fitting a Gaussian profile to the pixels clustered around zero. The standard deviation of the distribution of the pixel valuess is $\sigma_{\rm rms}$ = 150\,$\mu$Jy beam$^{-1}$. However, as can be seen from the lower panel of Fig.\,\ref{fig.integrated_map} there are strongly localized variations in the noise across the map particularly towards prominent complexes (e.g., W43, W44, G34.26+0.15), most of which are concentrated towards the mid-plane. This is a consequence of the poor $uv$-coverage of the observations themselves. These large {variations} of the noise across of the GLOSTAR map produce the non-Gaussian profile of the pixels values in Fig.\,\ref{fig.noise_histogram}. In regions free of extended emission the noise level is around 60 $\mu$Jy beam$^{-1}$. In Fig.\,\ref{fig.noise_histogram_fn_lat} we show the noise as a function of Galactic latitude, which reveals a steep increase in noise closer to the Galactic mid-plane from $\sim$100\,$\mu$Jy\,beam$^{-1}$ to $\sim$450\,$\mu$Jy\,beam$^{-1}$. 

\begin{figure}
\centering
\includegraphics[width=0.49\textwidth, trim= 0 0 0 0, angle=0]{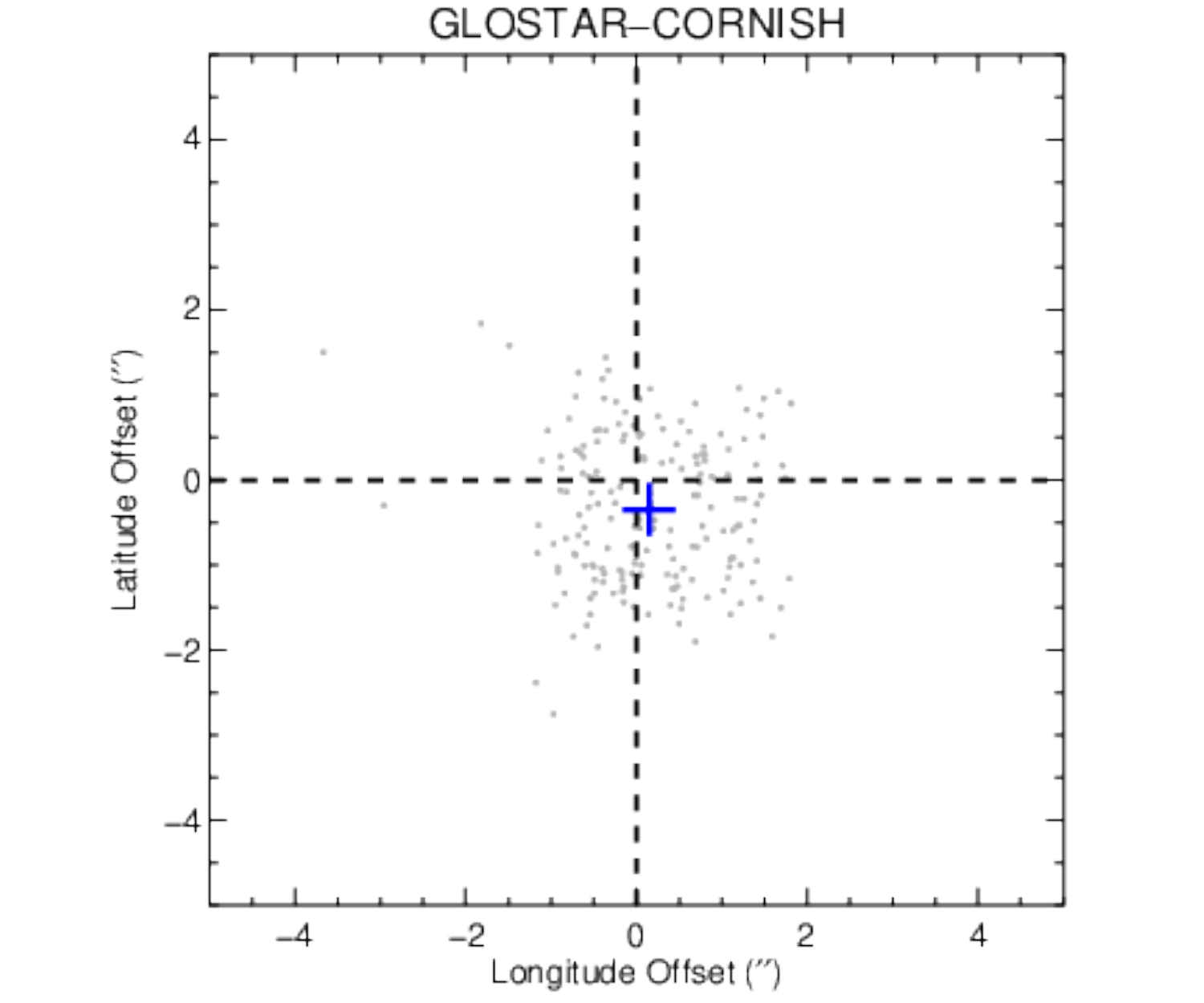}
\caption{Position offsets between GLOSTAR compact sources ($Y$-factor < 1.2) with their CORNISH counterparts. The blue cross indicates the mean value of the offsets and the dashed lines indicate where the longitude and latitude offsets are equal to zero.}
\label{fig:offsets}
\end{figure}

To measure the accuracy of the astrometry of the GLOSTAR map, we have matched the GLOSTAR unresolved sources ($Y$-factor < 1.2; this factor is the ratio of the integrated and peak flux densities, will be fully described in Sect.\,\ref{sect:catalogue_properties}) with more compact radio sources detected by the 5\,GHz CORNISH radio continuum survey (\citealt{purcell2013}) and measured the positional offsets between them. The  offset in $\ell$ and $b$ are plotted in Fig.\,\ref{fig:offsets}, which are clearly clustered around the centre; the mean values are  0\as2  $\pm$  0\as9 and  -0\as4  $\pm$ 0\as9, respectively, where the uncertainties are the standard deviations in the measurements. We therefore conclude that the GLOSTAR astrometry is reliable. We estimate the combined uncertainty on position, by adding the standard deviations in quadrature, to be 1\as2, which is {an} order of magnitude lower than our angular resolution (18\arcsec). 

\subsection{Source extraction} 
\label{sect:source extraction}

The source extraction has been performed using the \blobcat\ software package developed by \citet{hales2012}. \blobcat\ is a Python script that utilizes a flood fill algorithm to detect and identify blobs, or islands of pixels representing sources, in two-dimensional astronomical radio-wavelength images. The package has been successfully used to create the source catalog for The HI/OH/Recombination line survey of the inner Milky Way (THOR; \citealt{Beuther2016}) survey, and has been adopted by us to facilitate the comparison with this complementary VLA continuum and line survey.  We have followed the same method used by the THOR team at 1.4~GHz, which is outlined in \citet{bihr2016}.

The GLOSTAR radio continuum image has position-dependent noise and so we need to produce an independent noise map of the region to perform the automatic {source} extraction. We use the \emph{rms} estimation algorithm within the \sex\ package that has been proven to be trustworthy in creating noise maps from radio data \citep{Bondi2003,Huynh2005}. This algorithm defines the \emph{rms} value for each pixel in an image by determining the distribution of pixel values within a local mesh until all values are around a chosen $\sigma$ value. Most real emission is removed from the noise image and the determined noise map contains the correct noise level. We use a detection and analysis threshold of 5$\sigma$, a minimum size of 5 pixels, and a mesh size of $80 \times 80$\,pixels$^2$  (following the calculation of \citealt{hales2012}). The resulting noise map is shown in the lower panel of Fig.\,\ref{fig.integrated_map}, its large scale emission is consistent with the large scale emission previous reported by \citet{bihr2016} at 1.4 GHz.

To perform the automatic source extraction, we use the noise map as an input into \blobcat\ along with the FITS image of the region under study. We applied a detection threshold of (dSNR) of 4 and a minimum source size of 5 pixels in diameter ($\sim$13\arcsec). This resulted in the extraction of all sources that have flux above 4 times the local noise (i.e., the ratio between the input image and the \emph{rms} map) and have a size comparable or larger than the beam; these criteria reduce the number of false detections, particularly in complex regions. 

These input parameters resulted in the detection of \blobs sources within the 16 square degrees map. This sample is, however, likely to contain a significant number of artifacts and emission from large scale structures that have been over-resolved, resulting in the emission being separated into multiple components (e.g., the W44 supernova remnant at $\ell =34.6$\degr, $b =0.5$\degr), and from the blending of discrete sources towards star formation regions found towards the Galactic mid-plane. In the following subsections, we will describe how we have dealt with some of these issues while in Sect.\,\ref{sect:source_verification} we will describe the steps taken to identify and remove artifacts, and check the consistency of our catalog.



\begin{figure}
\centering
\includegraphics[width=0.49\textwidth, trim= 0 0 0 0, angle=0]{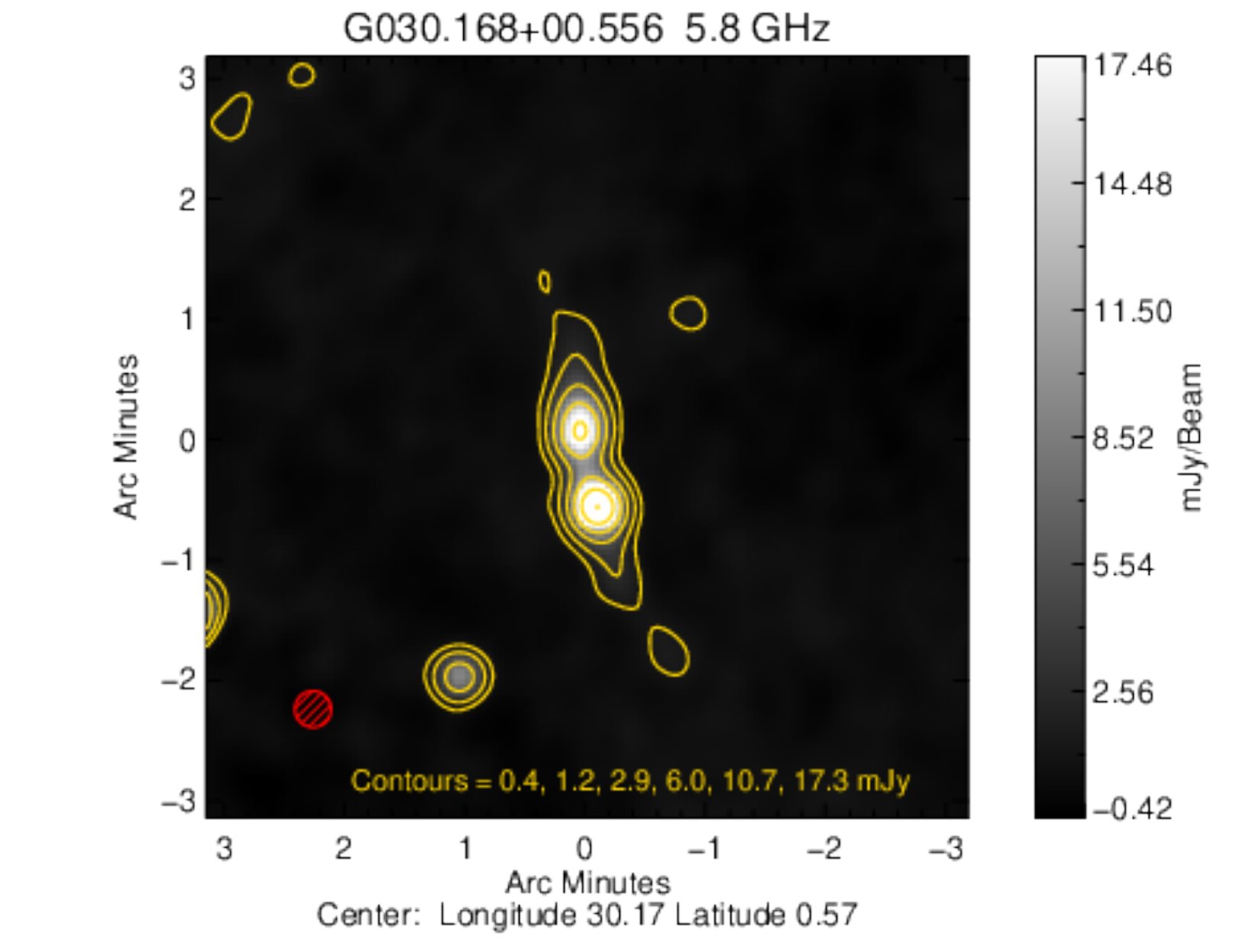}
\includegraphics[width=0.49\textwidth, trim= 0 0 0 0, angle=0]{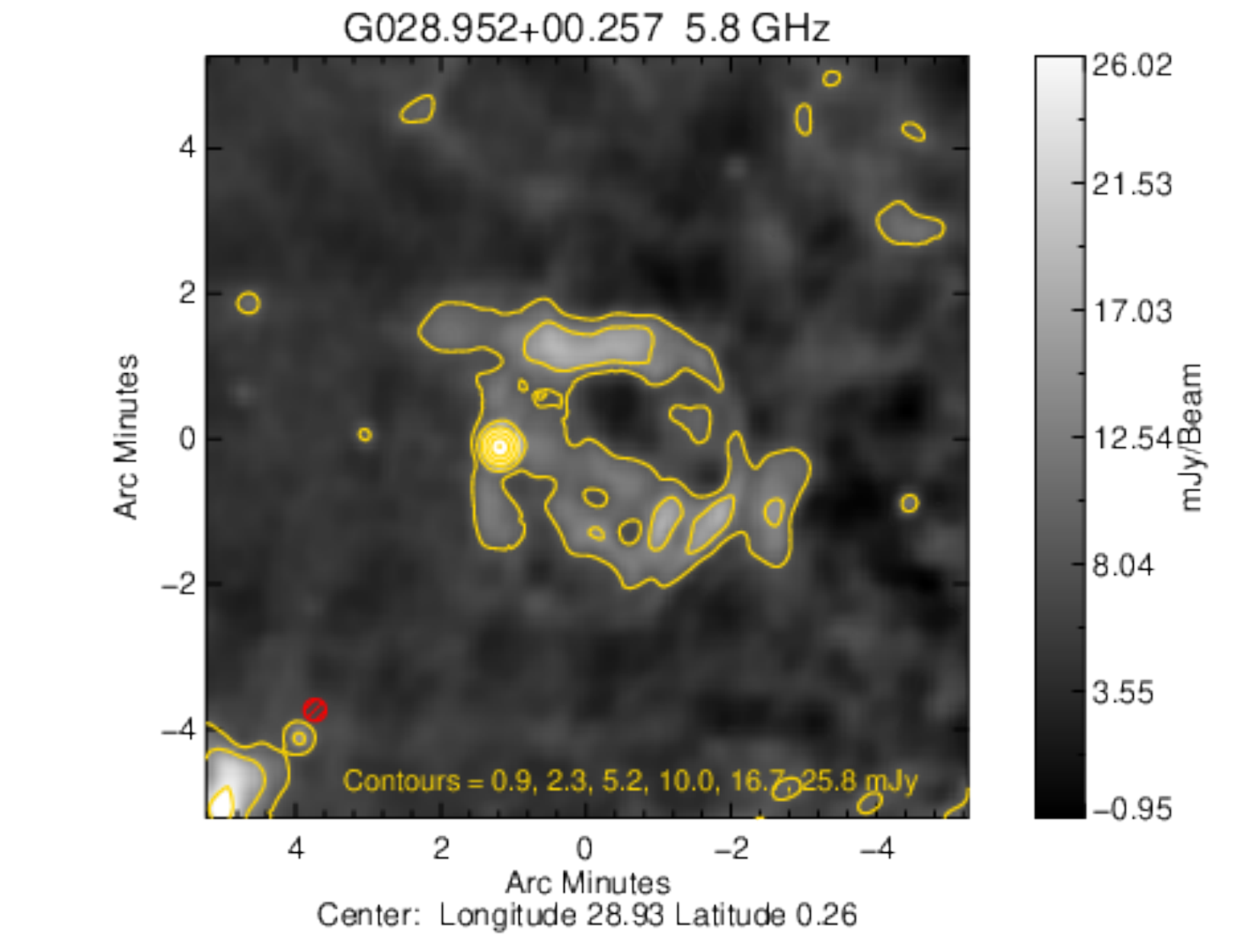}
\includegraphics[width=0.49\textwidth, trim= 0 0 0 0, angle=0]{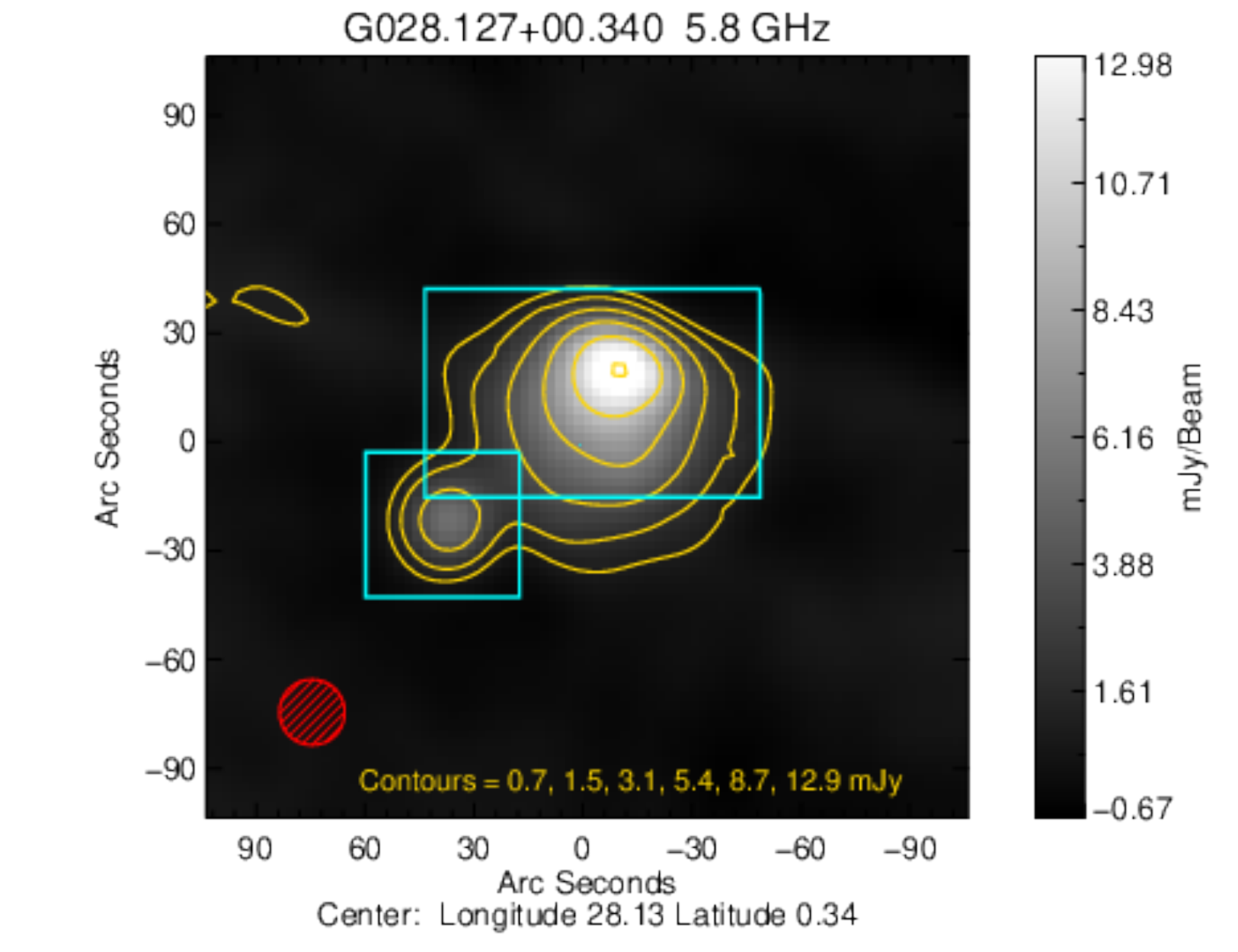}

\caption{Examples of various types of radio sources (see text). The contour levels  start at 3$\sigma$ and increase in steps determined using the dynamic range power-law fitting scheme where $D = N^i +3$ where $D$ is the dynamic range, $S_{\rm peak}/\sigma$, $N$ is the number of contour, in this case 6, and $i$ is the power-law that determines the separation between consecutive contours; this is a modified version of a scheme developed by \citealt{thompson2006}. The red hatched circle shown in the lower left corner of the map indicates the GLOSTAR beam size. {The cyan boxes  indicate the two sources identified and the region from which the flux has been estimated}.}
\label{fig.issues}
\end{figure}

\subsubsection{Superposition of distinct radio sources}
\label{sect:split}

In the visual inspection process, we flagged any sources that appeared to consist of two or more distinct emission peaks. We have grouped these into three categories: 1) elongated  `double lobed' sources; these look like extended emission regions powered by twin jets from active galactic nuclei (AGN; \citealt{treichel2001,malarecki2015,neff2015}); 2) superposition of a compact bright radio source and a larger more diffuse radio source; 3) superposition of spherical radio sources; these are likely to be \hii\ regions in the same star forming complexes. An example of each of these three types is shown in the upper, middle and lower panels of Fig.\,\ref{fig.issues}. 


We have made no attempt to separate the emission for the double lobed sources as it is likely to be associated with the same object. We have also made no attempt to separate the emission in cases in which a bright compact source is coincident with larger scale structures as there is not enough information to split them into reliable components as any measurement will be highly uncertain. There are a total of 51 of those cases. We have, however, separated the emission into two or more sources where the overlap is small and the integrated flux density can be reliably determined (see lower panel of Fig.\,\ref{fig.issues} as an example).

Inspection of the GLIMPSE 4.5, 8.0, and 24 $\mu$m mid-infrared images  of these blended sources identified 20 cases for which the emission peaks appear to be associated with distinct sources and could be separated. We split them into two or three sources according to the number of infrared counterparts.  We use the \casa\ software to select and split the regions in which emission is above 4$\sigma$ and then calculate their parameters such  as flux and size from their emission area. For some sources, e.g. uniform and compact radio emission, we make a Gaussian fit using the task \imfit\, also from the \casa\ software suite. An example of these type of sources  is shown in the lower panel of the Fig.\,\ref{fig.issues}. This process resulted in 44 sources being identified from the 20 blended sources examined. These split sources are treated in the source catalog (see Table\,\ref{tbl:glostar_cat}) as individual sources but we also include a flag to the entry so these can be easily identified. The peak and centroid position values are considered the same for these cases.

\subsubsection{Compact sources in noisy regions}
\label{sect:recovered}

\begin{figure}
\centering
\includegraphics[width=0.49\textwidth, trim= 0cm 0cm 0cm 0cm, clip]{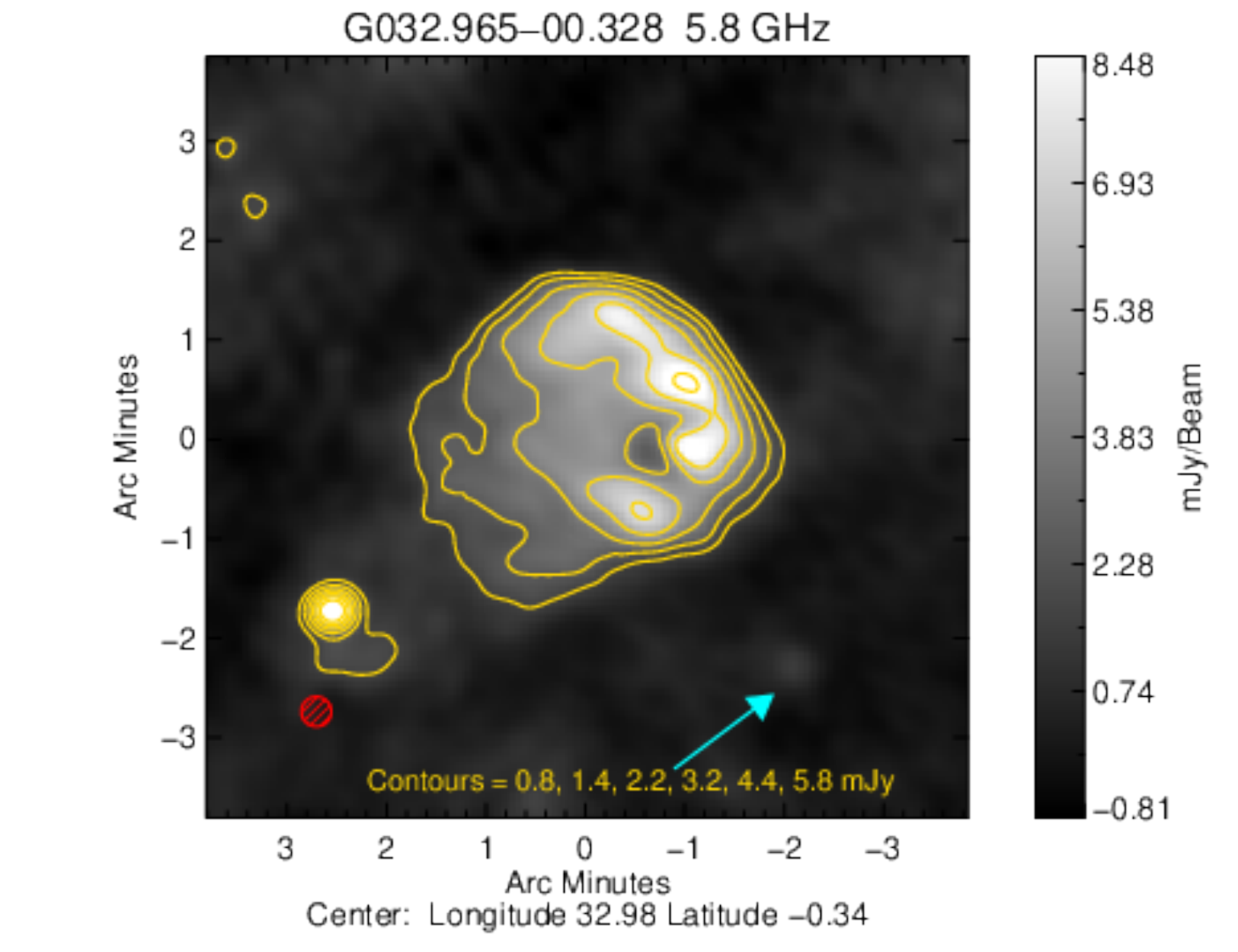}
\includegraphics[width=0.49\textwidth, trim= 0cm 0cm 0cm 0cm, clip]{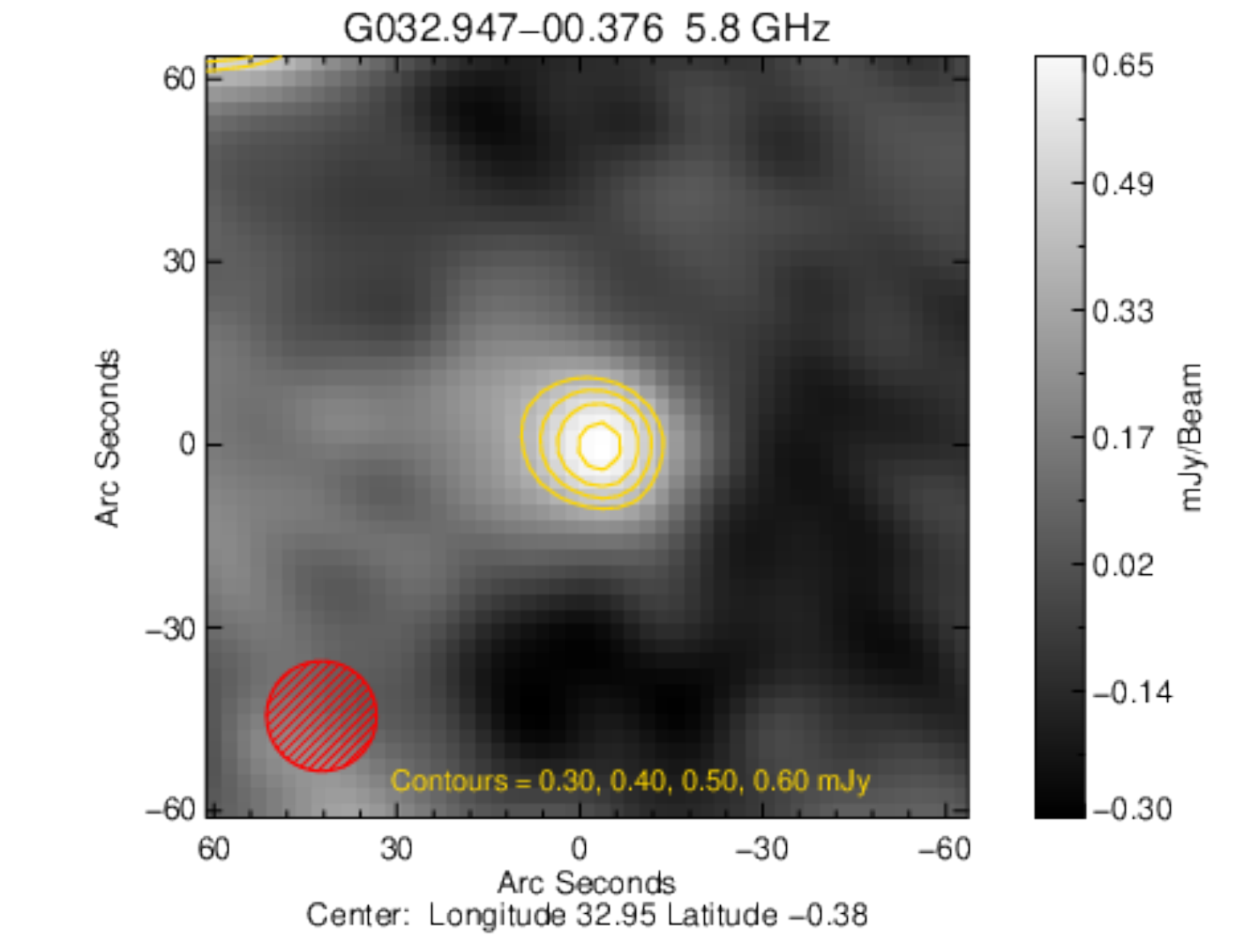}

\caption{In the upper panel we show an example of an instance where a bright extended source has increased the local noise such that a nearby compact 7$\sigma$ source located to the south-west has been missed. The position of this source is indicated by the cyan arrow. In the lower panel we show a smaller region centred on the missed compact source.}
\label{fig:misses_sources}
\end{figure}

Visual inspection of the extracted sources and comparison with other catalogs (some of which are discussed in Sect.\,\ref{sect:complementary_surveys}) revealed a small number of compact (15), low surface brightness sources (with peak flux $\sim$1\,mJy\,beam$^{-1}$) that are located near regions of bright extended sources. An example is the source in Fig.\,\ref{fig:misses_sources}. \blobcat\ does not extract these objects due to the higher localised noise surrounding them and poor dynamic range of snapshot interferometric observations. We have calculated their physical properties again using the \casa\ task \imfit. This performs an elliptical Gaussian fit to the region identified around the source. The source parameters for these objects are included in the final catalog presented in Table\,\ref{tbl:glostar_cat} but, as with the split sources, we also include a flag to the entry so these can be easily identified.

\subsubsection{Large scale structures}

\setlength{\tabcolsep}{1pt}
\begin{table}[!th]
\footnotesize
  \begin{center}
  \caption{Large scale structures. The source name is constructed from the central position of the box used to encapsulate the regions. The sources with * are associate with supernova remnants from the catalog of \cite{green2014}.}
  \label{tbl:LargeStruc}
 \begin{tabular}{l.....}\hline\hline
Name      & \multicolumn{1}{c}{$\ell_{\rm min}$} &\multicolumn{1}{c}{$\ell_{\rm max}$} &  \multicolumn{1}{c}{$b_{\rm min}$} & \multicolumn{1}{c}{$b_{\rm max}$} & \multicolumn{1}{c}{Int flux}\\ 
      & \multicolumn{1}{c}{(\degr)} & \multicolumn{1}{c}{(\degr)} &  \multicolumn{1}{c}{(\degr)} &  \multicolumn{1}{c}{(\degr)} & \multicolumn{1}{c}{(Jy)}\\ 
 \hline
G028.026$-$00.045	&	27.970	&	28.082	&	-0.100	&	0.010	&	0.32	\\
G028.520+00.132	&	28.470	&	28.570	&	0.049	&	0.215	&	0.30	\\
G028.6$-$00.1 (SNR)$^{\star}$	&	28.544	&	28.694	&	-0.188	&	-0.038	&	0.95	\\
G029.087$-$00.682	&	28.976	&	29.199	&	-0.756	&	-0.608	&	0.22	\\
G029.219+00.415	&	29.095	&	29.343	&	0.344	&	0.486	&	0.50	\\
G029.6+00.1 (SNR)$^{\star}$	&	29.512	&	29.612	&	0.071	&	0.165	&	0.13	\\
W43$-$south  center	&	29.805	&	30.131	&	-0.195	&	0.071	&	9.08	\\
G029.986$-$00.582	&	29.913	&	30.060	&	-0.650	&	-0.513	&	0.11	\\
G030.462+00.449	&	30.405	&	30.519	&	0.393	&	0.506	&	0.16	\\
W43 center	&	30.525	&	30.908	&	-0.238	&	0.080	&	13.43	\\
G031.053+00.482	&	30.992	&	31.115	&	0.435	&	0.529	&	0.59	\\
G031.166$-$00.106	&	31.117	&	31.214	&	-0.050	&	-0.162	&	0.42	\\
G031.411$-$00.234	&	31.355	&	31.467	&	-0.286	&	-0.183	&	0.60	\\
G031.5$-$00.6 (SNR)$^{\star}$	&	31.400	&	31.700	&	-0.790	&	-0.490	&	0.14	\\
G031.9+00.0 (SNR)$^{\star}$	&	31.821	&	31.920	&	-0.053	&	0.071	&	1.01	\\
G032.4+00.1 (SNR)$^{\star}$	&	32.322	&	32.488	&	0.027	&	0.172	&	0.10	\\
G032.586$-$00.172	&	32.534	&	32.637	&	-0.284	&	-0.059	&	0.26	\\
G032.8$-$00.1 (SNR)$^{\star}$	&	32.649	&	32.950	&	-0.269	&	0.115	&	0.67	\\
G033.179$-$00.010	&	33.118	&	33.240	&	-0.075	&	0.055	&	0.66	\\
G033.2$-$00.6 (SNR)$^{\star}$	&	33.033	&	33.327	&	-0.708	&	-0.395	&	0.09	\\
G033.6+00.1 (SNR)$^{\star}$	&	33.567	&	33.781	&	-0.067	&	0.132	&	0.93	\\
G034.260+00.125	&	34.177	&	34.343	&	0.043	&	0.207	&	10.17	\\
W44 $-$ G034.6$-$00.5 (SNR)$^{\star}$	&	34.390	&	34.930	&	-0.760	&	-0.093	&	2.96	\\
G035.032$-$00.283	&	34.994	&	35.070	&	-0.363	&	-0.204	&	0.04	\\
G035.506$-$00.026	&	35.316	&	35.696	&	-0.157	&	0.106	&	2.36	\\
G035.6$-$00.4 (SNR)$^{\star}$	&	35.502	&	35.692	&	-0.593	&	-0.258	&	2.36	\\
G035.680$-$00.868	&	35.610	&	35.750	&	-0.943	&	-0.792	&	0.19	\\
\hline
 
\end{tabular}
\end{center}
\end{table}

\setlength{\tabcolsep}{6pt}

\label{sect:large_scale_structures}

\begin{figure*}
\centering
\includegraphics[width=0.49\textwidth, trim= 0 0 0 0, angle=0]{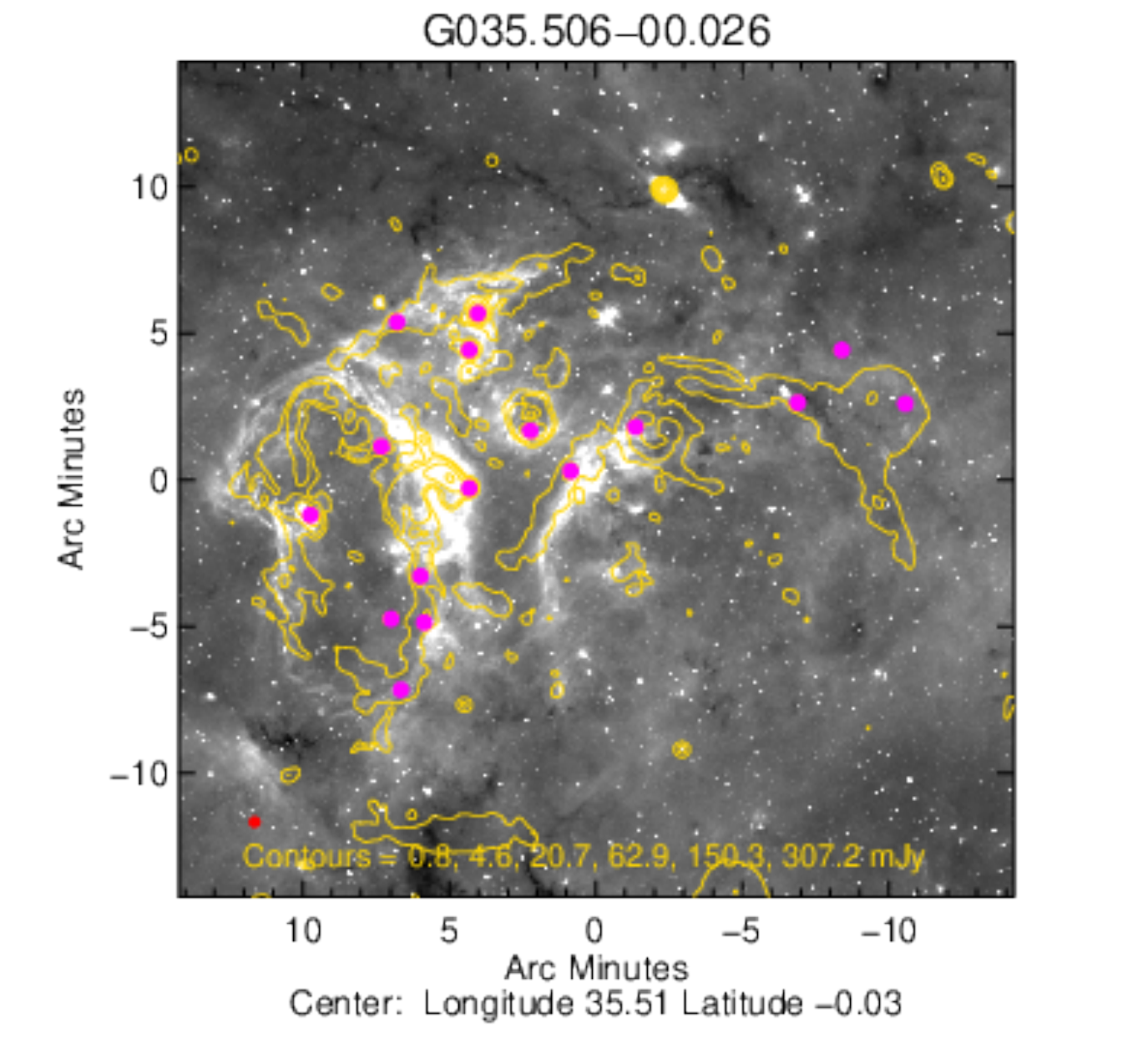}
\includegraphics[width=0.49\textwidth, trim= 0 0 0 0, angle=0]{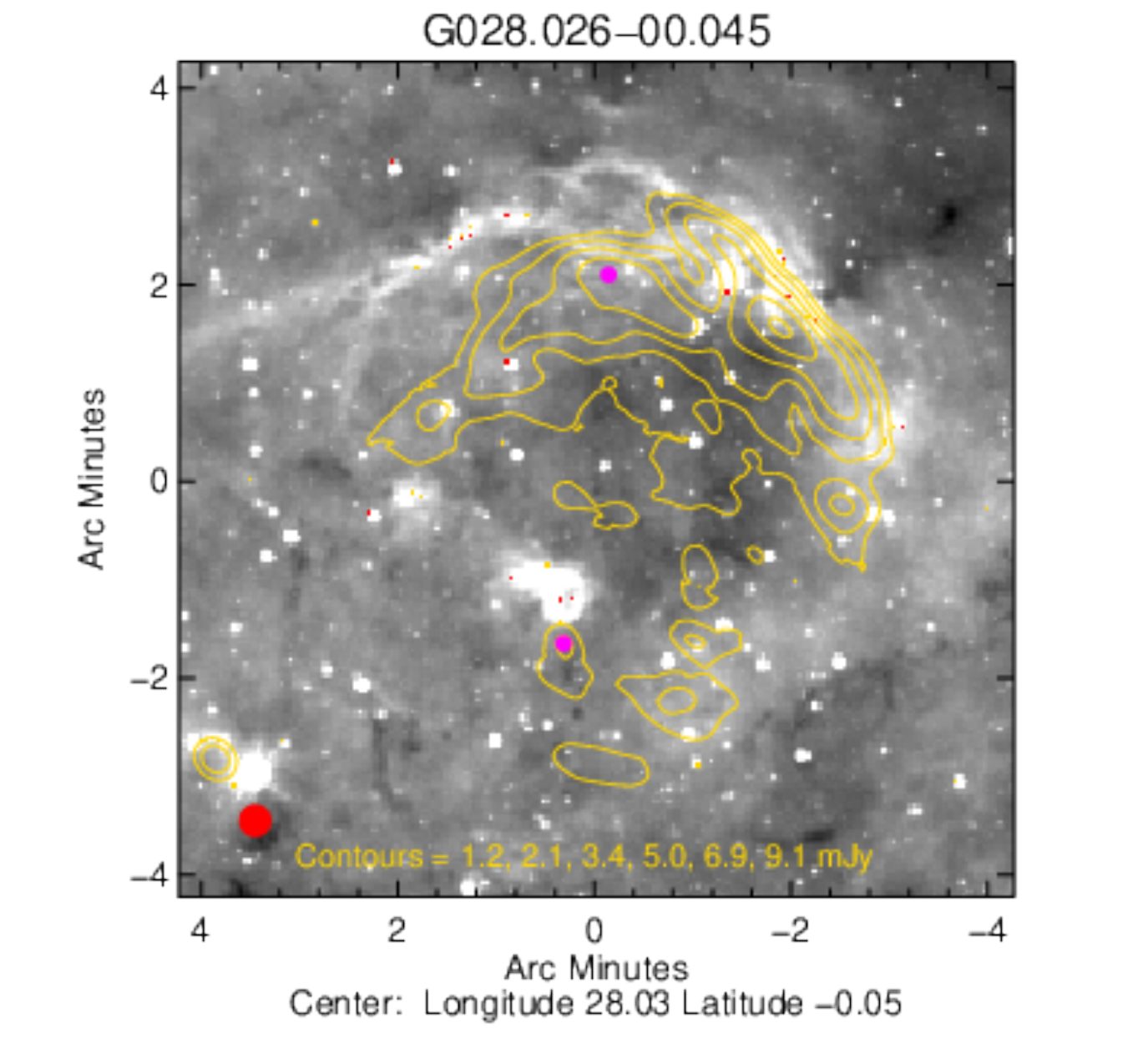}
\includegraphics[width=0.49\textwidth, trim= 0 0 0 0, angle=0]{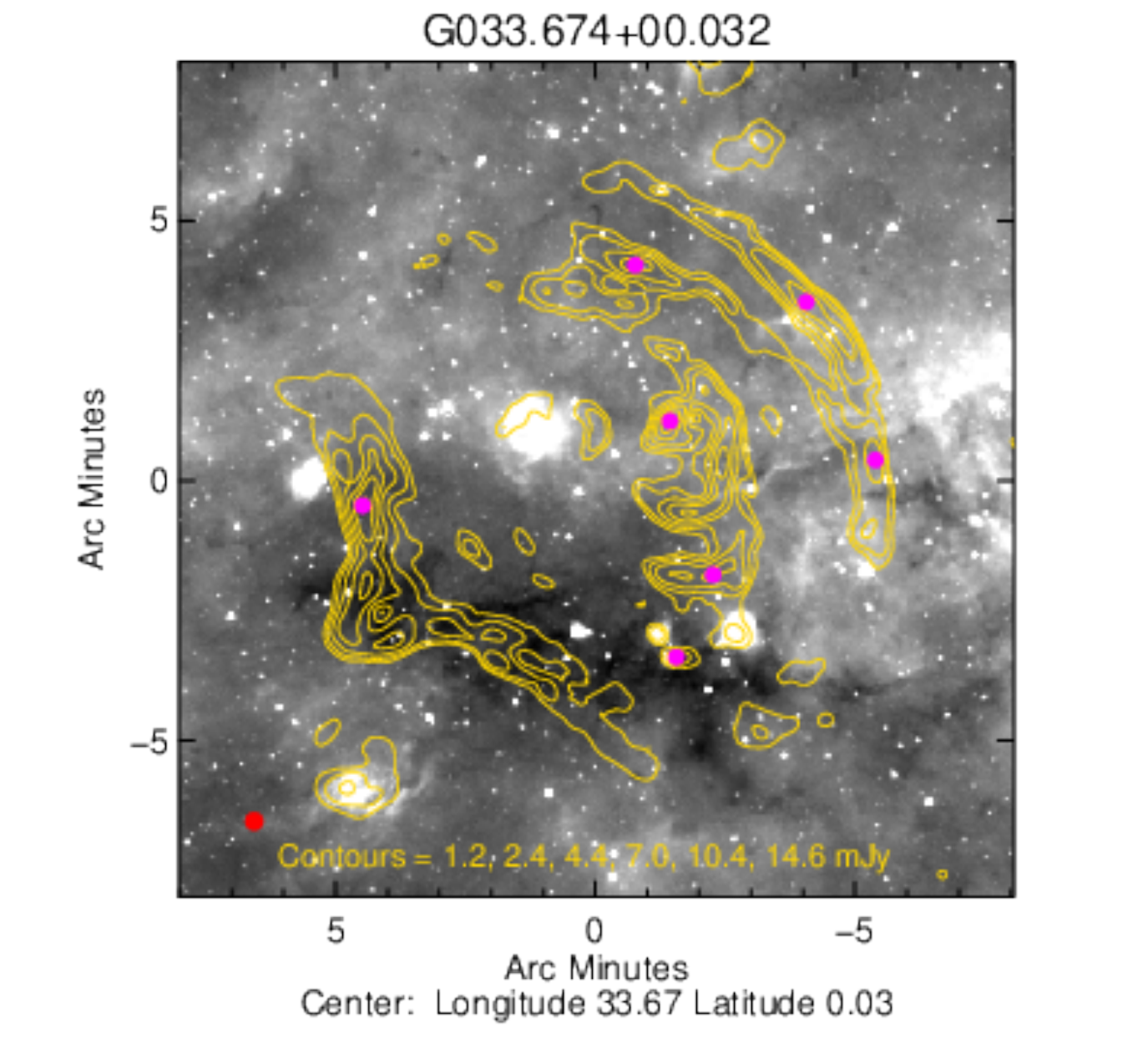}
\includegraphics[width=0.49\textwidth, trim= 0 0 0 0, angle=0]{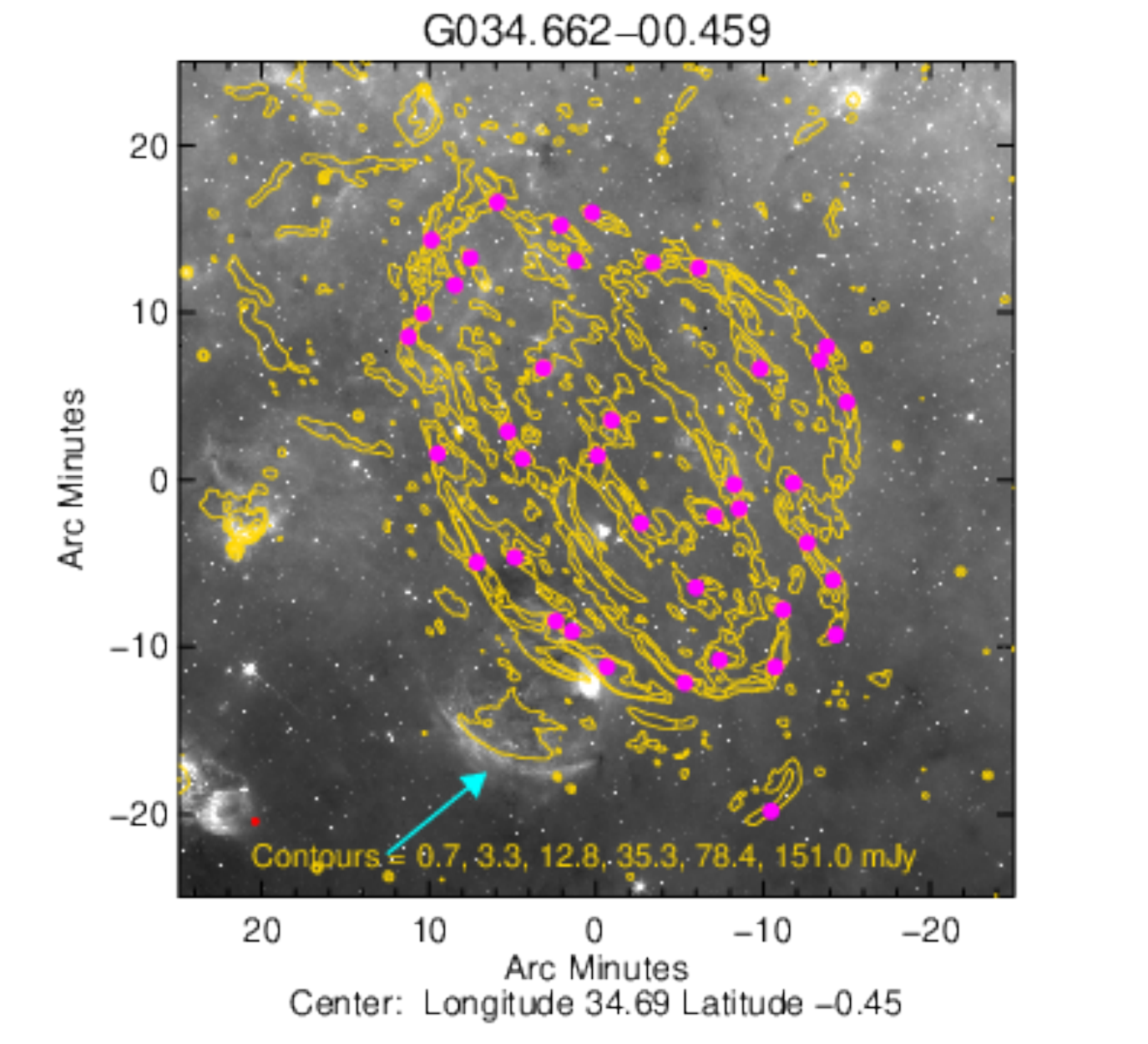}

\caption{Example of radio emission {associated with} large-scale structures. The background images are GLIMPSE 8\,$\mu$m while the yellow contours are the GLOSTAR 5.8\,GHz radio continuum emission. These regions often show coherent infrared structures that are morphologically correlated with {most of} the radio emission (see examples presented in the upper panels), however, the correlation is not always present but the radio emission is clearly correlated (see examples presented in the lower panels). These large-scale structures have been excluded from the final catalog of radio sources. The contours are determined as described in Fig.\,\ref{fig.issues} and the red filled circle shown in the lower left corners shows the GLOSTAR beam. The magenta circles indicate the peak positions of radio sources associated with these regions. In the lower-right panel the W44 SNR is shown; the cyan arrow points to the \hii\ region (G034.793$-$00.711) located close to the edge of this complex (see text for details).} 
\label{fig.zoom_maps}
\end{figure*}

We have identified 27 regions with large-scale emission by comparing the distribution of radio emission to that of the 8\,\mum\ maps extracted from the GLIMPSE archive. We have drawn a box around the coherent infrared structures that are morphologically correlated with the radio emission, but for which the radio emission has been broken up into 2 or more components. These 27 regions are indicated by squares in the upper and lower panels of Fig.\,\ref{fig.integrated_map}. It is clear that these regions are also correlated with the noisiest regions in the map where much of the large angular scale emission is poorly recovered by the observations. In Table\,\ref{tbl:LargeStruc} we give the name of each structure, the extent of the boxed region in $\ell$ and $b$ and the integrated flux of the total emission in the box minus the emission from any compact sources that are considered distinct sources.  The name of {each} structure has been constructed from the central coordinate of the enclosing box.

In Fig.\,\ref{fig.zoom_maps} we show some examples of these large scale structures. All extended radio sources in these boxed regions are considered to be part of the large-scale structure and have been excluded from the final catalog of radio sources. However, compact radio sources that are considered likely to be associated with discrete objects and not part of the larger-scale structure have been retained in the final catalog as distinct sources. 

In total, \larges sources are associated with the 27 large-scale structures identified. 
Of the 27 large scale regions identified, two stand out as being particularly significant; these are the W43 star forming complex ($29\degr \leq \ell \leq 32\degr$ and $|b| \leq 1\degr$; \citealt{Blum1999}) and the well known supernova remnant  W44 (\citealt{Clark1975}). Between them they contain 92 of the {195} individual fragments that have been associated with large scale emission regions. We note that radio recombination lines have been detected towards a radio source (G034.793$-$00.711) seen towards the south-eastern edge of W44 (\citealt{lockman1989}; G34.758-0.681). In the lower right panel of Fig.\,\ref{fig.zoom_maps} we present a map of the radio emission seen towards W44 and indicate the position of this nearby \hii\ region with an arrow. The association of RRL emission with this radio source provides strong evidence that this source is a \hii\ region that {is coincident with} the edge of this supernova remnant (G034.8$-$00.7; see \citealt{ortega2010} for detailed discussion of this \hii\ region) rather than part of W44 itself. Therefore, we treat this as a separate source, which we classify as a \hii\ region.\\

The W43 region has two prominent components: W43-main and W43-south. W43-main hosts a giant \hii\ region \citep{Blum1999} and has been classified as a mini star-burst region (\citealt{Motte2003}). Due to the complexity of the region and our limited  $uv$-coverage, the central parts of both regions are not well reconstructed in the radio emission map. That could {also be} a reason for the different integrated flux density measurement of the central part of W43 main between the THOR team, 55.55\,Jy at 2\,GHz (\citealt{bihr2016}), and our value, 13.43\,Jy at 6.0\,GHz. However, our observations have been able to recover very well much of the emission from the shell structure of W44 (see lower-right panel of Fig.\,\ref{fig.zoom_maps}). This extended object is $\sim$30\,pc across with thermal {x-ray} emission seen towards its center \citep{Clark1975}. We measure {an} integrated flux density of 2.96\,Jy, which is significantly larger than the value of 1.2\,Jy determined by \citet{Wynn-Williams1981} from 5\,GHz VLA observations; this new integrated flux density estimation is due to our higher sensitivity to more extended emission. 

Our sample of large-scale emission structures also includes three other well known regions. The first of these is the very bright ultra-compact \hii\ region G34.3$+$0.2 (G034.260+00.125 in Table\,\ref{tbl:LargeStruc}). {\citet{Garay1986} estimated the integrated} flux density at 5 and 8\,GHz, {to be} $\sim$1.5\,Jy and $\sim$2\,Jy respectively. We calculate the flux to be significantly higher at $\sim$10\,Jy but we are unable to fully resolve the structure of this object with the resolution of the current GLOSTAR data. Our higher integrated flux density determination is likely to be caused by our sensitivity to larger scale emission that its resolved out in the higher angular resolution data of \citet{Garay1986}. The other two regions worth mentioning are  G035.506$-$00.026 and G030.462+00.449, both of which have counterparts in the catalog of star-forming regions reported by \citet{RahmanMurray2010} (complexes 17 and 23 respectively). The  W44 and W43 regions have also counterparts in this catalog. Therefore, we have recovered  four of the most luminous free-free sources in the Galaxy.

Searching the literature we find that 10 of these large scale structures, including W44, have been previously identified as SNR (\citealt{green2014}); this corresponds to over a third of this sample. Where a match has been made we have adopted the catalog name from \citet{green2014} in Table\,\ref{tbl:LargeStruc}. 

\setlength{\tabcolsep}{3pt}
\begin{table}[!th]
\footnotesize
  \caption{Summary of detection categories. \label{tab:id_breakdown}}
 \begin{tabular}{lc}\hline\hline
Description   	   & Number \\ 
			       &  of sources\\ 
 \hline
Total number of sources extracted & \blobs \\
Number of sources assoc. with large structures & {195} \\
Number of split sources  & 20 (44)\\
Number of sources recovered near bright sources & 15 \\
Number of sources rejected  & 237 \\
\hline
Number of source in final catalog  & 1575  \\
\hline

\end{tabular}
\end{table}
\setlength{\tabcolsep}{6pt}

\subsection{Catalog completeness and reliability}\label{completeness}

The \blobcat\ source extraction method was verified by the THOR team (\citealt{bihr2015, wang2018}). They performed a detailed set of completeness tests. They choose a region of $0.5\times0.5$ degrees with a constant noise level and added artificial 2D Gaussian sources with the size of the beam and different peak intensities. They measured the fraction of sources recovered as a function of intensity and found that at a threshold of 7$\sigma$ the extraction algorithm was 95\% complete. Given that our data are similar to the THOR data (both are VLA continuum surveys and have similar resolutions) we have simply adopted their results. 

The CORNISH team used a similar flux threshold (7.2\,$\sigma$; \citealt{purcell2013}) as the minimum required for inclusion in their catalog. This was determined by calculating the number of spurious detections expected as a function of $\sigma$ and set at a level above which no spurious detections would be expected (see discussion by \citealt{Medina2018}). The CORNISH survey consisted of $5.6 \times 10^8$\,beams and so a high threshold was required to avoid any false detections. The region presented here consists of approximately 1000 times fewer beams ($\sim5.65\times 10^5$\,beams) and so we can set our threshold for reliable sources at a lower level. Based on Gaussian statistics, the number of spurious sources we would expect in our catalog using 5$\sigma$ flux threshold is $<$ 1 and so this is the level above which sources in the catalog are considered to be highly reliable. We find another $\sim$250 radio sources between 4 and 5$\sigma$ of which we estimate 7\% are likely to be spurious ($\sim$17 sources); we include these in the final catalog as the vast majority are likely to be real sources and have counterparts in other surveys, but these are considered to be less reliable.

\section{Verification of the Source Catalog}
\label{sect:source_verification}

As discussed in the previous section, only a small number of spurious sources are expected in our final catalog. However, due to the nature of these interferometric snapshot observations, the $uv$-plane is relatively sparsely sampled, which can result in significant {number of} imaging artifacts in the restored maps. As a consequence the number of spurious sources expected from Gaussian statistics is, therefore, a lower limit to the actual number of spurious sources {detected}.

To improve the reliability of the catalog produced by \blobcat\ we need to identify any spurious noise peaks and imaging artifacts (sidelobes and emission from over-resolved structures). We have already taken emission from over-resolved structures into account by identifying the large-scale structures as described in Sect.\,\ref{sect:large_scale_structures}; this reduces the number of sources in the catalog to {\color{black}1819}. Next we assume that any radio source that has a counterpart in a published radio or mid-infrared catalog is real; this is because it is extremely unlikely that a false detection would randomly appear in at same position in other independent surveys. 

\subsection{Complementary surveys}
\label{sect:complementary_surveys}

A number of complementary multi-wavelength surveys have produced {catalogs of sources located in} the Galactic plane that we have used to verify if the sources recovered from the GLOSTAR map are real and to help determine their nature. Below we provide a brief summary of the surveys used in this paper. In all cases we have used a 10\arcsec\ radius to identify matches with the exception of the ATLASGAL dust emission; this is because the dust emission tends to be more extended.  The surveys selected are either well matched in resolution (e.g. WISE, ATLASGAL, MAGPIS and THOR, with angular resolutions of 6-19\arcsec) or in wavelength (e.g. CORNISH at 5\, GHz).

\setlength{\tabcolsep}{6pt}
\begin{table*}[!th]
\footnotesize
  \begin{center}
  \caption{Statistics of the matches between GLOSTAR and other published surveys. A match radius of 10\arcsec\ has been used centred on the peak of the GLOSTAR emission except for the ATLASGAL dust emission which is more extended than the radio and mid-infrared catalogs.}
 \begin{tabular}{lrrrrrr}\hline\hline
Survey & Wavelength & Resolution &\emph{rms} level       & Num. of sources & Num. of  \\ 
       &            & (\arcsec)  &mJy beam$^{-1}$ & in region        & matches	\\ 

\hline
CORNISH          &6\,cm     & 1.5   &0.33 & 353     &   327 \\
MAGPIS        &  6\,cm		& 6-9   &0.3  & 218     &  128 \\
THOR              &20\,cm 	& 10-25 &0.3-1 & 1036    &   772 \\
ATLASGAL & 870\,$\mu$m 		& 19.2  &50-70 & 747    &   132 \\
WISE    &3.4-22\,$\mu$m 	&6-12   & $\cdots$ & $\cdots$    &   988 \\

\hline
\label{tab:cat_matches}
\end{tabular}
\end{center}
\end{table*}

\setlength{\tabcolsep}{6pt}

\subsubsection{CORNISH}
\label{sect:cornish_matches}

\begin{figure}
\centering
\includegraphics[width=0.49\textwidth, trim= 0 0 0 0, angle=0]{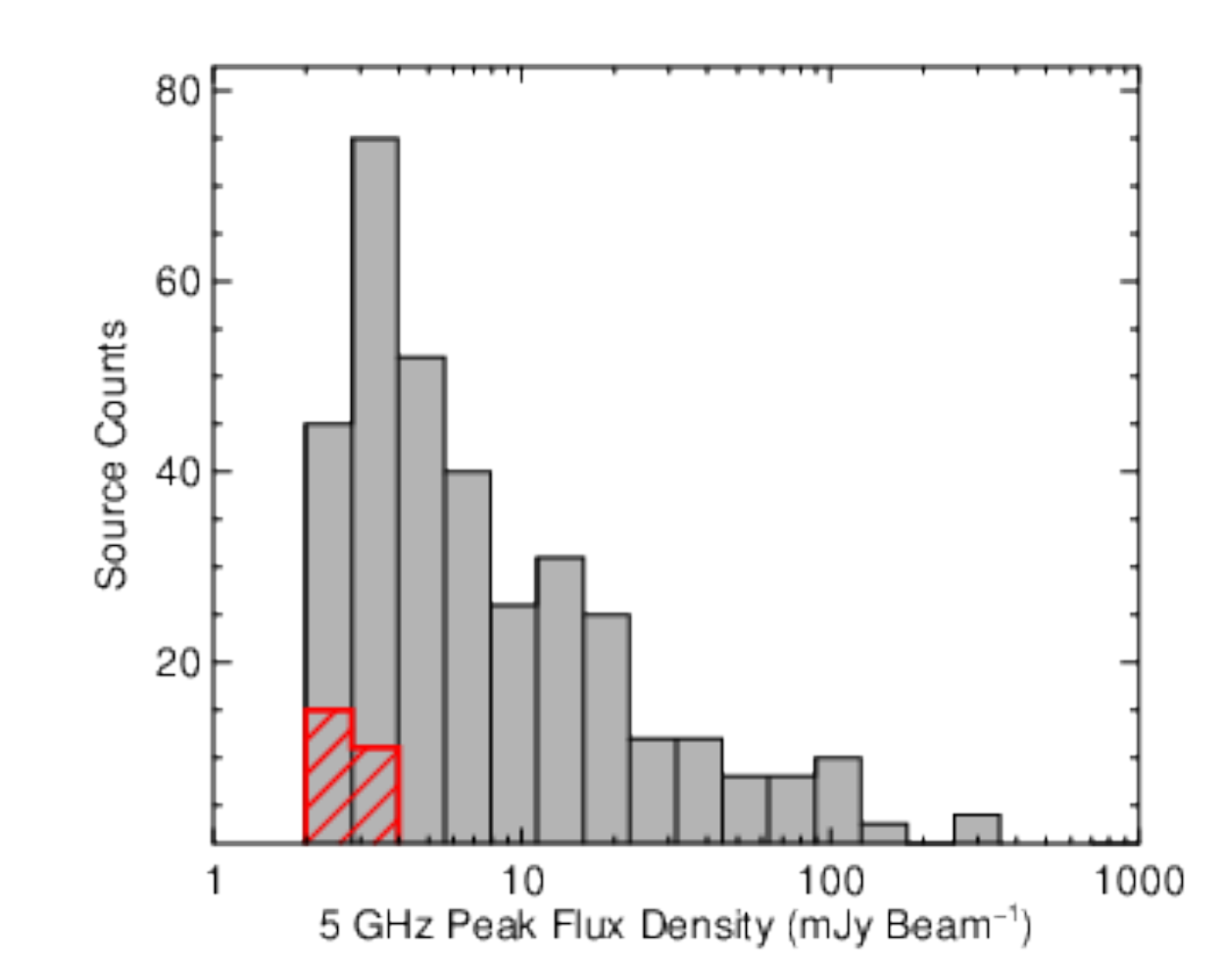}
\caption{Peak flux distribution of CORNISH sources located in the GLOSTAR field. The grey and red hatched histograms show the distribution of all CORNISH sources and those without a GLOSTAR 
counterpart, respectively. The bin width is 0.15\,dex.}
\label{fig:cornish_flux_distribution}
\end{figure}

The Co-Ordinated Radio 'N' Infrared Survey for High-mass star formation (CORNISH; \citealt{hoare2012}) is a sensitive and high-resolution 5\,GHz radio continuum survey that focuses on the northern GLIMPSE region ($10\degr < \ell<65\degr$ and $|b| < 1\degr$). The survey was conducted using the Very Large Array before its upgrade in B-configuration, which provides an angular resolution of $1\as5$ and has a 7.2$\sigma$ sensitivity of $\sim$2.7\,mJy\,beam$^{-1}$.  

\begin{figure}
\centering
\includegraphics[width=0.49\textwidth, trim= 0 0 0 0, angle=0]{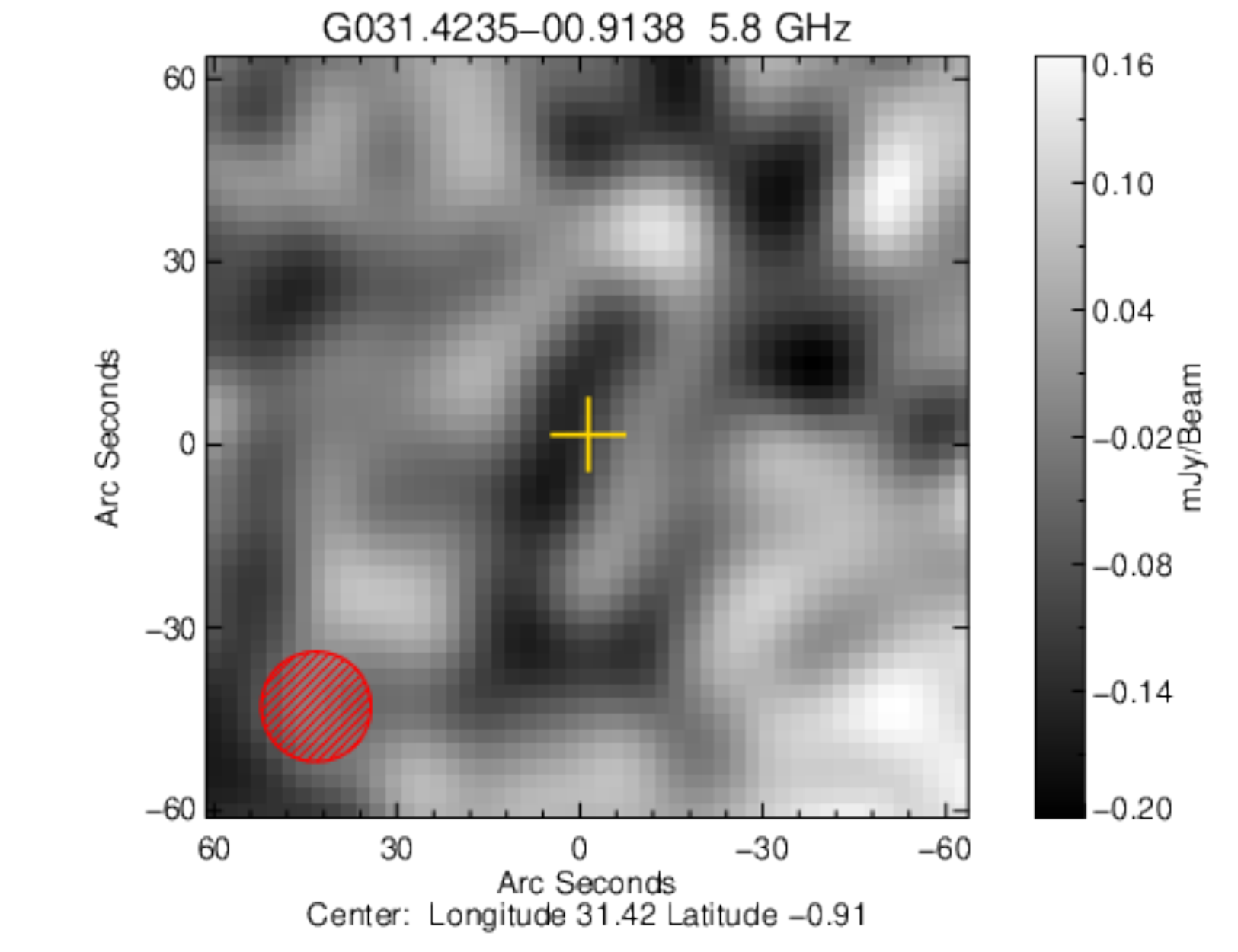}
\caption{Example of a CORNISH source not detected in GLOSTAR. The greyscale is the GLOSTAR map of a 2$\arcmin\times2$\arcmin\ region centered on the position of the CORNISH source G031.4235$-$00.9138 (as indicated by the yellow cross). The noise towards this source is 63\,$\mu$Jy\,beam$^{-1}$ and given that the integrated CORNISH flux is 3.47\,mJy it should have been easily detected in the GLOSTAR map. The red hatched circle shown in the lower left corner indicated the size of the GLOSTAR beam.}
\label{fig:cornish_source_example}
\end{figure}

The CORNISH catalog consists of $\sim$2600 sources above 7$\sigma$ (\citealt{purcell2013}), 353 of which are located in the GLOSTAR region presented in this paper. We find a GLOSTAR counterpart for 327 of these ($\sim$80\%). Inspection of the emission maps for the brighter CORNISH sources reveals that {20 more are} associated with extended or {multi-peak} emission sources found in GLOSTAR where the positions are less well constrained and consequently these sources were not matched.  In Fig.\,\ref{fig:cornish_flux_distribution} we show the peak flux distribution of all 353 CORNISH sources and the 26 sources where we have failed to find a GLOSTAR counterpart using either 10\arcsec\ search radius or through inspection of the emission maps. It is clear from this plot that all of the radio sources that are not picked up by GLOSTAR correspond to the weakest CORNISH sources. We have estimated the noise values from the GLOSTAR maps towards these 26 CORNISH position and find a mean noise value of 166\,$\mu$Jy beam$^{-1}$ and a standard deviation of 154\,$\mu$Jy beam$^{-1}$ (minimum and maximum noise values are 47 and 768\,$\mu$Jy beam$^{-1}$, respectively), which is significantly lower than the typical CORNISH 1$\sigma$ {(0.33 mJy beam$^{-1}$)} noise values (see Fig.\,\ref{fig:cornish_source_example}). 

\setlength{\tabcolsep}{3pt}

\begin{table}[!th]
\footnotesize
  \begin{center}
  \caption{Catalog names and fluxes for CORNISH sources not detected in GLOSTAR. The last column gives the measured \emph{rms} towards the position of the CORNISH source in the GLOSTAR map. The flux errors are of the order of the CORNISH noise level,  0.33 mJy beam$^{-1}$.} 
 \begin{tabular}{l...}\hline\hline
CORNISH        &  \multicolumn{1}{c}{$S_{\rm peak}$} & \multicolumn{1}{c}{$S_{\rm int}$} & \multicolumn{1}{c}{GLOSTAR rms} \\ 
  name        	&  \multicolumn{1}{c}{(mJy\,beam$^{-1}$)} & \multicolumn{1}{c}{(mJy)}		& \multicolumn{1}{c}{(mJy\,beam$^{-1}$)}			\\ 
\hline
G028.1875+00.5047	&	2.61	&	2.61	&	0.083	\\
G028.3660$-$00.9640	&	2.92	&	2.92	&	0.073	\\
G028.5690+00.0813	&	2.99	&	5.34	&	0.359	\\
G028.9064+00.2548	&	3.47	&	3.47	&	0.768	\\
G029.2555$-$00.8653	&	2.37	&	3.16	&	0.084	\\
G029.2620+00.2916	&	2.76	&	2.76	&	0.120	\\
G029.2648+00.9527	&	2.59	&	2.97	&	0.060	\\
G029.3096+00.5124	&	2.84	&	2.84	&	0.072	\\
G029.4302$-$00.9967	&	2.83	&	2.83	&	0.103	\\
G029.4404$-$00.3199	&	2.81	&	2.81	&	0.160	\\
G029.5184+00.9478	&	2.43	&	2.43	&	0.065	\\
G029.7805$-$00.2661	&	3.35	&	7.53	&	0.451	\\
G030.1884+00.1110	&	3.00	&	4.45	&	0.223	\\
G030.2193+00.6501	&	2.80	&	2.80	&	0.122	\\
G030.4543+00.3223	&	2.69	&	3.74	&	0.195	\\
G030.6328$-$00.7232	&	2.57	&	2.57	&	0.163	\\
G031.2859$-$00.2095	&	2.43	&	3.76	&	0.286	\\
G031.4235$-$00.9138	&	2.66	&	3.47	&	0.063	\\
G032.2408+00.1667	&	3.06	&	3.66	&	0.218	\\
G032.2783$-$00.1705	&	2.84	&	4.80	&	0.152	\\
G033.9622$-$00.4966	&	2.81	&	2.81	&	0.062	\\
G034.1382+00.3805	&	2.61	&	3.23	&	0.139	\\
G034.2171$-$00.6886	&	2.16	&	2.37	&	0.077	\\
G034.3555$-$00.0876	&	2.98	&	4.61	&	0.087	\\
G034.3852+00.3526	&	3.89	&	4.66	&	0.105	\\
G035.0605+00.6208	&	2.76	&	2.76	&	0.047	\\

\hline
 \label{tab:cornish_nondectections}
\end{tabular}

\end{center}
\end{table}

There are two possible explanations for these missing sources: 1) the flux density is  variable and has decreased between the time the CORNISH observations were completed and the GLOSTAR observations were made; and 2) these are spurious CORNISH detections. The variability of radio sources has been investigated by \citet{kalcheva2018} from an analysis of the CORNISH and MAGPIS integrated flux densities (we will discuss the MAGPIS survey in the next subsection).  \citet{kalcheva2018} estimate that $\sim$5\% of the CORNISH \uchii\ region sample are potentially variable sources (flux increases of over 50\,\% over a 15 year time period), which is consistent with the number of CORNISH detections that do not have a matching GLOSTAR counterpart ($\sim$\,7$\%$). In Table\,\ref{tab:cornish_nondectections} we provide the source name and integrated and peak flux densities of the unmatched CORNISH sources (taken from \citealt{purcell2013}) along the local \emph{rms} noise determined from the GLOSTAR maps.

In the case of variable sources, we should expect to find {approximately} as many sources with significantly larger fluxes in CORNISH than measured in GLOSTAR and comparing the fluxes we find 7 matches for which this is the case (following \citealt{kalcheva2018} we set this criterion to be $>$ 50\% more flux). The radio names of these sources and the flux ratio are given in Table\,\ref{tab:variable_sources}.  The discrepancy between this number and that of the number of CORNISH sources not detected in GLOSTAR (Table\,\ref{tab:cornish_nondectections}) suggests that variability alone cannot fully account for the 26 CORNISH source not detected in GLOSTAR. The highest flux ratio is 3 and so the differences are relatively modest compared to those reported by \citet{kalcheva2018} (range between 2.4 and 73.9). All of the sources are either marginally detected or unresolved, have integrated fluxes below 10\,mJy, and all but one have negative spectral indices, and so the variability appears to be restricted to weak and unresolved extragalactic sources. We note that these radio properties are similar to those reported for T Tauri stars (e.g., \citealt{Dzib2013}), however, given these radio sources appear to be mid-infrared dark this possibility is considered less likely. 


\citet{kalcheva2018} considered only \hii\ regions in their analysis, whereas none of the variable candidates we have identified have been classified as being \hii\ regions, PNe or radio stars. However, \citet{kalcheva2018} only found one variable \hii\ region in the region being considered here and given the shorter temporal baseline ($\sim$7\,yrs) the lack of variable \hii\ regions and the lower levels of variation are not surprising.


\setlength{\tabcolsep}{3pt}

\begin{table}[!th]
\footnotesize
  \begin{center}
  \caption{Variable source candidates where the integrated CORNISH flux is more than 50\% higher than the peak GLOSTAR flux.}
 \begin{tabular}{ll.}\hline\hline
GLOSTAR name& CORNISH   name     &  \multicolumn{1}{c}{$S_{\rm peak, GLOSTAR}/S_{\rm int, CORNISH}$}  \\ 
\hline
G029.589+00.579	&	G029.5893+00.5789	&	2.17	\\	
G030.104+00.399$^*$	&	G030.1039+00.3983	&	1.53	\\	
G031.585$-$00.063	&	G031.5854$-$00.0635	&	1.94	\\	
G032.453+00.368	&	G032.4535+00.3679	&	2.02	\\	
G032.599+00.826	&	G032.5996+00.8265	&	1.86	\\	
G034.178+00.257	&	G034.1782+00.2564	&	1.51	\\	
G034.265+00.719	&	G034.2655+00.7195	&	3.05	\\	
\hline
 \label{tab:variable_sources}
\end{tabular}

\end{center}
$^*$ The source was previously reported as a variable radio source by \citet{Becker2010}.

\end{table}

\setlength{\tabcolsep}{6pt}

\subsubsection{MAGPIS}

One of the first VLA surveys of a large fraction of the Galactic plane at 5\,GHz was conducted in the early 1990s ($350\degr < \ell < 40\degr$ and $|b| < 0.4\degr$; \citealt{becker1994}). This survey is now part of a larger collection of re-reduced radio surveys known as the Multi-Array Galactic Plane Imaging Survey (MAGPIS; \citealt{helfand2006}).\footnote{https://third.ucllnl.org/gps/} This part of the MAGPIS survey has a resolution of 6-9\arcsec\ and a 5$\sigma$ sensitivity of $\sim$2.5\,mJy\,beam$^{-1}$. This reprocessing has resulted in the identification of 1272 discrete sources including 218 that are located within the GLOSTAR region presented in this paper. Of these, we have found that {\color{black}128} are coincident with 125 GLOSTAR sources (the lower resolution of GLOSTAR results in three cases where {a} GLOSTAR source is associated with 2 MAGPIS sources). A further 22 MAGPIS sources are coincident with sources associated with large scale structures discussed in Sect.\,\ref{sect:large_scale_structures}. In total, 150 MAGPIS sources are matched in GLOSTAR, which corresponds to approximately 70\% of the MAGPIS 5\,GHz sample. We have investigated a few of the MAGPIS sources not matched in GLOSTAR and find  many of these to be associated with sources that are over-resolved in MAGPIS. The offsets between the partially recovered fragments and the GLOSTAR source are larger than the 10\arcsec\ used to identify counterparts.    

\subsubsection{THOR}

The \hi, OH, Recombination line survey (THOR; \citealt{Beuther2016,bihr2015,wang2018}) is a 1.4\,GHz radio continuum and line survey of $\sim$100 deg$^2$ of the Galactic plane in the first quadrant of the Milky Way. This survey covers the 21\,cm \hi\ line, 4 OH transitions, 19 radio recombination lines, and continuum emission between 1 to 2\,GHz. The survey was conducted using the Karl G. Jansky Very Large Array (VLA) in C-array and has a spatial resolution of 10-25\arcsec. The data reduction strategy and source catalog for the first half of the survey ($\ell=14\degr$ -$37.9\degr$, $\ell=47.1\degr$ - $51.2\degr$, and $|b| \leq 1.1\degr$) are described in \citet{bihr2016}. The catalog contains a total of $\sim$4,400 sources, of which $\sim$1200 are spatially resolved and $\sim$1,000 are possible artifacts.\footnote{These sources have been incorporated in the complete THOR source catalog that consists of $\sim$10,000 sources (\citealt{wang2018}).} 

\begin{figure}
\centering

\includegraphics[width=0.49\textwidth, trim= 0 0 0 0, angle=0]{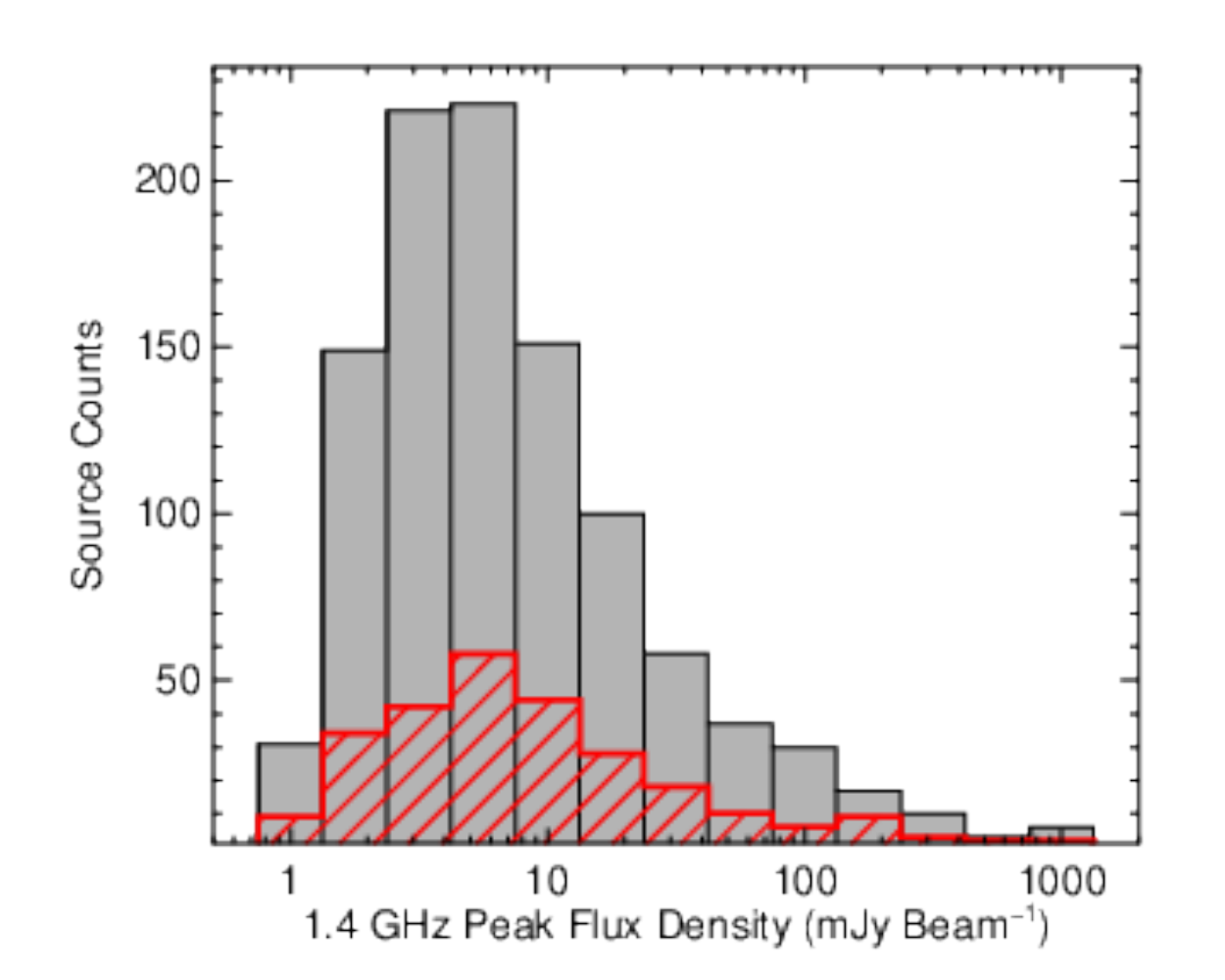}
\caption{Peak flux distribution of THOR sources located in the GLOSTAR field. The grey and red-hatched histograms show the distribution of all THOR sources and those without a GLOSTAR counterpart, respectively. The bin width is 0.25\,dex.}
\label{fig:thor_flux_distribution}
\end{figure}

A total of 1036 THOR sources are located in the region of the GLOSTAR survey described here; we are able to find a GLOSTAR counterpart for 711 of these. We have matched a further 61 THOR sources with GLOSTAR sources that have been identified as being part of large scale structures. In Fig.\,\ref{fig:thor_flux_distribution} we show the 1.4 GHz peak flux distribution of all THOR sources located in the GLOSTAR region and the flux distribution of those without a counterpart in GLOSTAR (grey and red-hatched histograms, respectively). It is clear from this plot that a higher proportion of the weaker THOR sources have not been matched. Spectral index measurements are only available in the THOR catalog for the brightest of these unmatched sources (203 of the 264 unmatched) and the majority of these have negative values (71\% have $\alpha < 0$, with a mean value = $-0.85$), which means they are weaker at 5 GHz than at 1.4 GHz and are hence less likely to be detected in the GLOSTAR map. The smaller proportion of unmatched sources that do have a positive spectral index (29\% have $\alpha > 0$, with a mean value = $0.28$) are a little harder to explain. However, we note that the THOR team acknowledge that $\sim$23\% of their catalog are possibly artifacts, which corresponds well with the fraction of unmatched sources we have found here.

\subsubsection{ATLASGAL}
\label{sect:atlasgal}

The APEX Telescope Large Area Survey of the Galaxy (ATLASGAL; \citealt{schuller2009,beuther2012}) is the first high-resolution ($\approx 20''$ FWHM) ground-based submillimetre (870\,$\mu$m) survey of the thermal dust emission in the inner Galactic plane. It has produced an unbiased  catalog of $\sim$10000 massive pre- and proto-stellar clumps (ATLASGAL Compact Source Catalog (CSC); \citealt{contreras2013, urquhart2014_csc}). Correlating the GLOSTAR and ATLASGAL catalogs is an excellent way to identify embedded or dust enshrouded objects such as \uchii\ regions and planetary nebula (e.g., {\citealt{irabor2018}}, \citealt{urquhart_radio_south,urquhart_radio_north,purcell2013,urquhart2013_cornish}).  

As previously mentioned, the dust emission is more extended than the radio sources, the majority of which are unresolved and so rather than using a 10\arcsec\ matching radius we identify matches where the radio source falls within the 4$\sigma$ boundary of the ATLASGAL dust emission. There are {\color{black}747} ATLASGAL sources located in the GLOSTAR region of which {\color{black}132} are positionally coincident with a radio source. Of these, {\color{black}92} have been previously identified as \hii\ regions (\citealt{kalcheva2018,anderson2014,giveon2005a}) and one has been identified as a PN and one as a radio star by the CORNISH team (\citealt{purcell2013}). The relatively low fraction of ATLASGAL matches is to be expected as only clumps hosting \hii\ regions will have a radio counterpart and \hii\ regions are only present in the most massive and evolved clumps.

\subsubsection{WISE} 
\label{sect:wise}

The Wide-field Infrared Survey Explorer (WISE) is a NASA infrared-wavelength astronomical space telescope mission that mapped the entire sky in four infrared bands W1, W2, W3, and W4  centered  at  3.4,  4.6,  12,  and  22\,$\mu$m using a 40\,cm telescope feeding arrays with a total of 4 million pixels; these wavelengths correspond to angular resolutions of 6.1, 6.4, 6.5 and 12\arcsec. For a full description of the mission see \cite{Wright2010}. 

The WISE ALL-sky Release Source Catalog contains astrometry and photometry for over half billion objects. We have searched this data repository to identify and extract the mid-infrared fluxes towards {988} 
GLOSTAR sources; where two or more WISE sources are found within the search radius, we have picked the {infrared counterpart to be the source} with the smallest angular offset to the peak radio emission. 

The WISE catalog is optimised for compact sources and so may overlook more extended mid-infrared emission associated with more evolved \hii\ regions. We have therefore also cross-matched the GLOSTAR sample with a catalog of \hii\ region candidates identified  from a combination of visual and automatic search of WISE data based on their MIR emission morphology \citep{anderson2014}. This catalog contains $\sim$8000 \hii\ region candidates, of which 667 are located in the GLOSTAR field. We find 130 of these mid-infrared sources to be coincident with a radio source, including 26 not found in the search made of the WISE catalog. Although we have only found a radio counterpart for $\sim$20\% of the \hii\ region candidates \citet{anderson2014} estimates that they are sensitive to \hii\ regions with radio fluxes greater than 0.2\,mJy, which is lower than our 4$\sigma$ sensitivity, which can be as poor as 1.6\,mJy towards the {Galactic} mid-plane (i.e., see Fig.\,\ref{fig.noise_histogram_fn_lat}) where most of the \hii\ regions are likely to be located. Also, objects being resolved out (> 4\arcmin) could only explain a small proportion (10\%) of unmatched WISE \hii\ regions and so the rest are weak \hii\ regions. In total, we have matched 988 radio sources with a WISE counterpart.

\subsection{Correlation with other surveys}
\label{sect:corr_other_surveys}

We have searched for counterparts in the THOR,  CORNISH and MAGPIS radio continuum surveys and the ATLASGAL dust and WISE mid-infrared surveys. 
The correlation process with other catalogs allow us to find any large scale structures and all compact sources that have a counterpart. This analysis has identified counterparts for {\color{black}1315} radio sources and these are therefore considered to be reliable detections.

The remaining {\color{black}465} radio sources are a mix of compact, moderately extended sources, imaging artefacts and spurious detections. The extended sources are also likely to be associated with extended \hii\ regions or supernova remnant. This can be confirmed by comparing the morphology of the radio emission with that at mid-infrared wavelengths. This can also identify weak mid-infrared counterparts to the compact radio sources that were not picked up in the mid-infrared catalogs. The next step is, therefore, to compare the radio and mid-infrared emission maps for these 465 sources; sources for which mid-infrared counterparts can be identified can be considered real. This visual inspection is also useful in identifying and removing imaging artifacts. 

We have compared the radio emission with mid-infrared maps extracted from the GLIMPSE  4.5 and 8.0\,$\mu$m bands, and MIPSGAL 24\,$\mu$m band archives. We have discarded any compact radio sources that appear to be localised peaks in high-noise regions, sources that are smaller than the beam (these tended to be clustered towards the detection threshold) and extended emission that appears to be associated with large-scale structures that, due to poor uv-coverage, have not been reliably imaged (these have a wider range of fluxes). This process has resulted in {\color{black}237} sources being excluded from the {\color{black}465} without a counterpart in one of the other surveys. 

Combined with the split sources and recovery (by hand) of compact sources in noisy regions, previously discussed, the final radio catalog therefore consists of  {\color{black}1575} sources. In Table\,\ref{tab:id_breakdown} we present a summary of the various steps taken in the source verification process and in Table\,\ref{tab:cat_matches} the summary of the search for counterparts with continuum, dust and mid-infrared surveys.

\begin{figure}
\centering
\includegraphics[width=0.49\textwidth, trim= 0 0 0 0, angle=0]{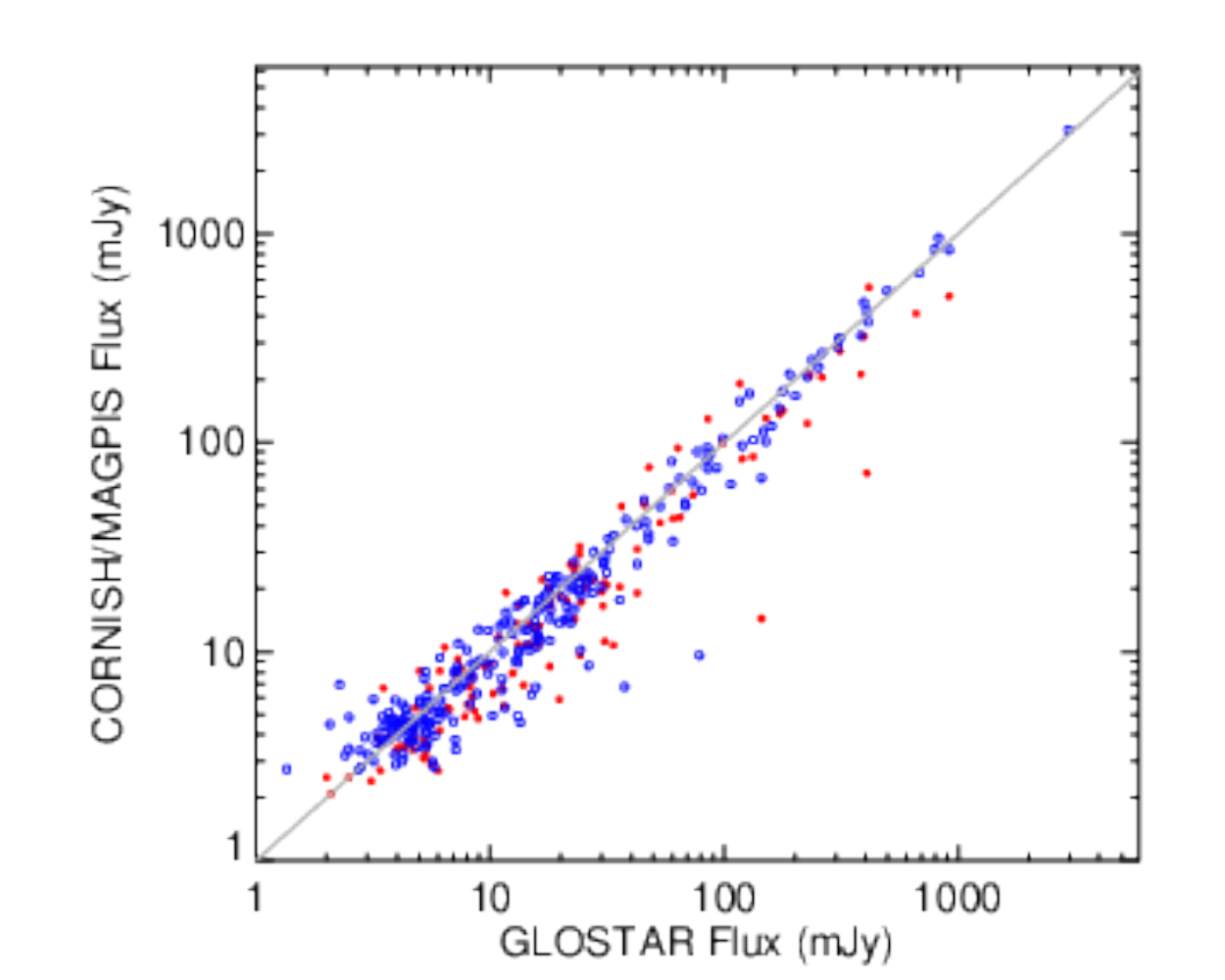}
\includegraphics[width=0.49\textwidth, trim= 0 0 0 0, angle=0]{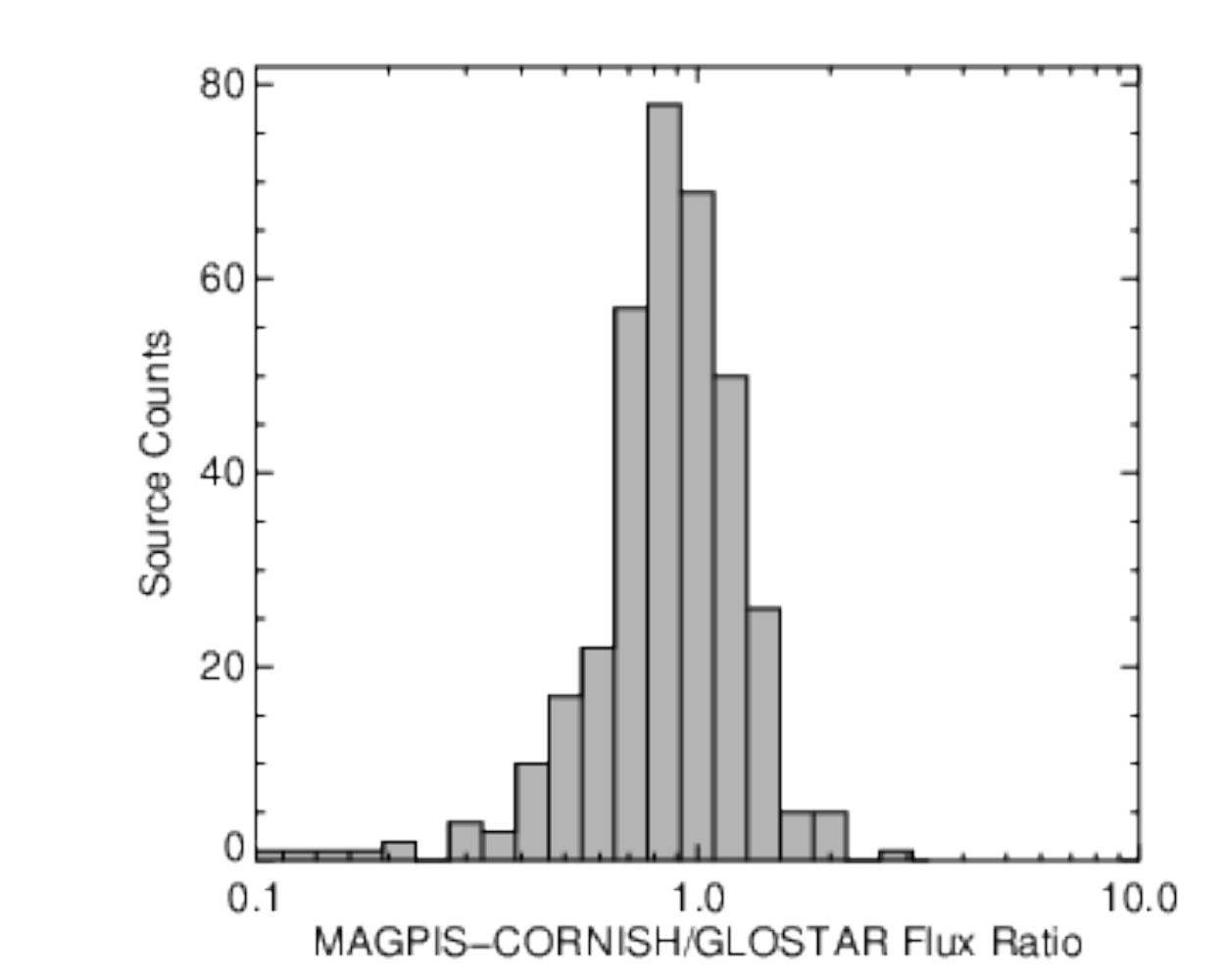}

\caption{Upper panel: Comparison of the GLOSTAR flux measurements with the \citet{becker1994} survey (MAGPIS) and CORNISH; data from these are represented as red (filled) and blue (open) circles, respectively. Due to the difference in resolution between the surveys we compare the integrated CORNISH flux densities with the peak GLOSTAR flux densities, while for the \citet{becker1994} survey we compare the integrated flux densities with the peak GLOSTAR flux densities, when the \citet{becker1994} source size is less than 20\arcsec\ and the integrated flux from both surveys if the source is larger than 20\arcsec. The gray line is the line of equality. Lower panel: Flux ratio distribution (as described above), which has a mean of 0.9 and standard deviation of 0.34.  }
\label{fig:flux_comparison}
\end{figure}

\subsection{Comparison of fluxes with other 5\,GHz surveys}

As mention earlier, we have identified matched counterparts of GLOSTAR sources in the CORNISH and MAGPIS catalogs. Although these surveys were conducted at a higher resolution than the part of the GLOSTAR survey reported here, they were conducted at a similar frequency and can hence provide a useful consistency check on the measured fluxes. Due to the different resolutions we have compared the GLOSTAR peak fluxes with the integrated fluxes from CORNISH and MAGPIS, except when the sources in those surveys are larger than the GLOSTAR beam in which case we compare the integrated fluxes from both surveys; CORNISH filters out any structures larger than $\sim$15\arcsec\ and so this is only a consideration for some of the MAGPIS sources. 

In Fig.\,\ref{fig:flux_comparison} we show the results of this flux comparison. It is clear that the fluxes from all three surveys are in very good agreement, however, it is noticeable that there is bias towards higher fluxes measured by GLOSTAR. This is to be expected given the lower resolution resulting in more flux being present in the beam and a result of some of the flux being filtered out by the high-resolution surveys due to their poorer sensitivity to larger angular scales.  We also note that a handful of CORNISH sources have a significantly higher flux than measured in the GLOSTAR data; these appear as a small cluster of sources above the line of equality towards the lower left corner of the plot and have been previously discussed in Sect.\,\ref{sect:cornish_matches}.

\section{Catalog properties}
\label{sect:catalogue_properties}

\setlength{\tabcolsep}{2pt}

\begin{table*}

\begin{center}
\footnotesize
\renewcommand{\arraystretch}{1.1}
\caption{GLOSTAR source catalog. The source names are appended with a $\ddagger$ to indicate if a source has been split (as described in Sect.\,\ref{sect:split}) and a $\dagger$ if the source has been recovered (as described in Sect.\,\ref{sect:recovered}).}
\label{tbl:glostar_cat}
\begin{minipage}{\linewidth}
\begin{tabular}{l........c.cc.cc}
\hline \hline
\multicolumn{1}{c}{GLOSTAR name}&  \multicolumn{1}{c}{$\ell$}&\multicolumn{1}{c}{$b$}&	\multicolumn{1}{c}{SNR}  &\multicolumn{1}{c}{$S_{\rm peak}$}  &  \multicolumn{1}{c}{$\Delta S_{\rm peak}$}  &\multicolumn{1}{c}{$S_{\rm int}$} &\multicolumn{1}{c}{$\Delta S_{\rm int}$} & \multicolumn{1}{c}{$Y$-factor} & \multicolumn{1}{c}{Radius} & \multicolumn{4}{c}{Spectral index}  & \multicolumn{2}{c}{Classification} \\



\multicolumn{1}{c}{}&  \multicolumn{1}{c}{(\degr)}&\multicolumn{1}{c}{(\degr)}&	\multicolumn{1}{c}{}& \multicolumn{2}{c}{(mJy\,beam$^{-1}$)}  & \multicolumn{2}{c}{(mJy)}  & \multicolumn{1}{c}{} & \multicolumn{1}{c}{(\arcsec)} &  \multicolumn{1}{c}{$\alpha$} & \multicolumn{1}{c}{$\Delta \alpha$} &  \multicolumn{1}{c}{$\alpha$*} & \multicolumn{1}{c}{$\Delta \alpha$*} & \multicolumn{1}{c}{Type} & \multicolumn{1}{c}{Ref.}\\

\multicolumn{1}{c}{(1)}&  \multicolumn{1}{c}{(2)}&\multicolumn{1}{c}{(3)}&	\multicolumn{1}{c}{(4)}& \multicolumn{1}{c}{(5)} & \multicolumn{1}{c}{(6)} & \multicolumn{1}{c}{(7)}  & \multicolumn{1}{c}{(8)}  & \multicolumn{1}{c}{(9)} & \multicolumn{1}{c}{(10)} &  \multicolumn{1}{c}{(11)} & \multicolumn{1}{c}{(12)} & \multicolumn{1}{c}{(13)} & \multicolumn{1}{c}{(14)} & \multicolumn{1}{c}{(15)} & \multicolumn{1}{c}{(16)}\\

\hline

G028.091$-$00.092&28.091&-0.093&11.4&3.12&0.32&4.26&0.35&1.1&16&\multicolumn{1}{c}{$\cdots$}&\multicolumn{1}{c}{$\cdots$}&$-$0.07&0.25&\multicolumn{1}{c}{$\cdots$}&\multicolumn{1}{c}{$\cdots$}\\
G028.097$-$00.951&28.097&-0.951&6.1&0.54&0.09&0.45&0.09&0.6&10&\multicolumn{1}{c}{$\cdots$}&\multicolumn{1}{c}{$\cdots$}&$-$0.36&0.69&\multicolumn{1}{c}{$\cdots$}&\multicolumn{1}{c}{$\cdots$}\\
G028.098$-$00.781&28.098&-0.781&186.9&17.80&0.96&22.06&1.11&1.2&25&-0.91&0.01&$-$1.08&0.03&Radio star&8\\
G028.100+00.644&28.099&0.644&6.5&0.87&0.14&0.76&0.14&0.6&10&\multicolumn{1}{c}{$\cdots$}&\multicolumn{1}{c}{$\cdots$}&$-$0.57&0.28&\multicolumn{1}{c}{$\cdots$}&\multicolumn{1}{c}{$\cdots$}\\
G028.107$-$00.301&28.109&-0.300&5.6&0.81&0.15&1.89&0.17&1.3&16&\multicolumn{1}{c}{$\cdots$}&\multicolumn{1}{c}{$\cdots$}&$-$0.57&0.62&\multicolumn{1}{c}{$\cdots$}&\multicolumn{1}{c}{$\cdots$}\\
G028.109$-$00.490&28.108&-0.490&20.0&2.84&0.21&2.70&0.20&0.9&15&0.06&0.09&$-$0.18&0.1&\multicolumn{1}{c}{$\cdots$}&\multicolumn{1}{c}{$\cdots$}\\
G028.116$-$00.368&28.116&-0.368&44.0&5.77&0.34&5.82&0.32&1.0&18&-0.1&0.04&$-$0.12&0.05& \multicolumn{1}{c}{$\cdots$}&\multicolumn{1}{c}{$\cdots$}\\
G028.120+00.419&28.120&0.419&4.9&0.87&0.19&0.64&0.18&0.4&8&\multicolumn{1}{c}{$\cdots$}&\multicolumn{1}{c}{$\cdots$}&0.15&0.43&\multicolumn{1}{c}{$\cdots$}&\multicolumn{1}{c}{$\cdots$}\\
G028.126+00.742&28.126&0.742&33.6&4.26&0.26&3.86&0.23&0.9&17&-0.28&0.04&$-$0.81&0.1&\multicolumn{1}{c}{$\cdots$}&\multicolumn{1}{c}{$\cdots$}\\
G028.127+00.341$\ddagger$&28.127&0.341&58.2&12.98&0.22&51.78&0.45&4.0&\multicolumn{1}{c}{$\cdots$}&\multicolumn{1}{c}{$\cdots$}&\multicolumn{1}{c}{$\cdots$}&\multicolumn{1}{c}{$\cdots$}&\multicolumn{1}{c}{$\cdots$}&\multicolumn{1}{c}{$\cdots$}&\multicolumn{1}{c}{$\cdots$}\\
G028.137$-$00.194&28.138&-0.194&4.2&0.77&0.19&0.68&0.19&0.4&8&\multicolumn{1}{c}{$\cdots$}&\multicolumn{1}{c}{$\cdots$}&0.53&0.37&\multicolumn{1}{c}{$\cdots$}&\multicolumn{1}{c}{$\cdots$}\\
G028.137$-$00.567&28.137&-0.566&5.7&1.15&0.21&1.29&0.21&0.7&11&-0.58&0.15&$-$1.14&0.48&\multicolumn{1}{c}{$\cdots$}&\multicolumn{1}{c}{$\cdots$}\\
G028.140+00.329$\ddagger$&28.140&0.329&11.0&4.50&0.43&7.30&1.00&1.6&\multicolumn{1}{c}{$\cdots$}&0.23&0.12&-0.52&0.19&HII region&3\\
G028.144$-$00.378&28.144&-0.379&8.0&1.08&0.15&1.11&0.15&0.8&12&\multicolumn{1}{c}{$\cdots$}&\multicolumn{1}{c}{$\cdots$}&$-$0.46&0.55&\multicolumn{1}{c}{$\cdots$}&\multicolumn{1}{c}{$\cdots$}\\
G028.146+00.517&28.145&0.517&8.7&1.21&0.15&1.86&0.17&1.2&16&\multicolumn{1}{c}{$\cdots$}&\multicolumn{1}{c}{$\cdots$}&-0.66&0.27&\multicolumn{1}{c}{$\cdots$}&\multicolumn{1}{c}{$\cdots$}\\
G028.149$-$00.998&28.149&-0.997&76.9&11.93&0.66&19.71&1.00&1.6&25&\multicolumn{1}{c}{$\cdots$}&\multicolumn{1}{c}{$\cdots$}&$-$1.05&0.05&\multicolumn{1}{c}{$\cdots$}&\multicolumn{1}{c}{$\cdots$}\\
G028.151+00.164&28.148&0.149&38.6&10.38&0.62&194.53&9.73&18.0&95&\multicolumn{1}{c}{$\cdots$}&\multicolumn{1}{c}{$\cdots$}&\multicolumn{1}{c}{$\cdots$}&\multicolumn{1}{c}{$\cdots$}&HII region&1\\
G028.151$-$00.958&28.151&-0.957&16.0&2.00&0.16&2.25&0.17&1.0&17&-0.37&0.09&$-$0.4&0.28&\multicolumn{1}{c}{$\cdots$}&\multicolumn{1}{c}{$\cdots$}\\
G028.159$-$00.024&28.157&-0.020&8.0&2.91&0.40&28.85&1.49&6.6&38&\multicolumn{1}{c}{$\cdots$}&\multicolumn{1}{c}{$\cdots$}&\multicolumn{1}{c}{$\cdots$}&\multicolumn{1}{c}{$\cdots$}& Extended/Diffuse&1\\
G028.159$-$00.798&28.159&-0.798&29.1&2.97&0.19&2.67&0.17&0.9&16&-0.35&0.05&$-$0.36&0.23&\multicolumn{1}{c}{$\cdots$}&\multicolumn{1}{c}{$\cdots$}\\
G028.160$-$00.046&28.160&-0.046&4.8&1.68&0.36&1.59&0.36&0.5&9&\multicolumn{1}{c}{$\cdots$}&\multicolumn{1}{c}{$\cdots$}&$-$0.15&0.52& HII region&3\\
G028.163$-$00.163$\dagger$&28.163&-0.163&7.0&0.50&0.07&2.00&0.34&4.0&\multicolumn{1}{c}{$\cdots$}&\multicolumn{1}{c}{$\cdots$}&\multicolumn{1}{c}{$\cdots$}&\multicolumn{1}{c}{$\cdots$}&\multicolumn{1}{c}{$\cdots$}&HII region&3\\
G028.172$-$00.605&28.173&-0.606&4.9&1.13&0.24&14.55&0.76&5.5&32&\multicolumn{1}{c}{$\cdots$}&\multicolumn{1}{c}{$\cdots$}&\multicolumn{1}{c}{$\cdots$}&\multicolumn{1}{c}{$\cdots$}&Ionization Front&1\\
G028.178+00.074&28.178&0.072&4.3&1.49&0.35&9.56&0.59&2.5&21&\multicolumn{1}{c}{$\cdots$}&\multicolumn{1}{c}{$\cdots$}&\multicolumn{1}{c}{$\cdots$}&\multicolumn{1}{c}{$\cdots$}& \multicolumn{1}{c}{$\cdots$}&\multicolumn{1}{c}{$\cdots$}\\
G028.187+00.583&28.187&0.583&22.3&3.02&0.21&2.75&0.19&0.9&15&-0.01&0.06&$-$0.71&0.11& \multicolumn{1}{c}{$\cdots$}&\multicolumn{1}{c}{$\cdots$}\\
G028.189$-$00.744&28.189&-0.744&17.0&1.74&0.14&1.36&0.12&0.7&13&-0.33&0.08&$-$1.09&0.2&\multicolumn{1}{c}{$\cdots$}&\multicolumn{1}{c}{$\cdots$}\\
G028.193$-$00.785&28.193&-0.785&8.6&0.87&0.11&1.29&0.12&1.1&16&-0.68&0.12&$-$0.22&0.26&Pulsar&5\\
G028.197$-$00.891&28.198&-0.892&4.6&0.48&0.11&0.96&0.11&1.0&13&\multicolumn{1}{c}{$\cdots$}&\multicolumn{1}{c}{$\cdots$}&$-$0.94&0.47&\multicolumn{1}{c}{$\cdots$}&\multicolumn{1}{c}{$\cdots$}\\
G028.200$-$00.050&28.191&-0.050&771.1&332.67&17.92&396.10&19.81&1.2&48&\multicolumn{1}{c}{$\cdots$}&\multicolumn{1}{c}{$\cdots$}&\multicolumn{1}{c}{$\cdots$}&\multicolumn{1}{c}{$\cdots$}& HII region&3\\
G028.209+00.226&28.209&0.226&5.8&1.07&0.19&0.95&0.19&0.6&10&\multicolumn{1}{c}{$\cdots$}&\multicolumn{1}{c}{$\cdots$}&0.22&0.4&\multicolumn{1}{c}{$\cdots$}&\multicolumn{1}{c}{$\cdots$}\\
\hline

\end{tabular}\\
References: (1) this work, (2) \citealt{kalcheva2018}, (3) \citet{anderson2014}, (4) \citet{giveon2005a}, (5) SIMBAD, (6) \citet{irabor2018}, (7) \citet{ortega2010}, (8) \citealt{purcell2013}\\
* Values calculated using the 8 sub images are described in  { Sect.\,\ref{sect:glostar_vla_survey}}.\\ 
Notes: Only a small portion of the data is provided here, the full table is available in electronic form at the CDS via anonymous ftp to cdsarc.u-strasbg.fr (130.79.125.5) or via http://cdsweb.u-strasbg.fr/cgi-bin/qcat?J/A\&A/.\\ 
\end{minipage}
\end{center}
\end{table*}
\setlength{\tabcolsep}{6pt}

The final catalog consists of {\color{black}1575} distinct sources with flux densities above 4$\sigma$, where $\sigma$ corresponds to the local noise. We give the names of the sources and the observed parameters in Table\,\ref{tbl:glostar_cat}. The source names are constructed from the peak flux density positions. The position given in Cols.\,2 and 3 are the barycentric coordinates determined from the first order moments of the longitude and latitude profiles. The barycentric position generally defines the centroid position of a source. However, if the emission profile of a source is extended and consists of multiple peaks, then this can be offset from the peak position. The mean uncertainties on the peak flux density positions is 1\as22 with a standard deviation of 0\as81; the positional uncertainties associated with individual sources are provided in the electronic version of Table\,\ref{tbl:glostar_cat}. Combined with the uncertainty in the map astrometry the mean uncertainty in the catalog positions is $\sim 2$\arcsec. The signal to noise values are given in Col.\,4; these are estimated by dividing the peak flux by the \emph{rms} at the same position. The flux density values given in Cols.\,5-8 have been corrected for clean bias and bandwidth smearing (see \cite{hales2012} for details) and in Col.\,9 we give the $Y$-factor, which is the ratio of the integrated and peak flux densities (i.e., $S_{\rm int}/S_{\rm peak}$); this gives an indication of how compact a source is with unresolved sources having a value of one and very extended sources having values of several 10s. The radius in Col.\,10 is determined from the number of pixels associated with the source and is therefore a rather crude estimate of the size of a source particularly for extended structures and should be used with caution. In Col.\,11 and 12 we give the spectral index for compact sources (i.e. $Y$-factor $<$ 2) that have a 20\,cm counterpart detected by the THOR survey while in Col.\,13 and 14 are the values calculated {from} the 8 sub images {discussed in} Sect.\,\ref{sect:RadioContiMap}; this will be described in Sect.\,\ref{sect:thor_spectral_index}. In Cols.\,15 and 16 we give the classification assigned to the source and a reference when available.\\

\begin{figure}
\centering
\includegraphics[width=0.49\textwidth, trim= 0 0 0 0, angle=0]{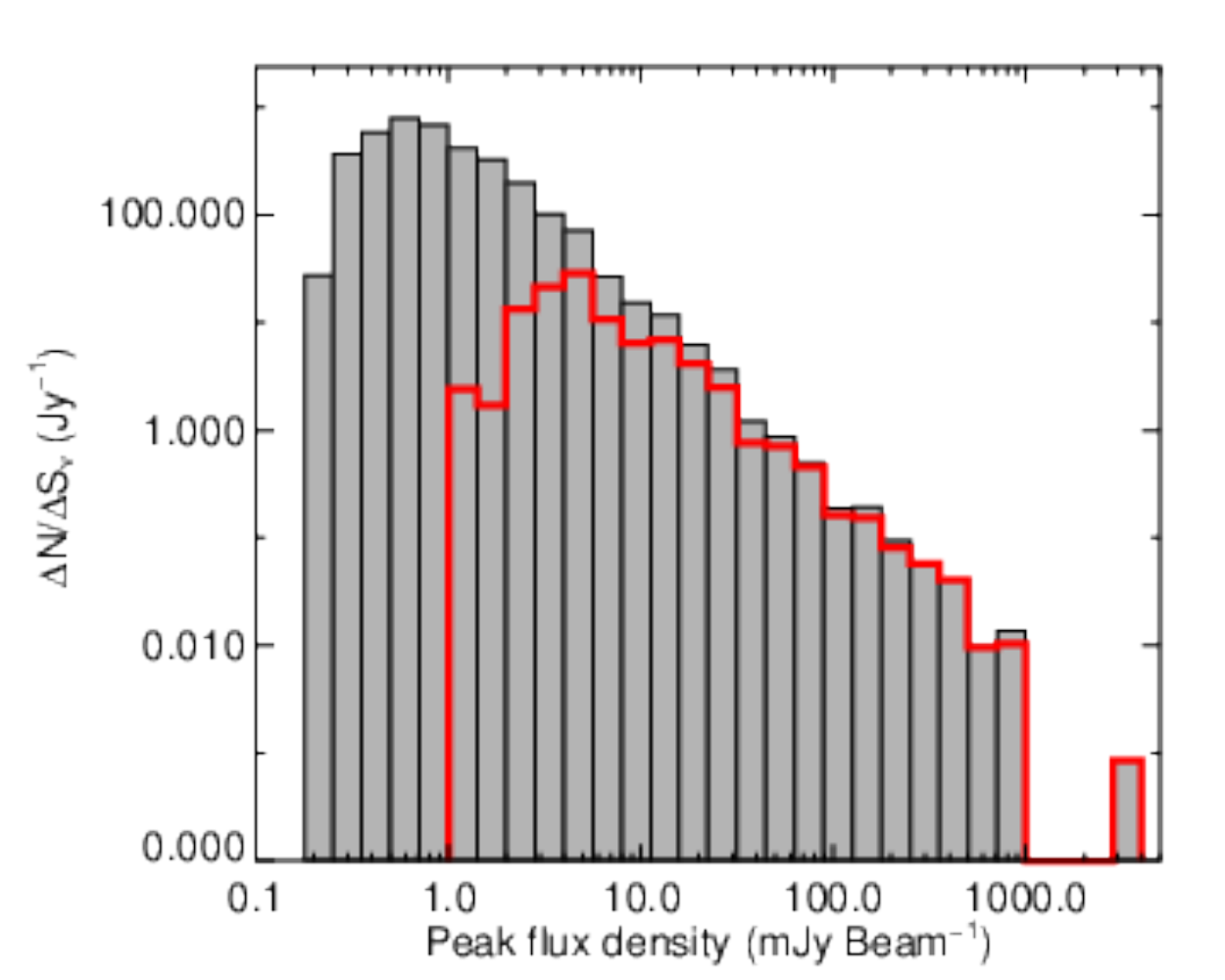}
\includegraphics[width=0.49\textwidth, trim= 0 0 0 0, angle=0]{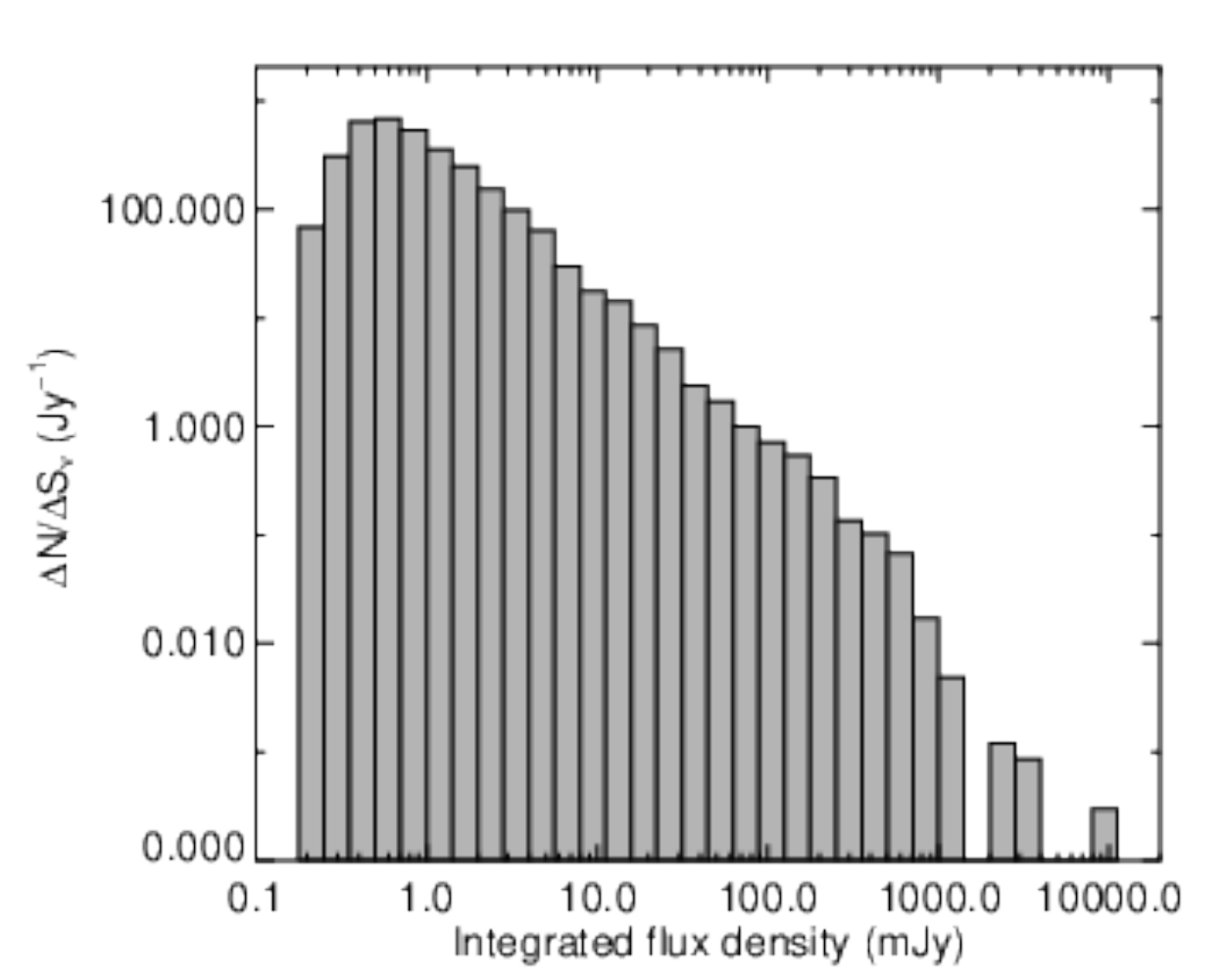}
\caption{Peak and integrated flux distribution of GLOSTAR sources are shown in the upper and lower panels, respectively. The thick red histogram  on the upper panel shows the flux distribution of sources previously identified by CORNISH and/or MAGPIS. The bin width is 0.15\,dex.}
\label{fig:glostar_flux_distribution}
\end{figure}

\subsection{Fluxes and angular sizes}

In Fig.\,\ref{fig:glostar_flux_distribution} we show the peak and integrated flux densities distribution of the detected sources. Together with the peak flux density distribution we also show the integrated flux density distribution of all GLOSTAR sources that are matched to sources previously identified by CORNISH and/or MAGPIS (red histogram overlaid on the upper panel); this shows clearly that the vast majority of 5\,GHz sources detected in the GLOSTAR map with a peak flux density above a few mJy\,beam$^{-1}$ have been detected before. The sources not picked up by the previous surveys are all extended and would have been filtered out by CORNISH or in regions not covered by MAGPIS. What is also clear from this plot is that the improved sensitivity of GLOSTAR has resulted in the detections of many previously unknown integrated low-flux density sources.

\begin{figure}
\centering
\includegraphics[width=0.49\textwidth, trim= 0 0 0 0, angle=0]{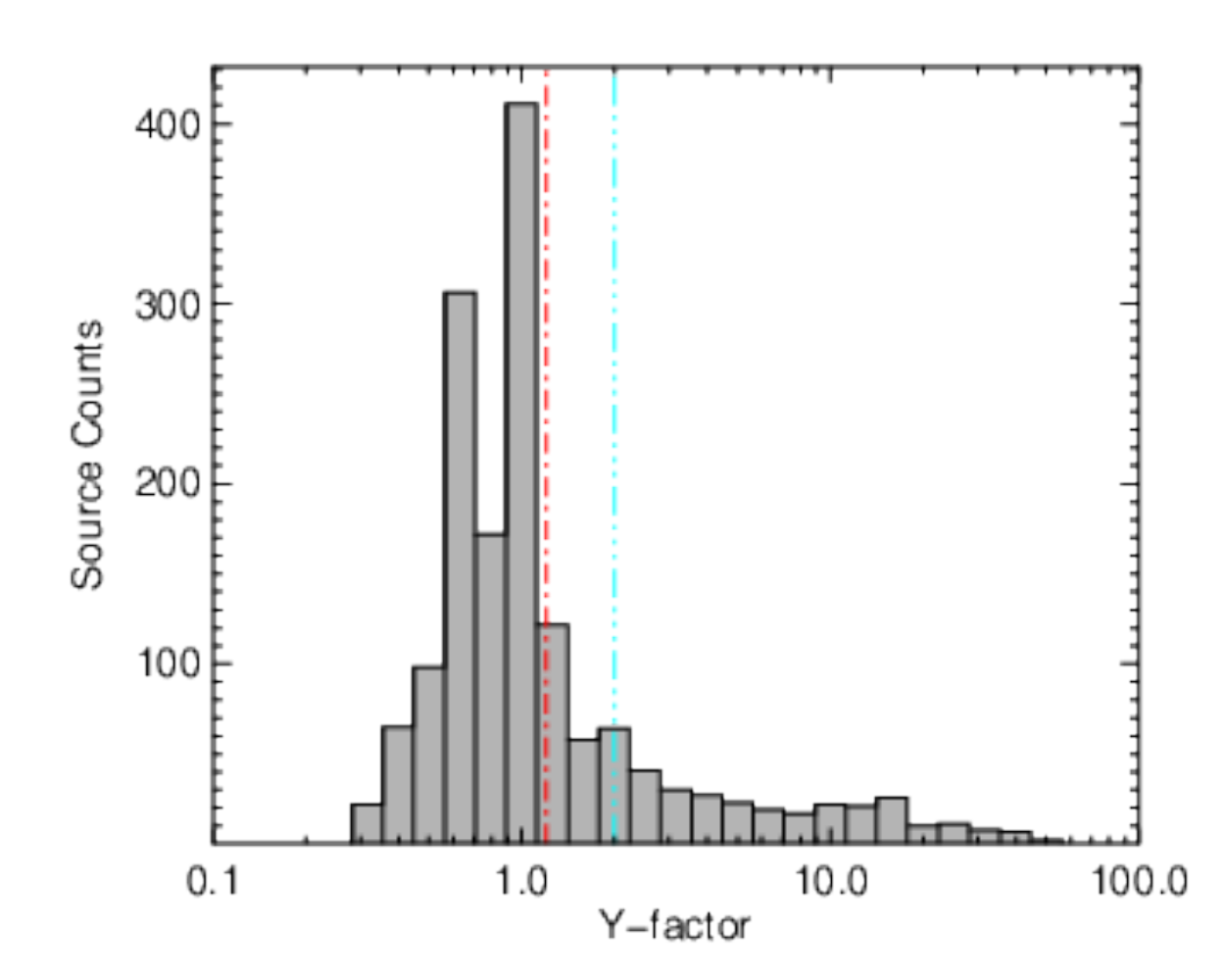}
\caption{Distribution of the $Y$-factor ($S_{\rm int}/S_{\rm peak}$). The red dash-dot ($Y$-factor = 1.2) and cyan dash-dotted lines ($Y$-factor = 2) indicate the criterion used to distinguish between unresolved, compact and extended sources, respectively.  The bin size used is 0.1\,dex.}
\label{fig:y-factor}
\end{figure}

In Fig.\,\ref{fig:y-factor} we show the distribution of $Y$-factor for the 
sample; this parameter is the ratio between integrated and peak flux density (i.e., $S_{\rm int}/S_{\rm peak}$) and gives an indicate of the source size. 
While for unresolved sources the $Y$-factor should be equal to 1, allowing for measurement uncertainties we classify all sources with a  $Y$-factor $<$ 1.2 as unresolved (cf. \citealt{bihr2016}). We make a further distinction between compact (which includes all of the unresolved sources) and extended sources using a $Y$-factor = 2 as dividing line. We note that a significant number of sources have $Y$-factors smaller than 1 and this {has} been discussed in detail by \citet{bihr2016} who identified three possible explanations for this unexpected result: 1) \blobcat\ does not fit enough pixels for weak sources, which leads to lower integrated flux densities; 2) unresolved sources are located in slightly negative side lobes from nearby strong sources; 3) weak sources are not cleaned properly. Bihr et al. suggest using the peak flux densities for unresolved sources in any analysis and we have also adopted this recommendation.

\blobcat\ gives source size in terms of the number of pixels (parameter $npix$) that are flood filled in the source extraction process. We convert this into a radial size using:


\[
R_{\rm source} = \sqrt{\frac{A}{\pi}}
\]

\begin{figure}
\centering
\includegraphics[width=0.49\textwidth, trim= 0 0 0 0, angle=0]{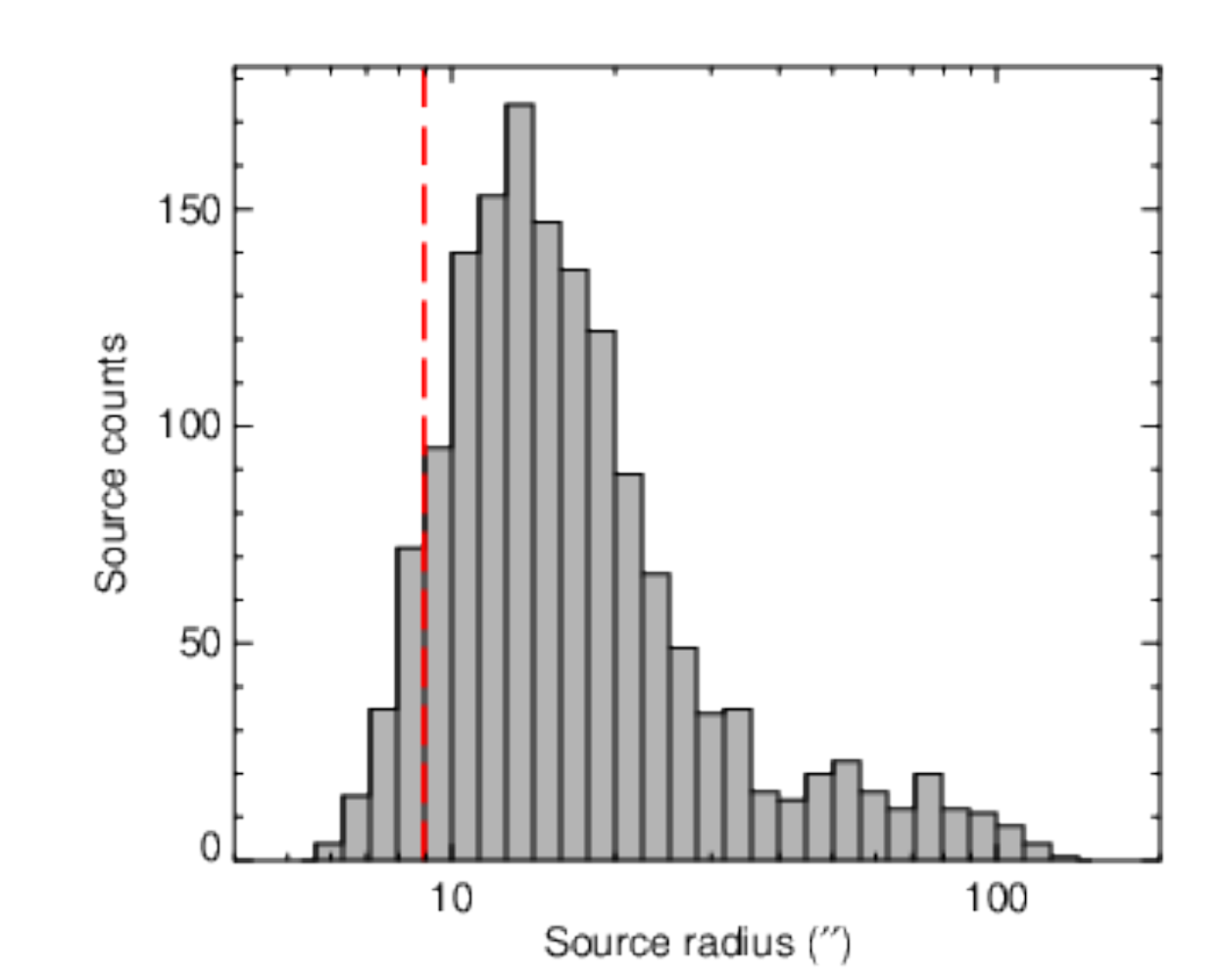}
\caption{Distribution of the source radii. The dashed-dotted red line indicates the resolution of the observations (radio beam). The bin size used is 0.1\,dex.}
\label{fig:angular_sizes}
\end{figure}

\noindent where $R_{\rm source}$ is the effective radius, $A$ is the area covered by the source calculated by $npix$ times the pixel size squared, which in this case is $2\as5 \times 2\as5$. Fig.\,\ref{fig:angular_sizes} shows the distribution of the angular radius of the sources. 

\subsection{Spectral index}
\label{sect:thor_spectral_index}

Astrophysical sources mainly emit two types of radio emission, thermal and non-thermal. The first refers to bremsstrahlung radiation and the latter is also known as synchrotron radiation (see, e.g., \citealt{Wilson2012}). Thermal bremsstrahlung radiation is produced in \hii\ regions and planetary nebulae and usually has spectral indices between $-$0.1 and 2 depending on whether the emission region is optically thin or thick, respectively. Active Galactic Nuclei (AGN) galaxies and supernova remnants give rise to synchrotron radiation, which has negative spectral indices, typically between  $-$1 and $-$0.5 (\citealt{Condon1984}; \citealt{Dzib2013}; \citealt{Rodriguez2012} and references therein). The spectral index can, therefore, help to broadly distinguish between different types of radio sources.  

As mentioned in Sect.\,\ref{sect:Calibration_redution}, the calibration process produced 8 sub images in the frequency range between 4 and 8\,GHz which we used to estimate the spectral index only of the compact sources (i.e. $Y$-factor $<$ 2). We used the peak flux density from each sub image, derived with the \casa\ task \imfit, and a $\chi^2$ linear fit to calculated the spectral index value.   

We also estimate the spectral index of the compact sources by combining their peak flux densities with the peak lower-frequency flux densities determined by THOR. As previously mentioned, we have matched 711 sources with a THOR source, of which 640 have $Y$-factors smaller than 2 and so can be assumed to be relatively compact. The mid-frequency for THOR and GLOSTAR are 1.5\,GHz and 5.8\,GHz, respectively.  Following \citet{kalcheva2018} we estimate the spectral index using the following:

\[
\alpha=\frac{{\rm ln}\left(S_{\rm GLOSTAR}/S_{\rm THOR} \right)}{{\rm ln}{\rm \left(5.8/1.5 \right)}},
\]

\noindent where $S_{\rm THOR}$ and $S_{\rm GLOSTAR}$ are the peak flux densities at 6\,cm and 20\,cm, respectively. The uncertainty is calculated using:

\[
\Delta \alpha = \frac{\sqrt{\left(\sigma_{\rm THOR}/S_{\rm THOR}\right)^2 + \left(\sigma_{\rm GLOSTAR}/S_{\rm GLOSTAR}\right)^2}}{{{\rm ln}\left(5.8/1.5 \right)}}
\]

\noindent where $\sigma_{\rm GLOSTAR}$ and $\sigma_{\rm THOR}$ are the uncertainties in the peak flux densities.

\begin{figure}
\centering
\includegraphics[width=0.49\textwidth, trim= 0 0 0 0, angle=0]{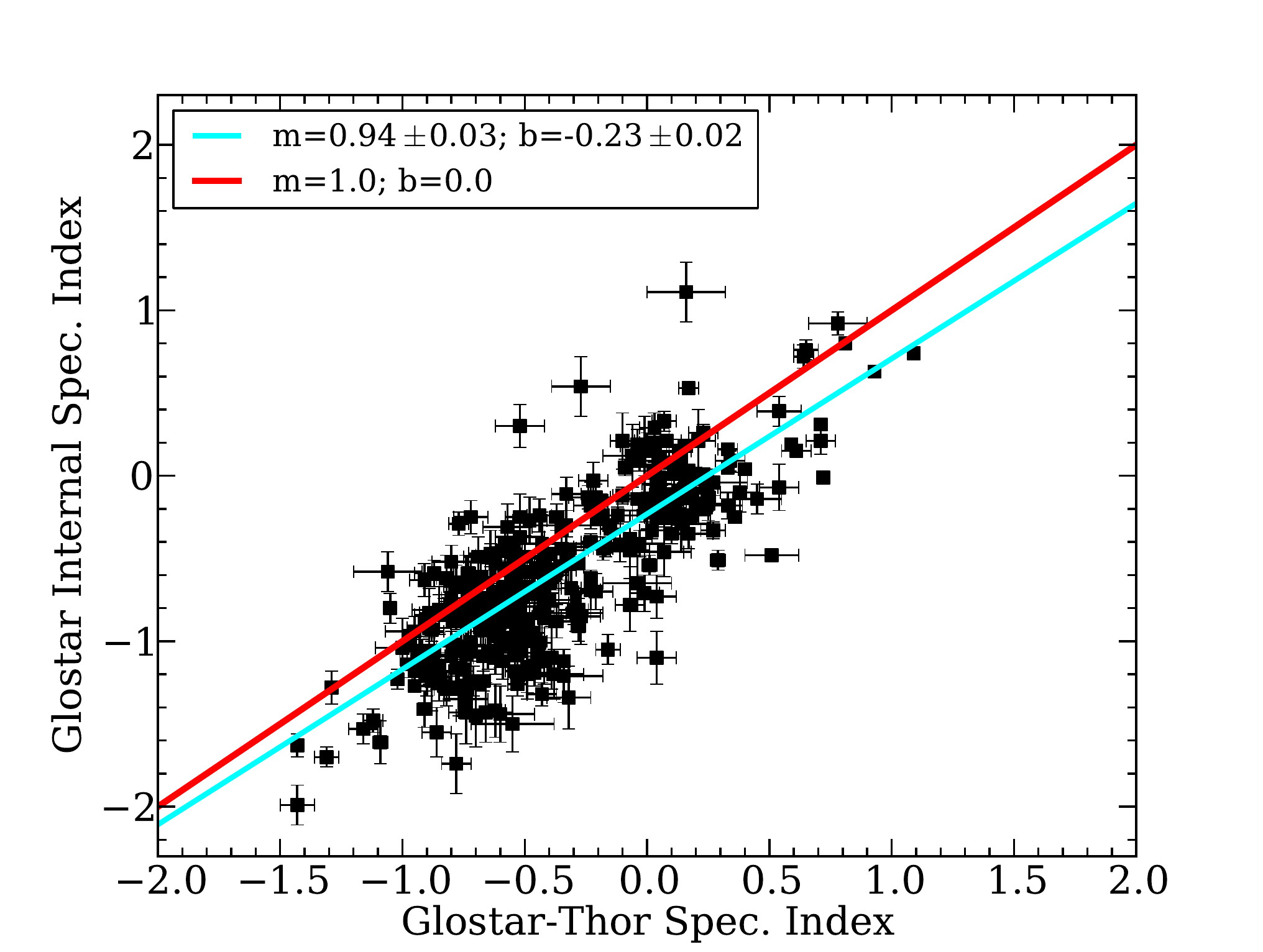}
\caption{ Comparison of spectral index values between only GLOSTAR data with uncertainties less than 0.2 and GLOSTAR combined with THOR data. The $m$ and $b$ are the slope and intersection of the linear fit, respectively, {and is represented with the blue line}. {The red line is the line of equality.} }
\label{fig:specIndex}
\end{figure}

\begin{figure}
\centering
\includegraphics[width=0.49\textwidth, trim= 0 0 0 0, angle=0]{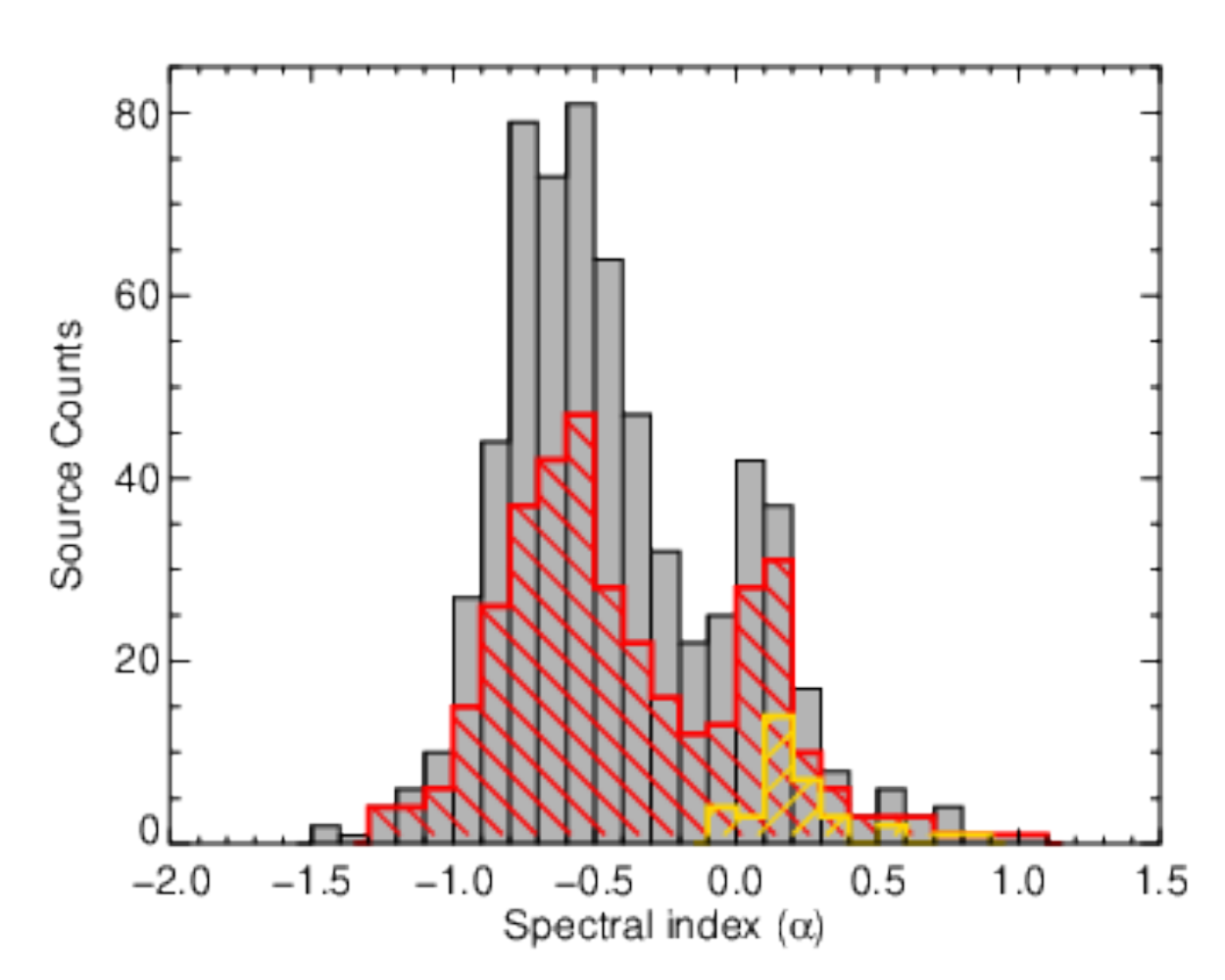}
\includegraphics[width=0.49\textwidth, trim= 0 0 0 0, angle=0]{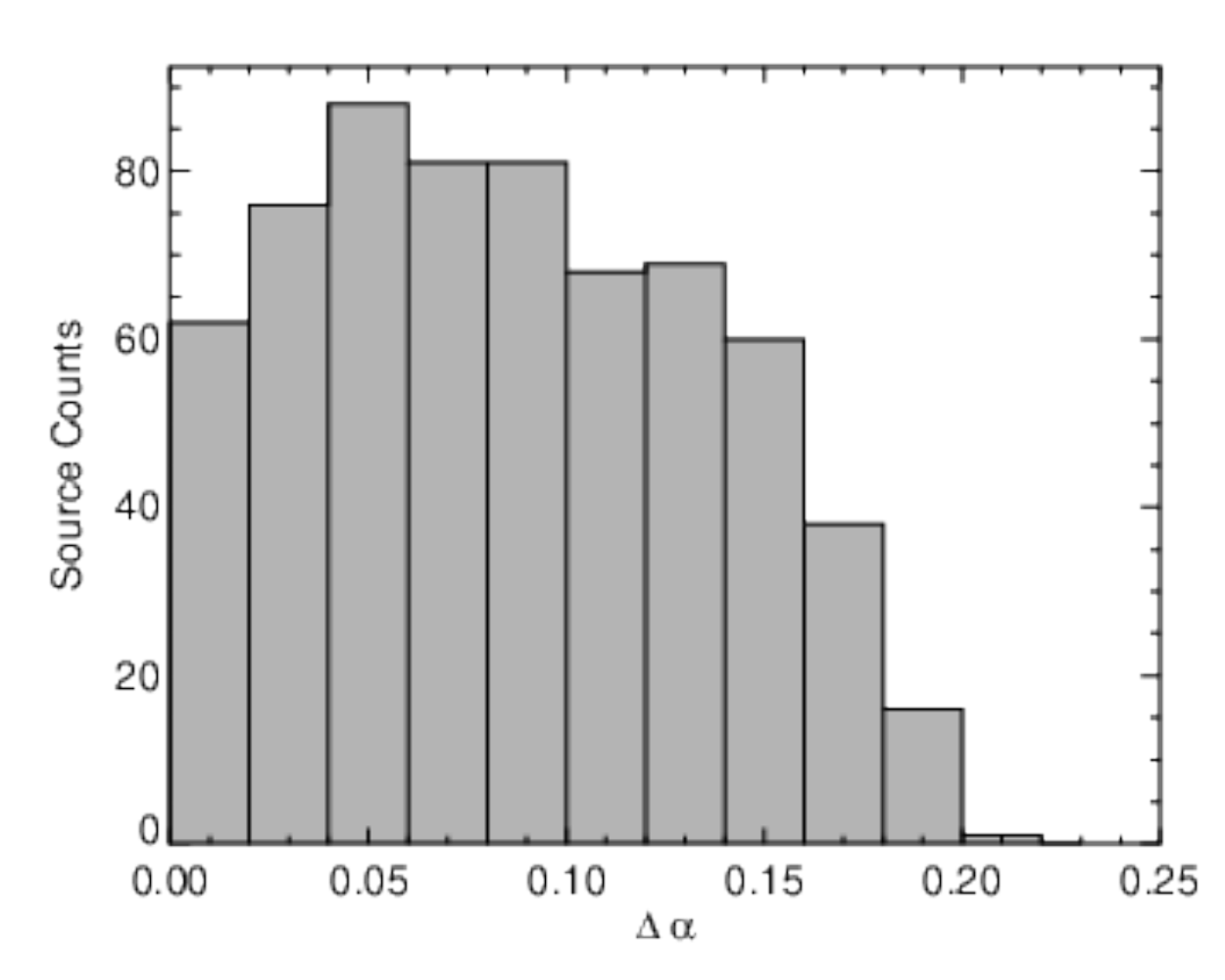}

\caption{In the upper panel we show the spectral index distribution of all compact GLOSTAR sources with $Y$-factor $<2$ and a counterpart detected at 1.5\,GHz in the THOR survey. The red and yellow hatched histograms show the distribution of compact radio sources that are associated with mid-infrared sources and dust emission, respectively. In the lower panel we show the distribution of the associated uncertainties to the spectral index measurements. The bin size used in the upper and lower panels are  0.1 and 0.02, respectively.}
\label{fig:spectral_index}
\end{figure}

For a comparison of both methods discussed above, we selected sources with reliable GLOSTAR internal spectral indices, $\sigma_{\alpha}<0.2$, and compare them with their GLOSTAR-THOR spectral indices (see Fig.\,\ref{fig:specIndex}).  A linear fit to these values
yields that the slope is $\mathbf{m=0.94\pm0.03}$, close to the slope of the equality line. In conclusion, the spectral index values derived 
with the two methods are statistically similar.



The spectral index derived from GLOSTAR and THOR data combination and their associated uncertainty are given in Cols.\,11 and 12 of Table\,\ref{tbl:glostar_cat}. In the upper panel of Fig.\,\ref{fig:spectral_index} we show the distribution of spectral indices; these cover a range of values between $-1.5$ and 1  and have a  bimodal distribution with peaks at approximately $-0.7$ and 0.1. The dip between these two peaks occurs at approximately $-$0.2, which we consider the transition from thermal and non-thermal emission. We also show the spectral indices of radio sources that are associated with mid-infrared emission (i.e., have a WISE counterpart) and associated with dust emission (i.e., have an ATLASGAL counterpart); these are shown in red and yellow hatched histograms, respectively. It is clear from a comparison of these distributions that the sources with mid-infrared counterparts span the whole range of spectral indices, while the sources associated with dust are almost exclusively thermal sources. So while sources associated with dust are likely to be a mix of \hii\ regions and PNe, the association of a particular source with mid-infrared emission does not provide any constraints on its nature. We will explore this issue of source types in the next section. In the lower panel of  Fig.\,\ref{fig:spectral_index} we show the distribution of the uncertainties associated with the spectral index measurements; these have typical values of $\pm$0.1 but can be as large as $\pm$0.2.


We have calculated the spectral indices for 671 GLOSTAR sources without a THOR counterpart. The values and their associated uncertainties are given in Cols.\,13 and 14 of Table\,\ref{tbl:glostar_cat}. We note that the limited frequency range or poor signal to noise results in large uncertainties for the majority (74\% have $\Delta \alpha > 0.2$). The spectral indices range between $-$1 and 1 for the sources for which they are considered to be reliable. Given that the sample of GLOSTAR-internal spectral index is relatively modest we decided not to perform a detailed statistical analysis of the spectral index.

\section{Classification of radio sources}
\label{sect:identification_emission_sources}

\subsection{Literature search}
\label{sect:literature}

A number of studies have investigated the nature of radio and mid-infrared sources. By correlating our catalog with these we can adopt their classifications. Cross-matching our radio sources with \hii\ region catalogs compiled by \citet{purcell2013} and \citet{kalcheva2018} using CORNISH data, \citet{giveon2005b} using MAGPIS data and \citet{ortega2010} we have identified {a total of} 46 previously known \hii\ regions. As described in Sect.\,\ref{sect:wise} we  also cross-matched our radio sources with a catalog of \hii\ region candidates identified by \citet{anderson2014} from their analysis of WISE colors. This revealed a match for 130 radio sources including 90 not previously matched to a CORNISH or MAGPIS source. 
Of these, 43 have spectral indexes determined from the THOR and GLOSTAR data;  42 of these have spectral indexes between $-$0.17 to 0.88, which supports their classification as \hii\ regions (\citealt{anderson2014}). Combined with the 46 sources identified from the literature brings the sample of reliable \hii\ regions in the GLOSTAR region to 136. 



A further {\color{black}28} sources have been identified as PNe by the CORNISH team (\citealt{purcell2013} and \citealt{irabor2018}),  and another 15 as radio stars (12 identified by \citealt{purcell2013} and 3 found in SIMBAD). We found, with a cross match {radius} of 10\arcsec,  that 3 of these candidates have a counterpart in the HASH (Hong Kong/AAO/Strasbourg H$\alpha$) planetary nebula database (\citealt{parker2016}). Our literature search has also identified another three interesting compact radio sources that are consistent in position with pulsars: G032.763+00.092 with PSR J1850-0006 (\citealt{Keith2009}), G028.880$-$00.938 with PSR B1844-04, and G028.193$-$00.785 with PSR B1842-04 (\citealt{Hobbs2004}). Radio continuum emission from energetic pulsars is usually associated to the integrated emission of their pulses, but other mechanisms such as a pulsar wind and magnetic activity have been recently discussed by \citet{dzib2018}. For the GLOSTAR data discussed here, the angular resolution is insufficient to decide {whether or not} the radio emission comes from the pulsar, from the surrounding medium, or from an unassociated object. This will be better constrained with the GLOSTAR VLA B-configuration data when it becomes available.

The importance of pulsars with radio continuum emission is that they may {targeted by} Very Long Baseline Interferometry (VLBI) observations {to} determine their astrometric parameters, such as proper motions and trigonometric parallaxes \citep{brisken2000}. This information may be used to associate them with supernova remnant (SNR) or to better constrain their luminosity, and Galactic electron distribution models \citep{brisken2005,deller2009}.

\begin{figure}
\centering

\includegraphics[width=0.49\textwidth, trim= 0 0 0 0, angle=0]{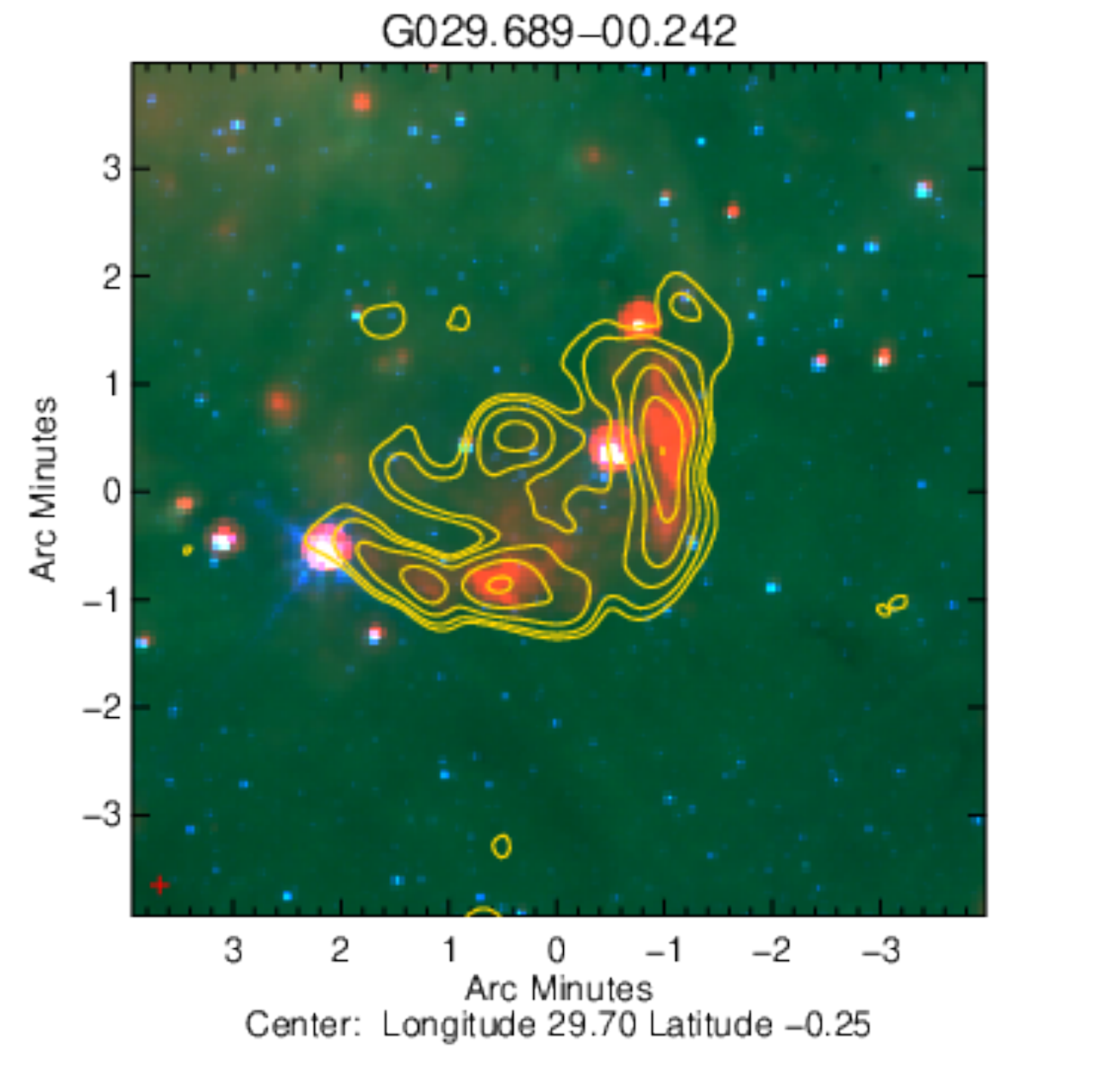}

\caption{The compact SNR G029.689$-$00.242 identified from the \cite{green2014} catalog (G029.7$-$00.3 in their catalog). The background image is a false color composite image produced by combining the GLIMPSE 5.8 and 8.0\,$\mu$m bands and the MIPSGAL 24\,$\mu$m band; these are shown in {blue, green and red}
, respectively. The yellow contours show the distribution of the 5.8\,GHz radio emission associated with the GLOSTAR source.}
\label{fig:G029.689}
\end{figure}

We also identified 2 supernova remnants from a cross-matching with a radius of 10\arcsec\  between the GLOSTAR sources and the THOR $\sim$20 supernova remnants candidates identified by \cite{anderson2017} located in the GLOSTAR field; these are G028.642+00.187 and G028.218$-$00.080. The low number of matches may be due the negative spectral index of the supernova remnants, which are better recover at lower frequencies, or may be due to the fact that SNRs tend to be extended and so many are likely to be either filtered out by the interferometer entirely or only partially detected and classified as extended emission. To investigate this further we also searched for SNR using the \citet{green2014} catalog. This identifies 12 SNR within the GLOSTAR region, 10 of which are matched with the large-scale emission regions identified in Table\,\ref{tbl:LargeStruc} and mentioned in Sect.\,\ref{sect:large_scale_structures}. Of the remaining 2 SNRs in the field 1 {is} not detected at all (G032.1$-$00.9), and one is detected as a compact radio source (G029.7$-$00.3); see Fig.\,\ref{fig:G029.689} for an image of this compact SNR. The one SNR not recovered in our map happens to be the largest identified by \citet{green2014} in the GLOSTAR field with a size of $\mathbf{40\arcmin \times 40\arcmin}$ and so its surface brightness may have fallen below the GLOSTAR detection threshold. 

\subsection{Analysis of mid-infrared images}

The correlation of compact radio emission with dust and mid-infrared emission is an indicator of the presence of \hii\ regions and there have been a number of studies that have used this correlation to identify samples of \hii\ regions (e.g., \citealt{urquhart2013_cornish}; \citealt{urquhart_radio_north}; \citealt{urquhart_radio_south}; \citealt{thompson2006}). In addition to identifying and removing spurious sources (as discussed in Sect.\,\ref{sect:corr_other_surveys}) we have, therefore, also used the mid-infrared images to distinguish between the nature of different kinds of radio emission.  In the following two subsections we will give a brief description of the rationale used to define these different kinds of sources.  To do this, we split the remaining unclassified radio source into extended sources ($Y$-factor $>$ 2) and compact sources ($Y$-factors $<$ 2) because the size and structure of the radio emission itself can provide some useful clues to their nature.\\

We indicate the emission type attributed to each source and a reference if the identification has come from the literature in the last column of Table\,\ref{tbl:glostar_cat}. In Table\,\ref{tab:statistics} we provide a summary of the number of each type identified.

\begin{table}


\begin{center}
\caption{Summary of emission types. The number in parenthesis in the SNR column indicates the number of SNR identified as large-scale structures in Sect.\,\ref{sect:large_scale_structures}.}
\label{tab:statistics}
\begin{tabular}{lc}\hline\hline
Description  & Number \\
\hline
Ionization fronts       & 37 \\
Extended emission sources        & 62\\
Planetary nebulae        &  46\\
\hii\ regions       & {231}\\
Radio stars        & {15}\\
Pulsars       & 3 \\
SNR& 3 (10) \\

\hline
\end{tabular}
\end{center}
\end{table}

\begin{figure*}
\centering

\includegraphics[width=0.44\textwidth, trim= 0 0 0 0, angle=0]{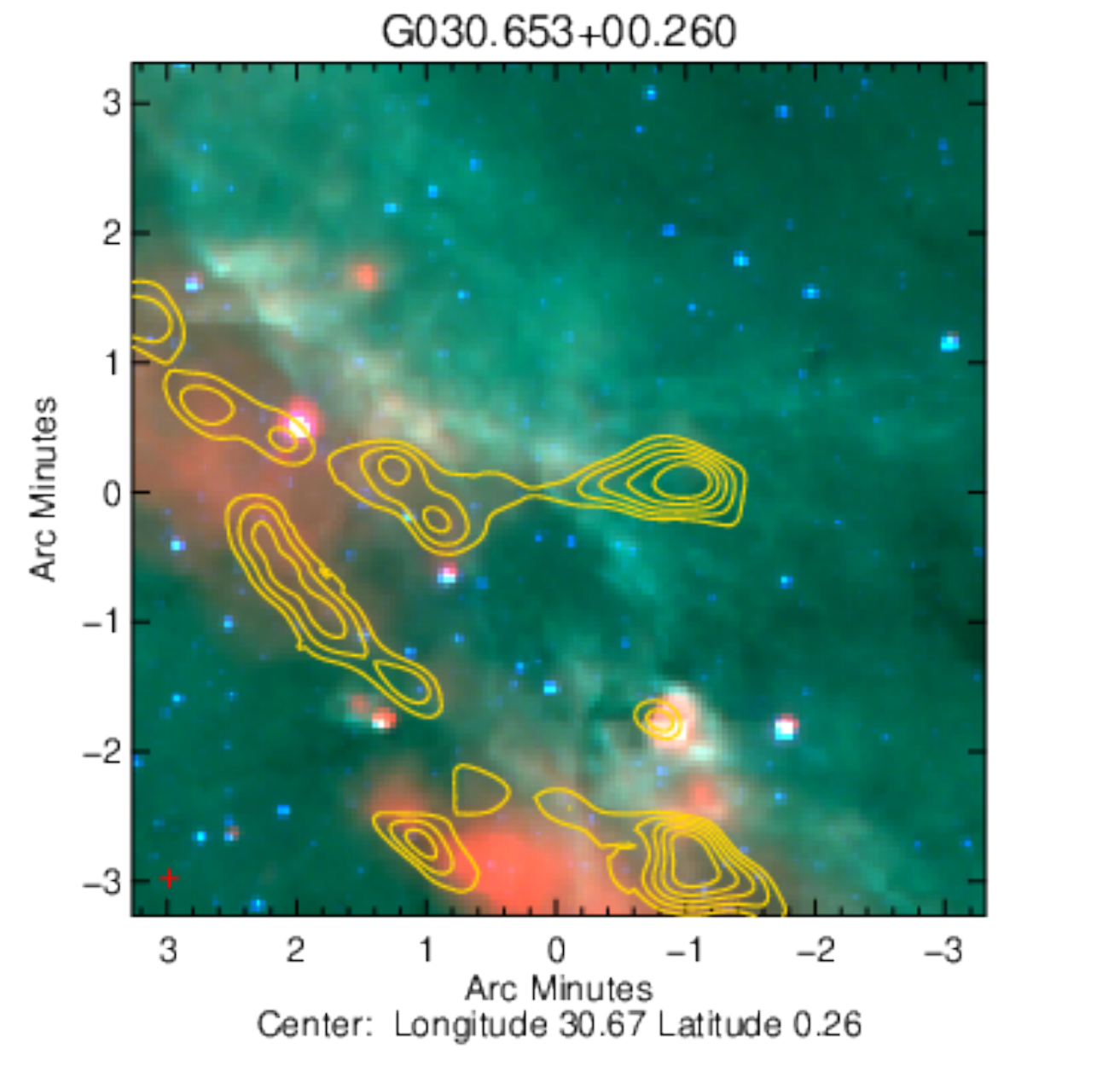}
\includegraphics[width=0.44\textwidth, trim= 0 0 0 0, angle=0]{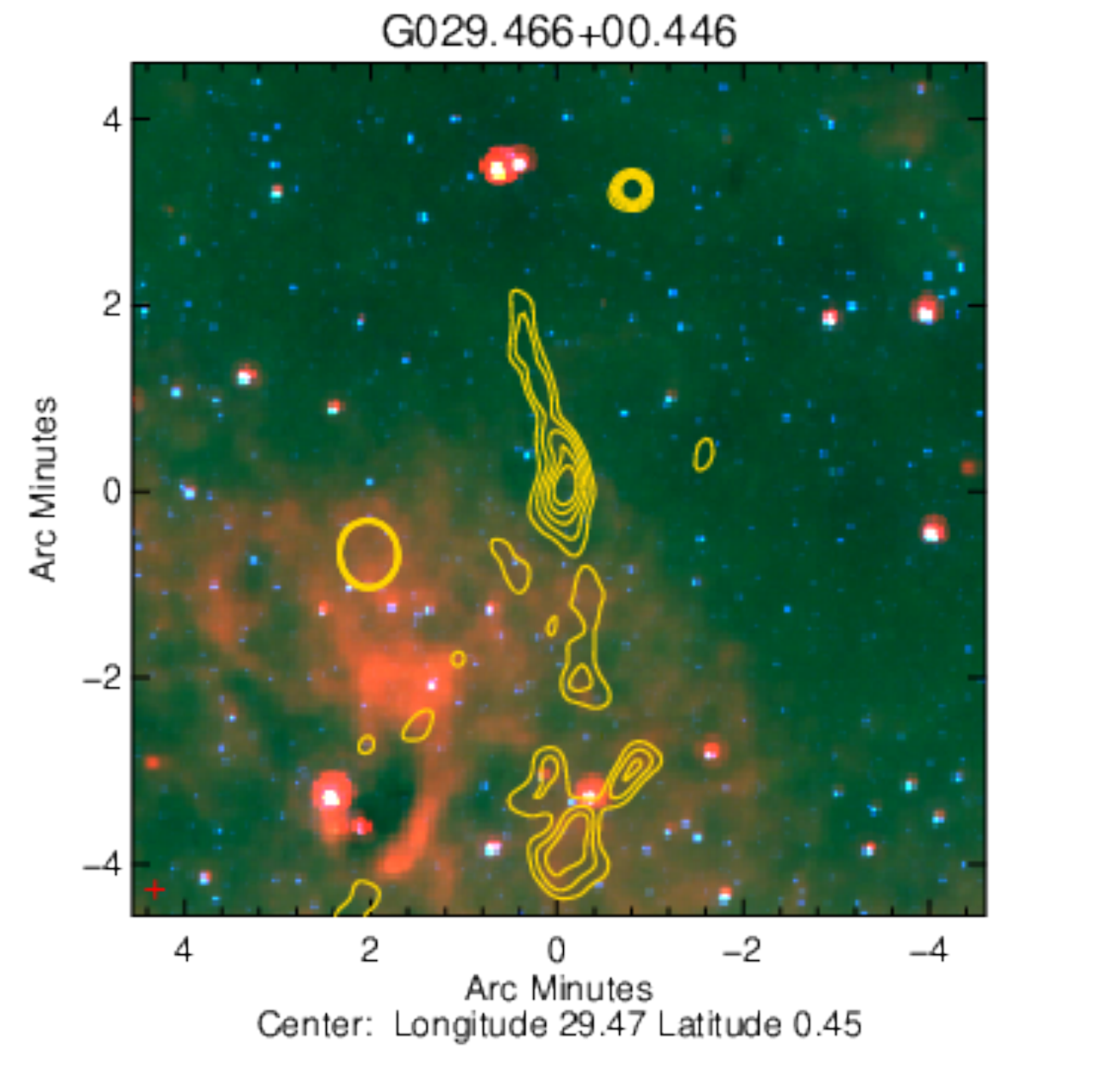}
\includegraphics[width=0.44\textwidth, trim= 0 0 0 0, angle=0]{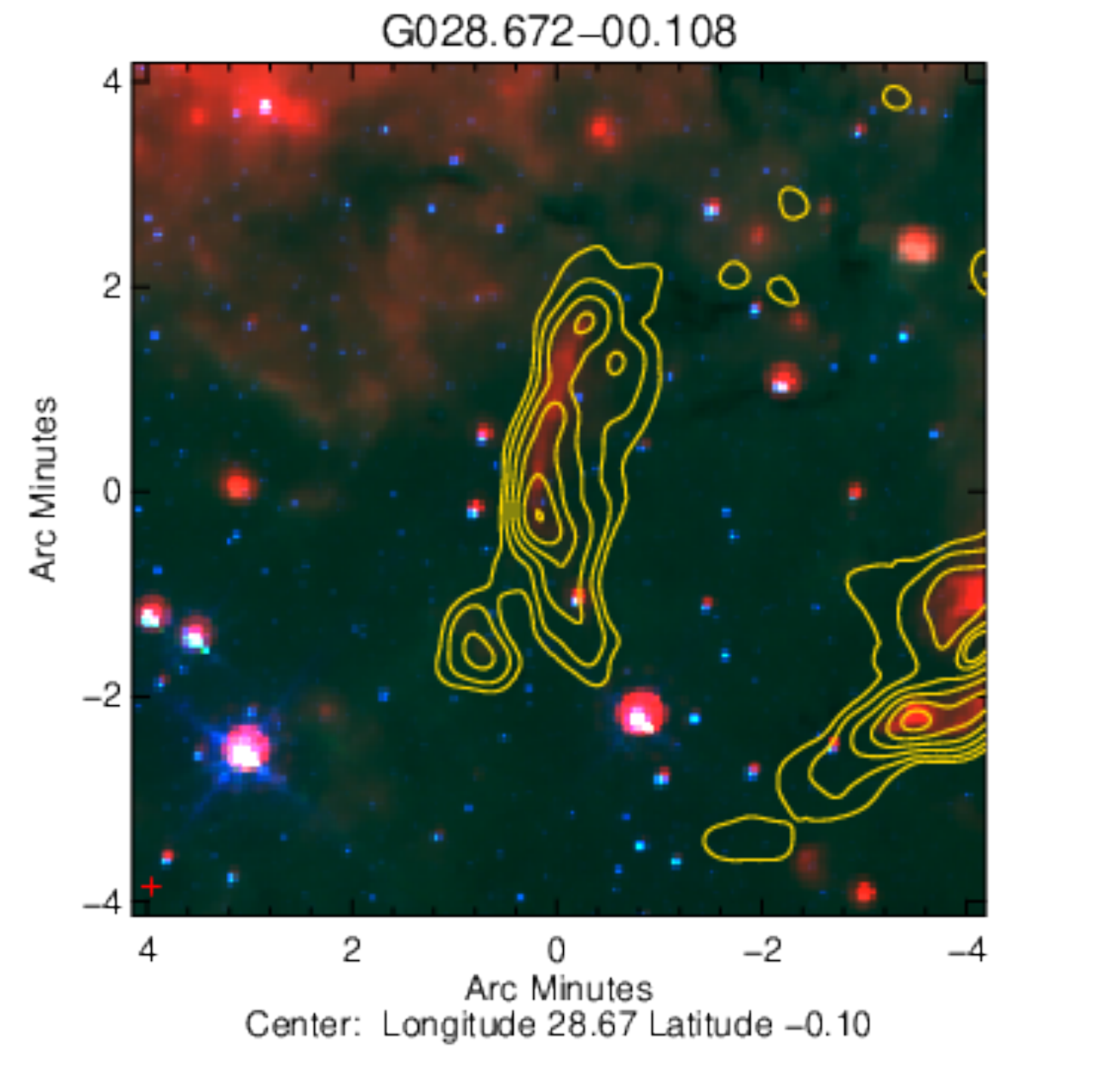}
\includegraphics[width=0.44\textwidth, trim= 0 0 0 0, angle=0]{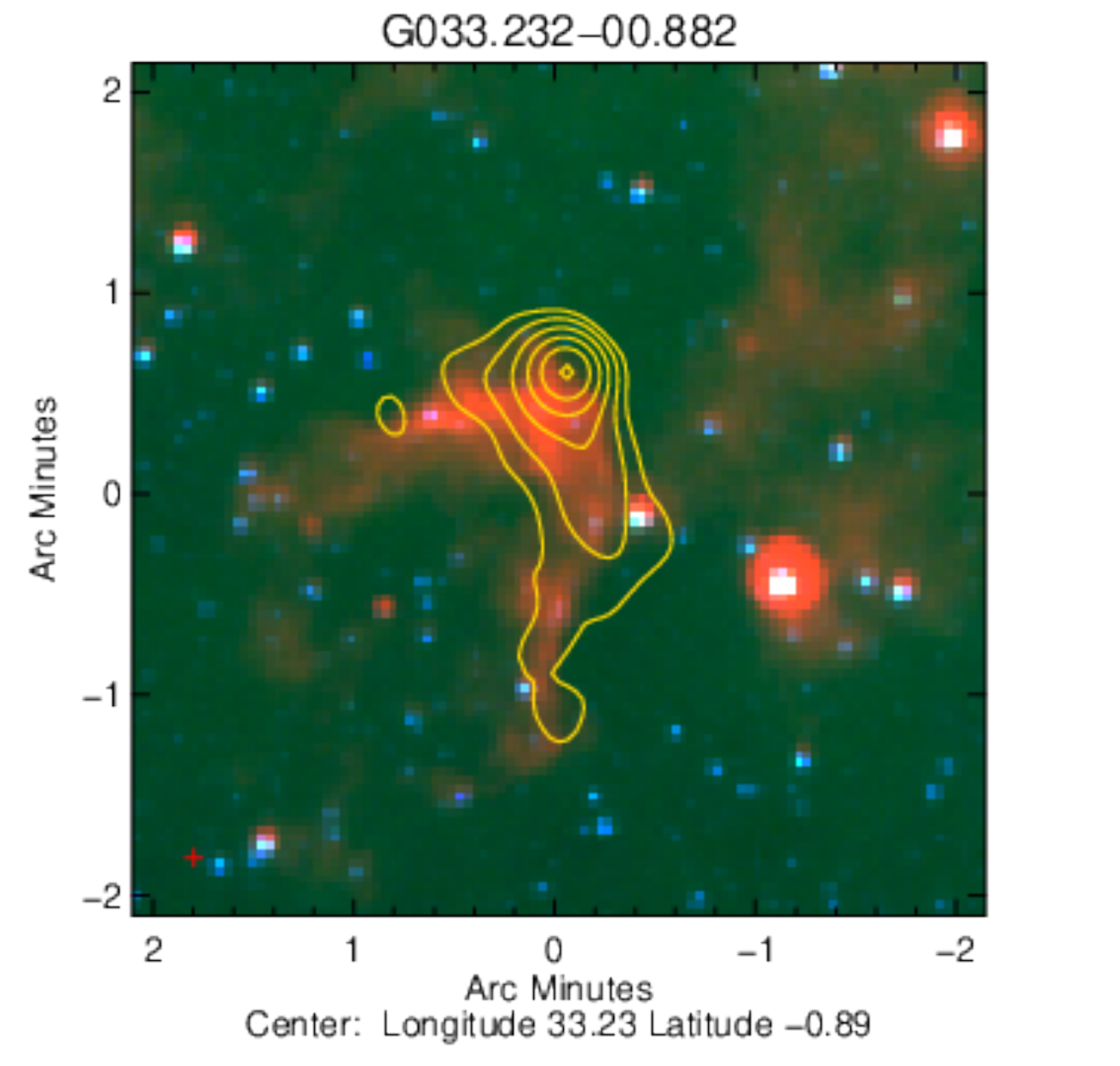}
\includegraphics[width=0.44\textwidth, trim= 0 0 0 0, angle=0]{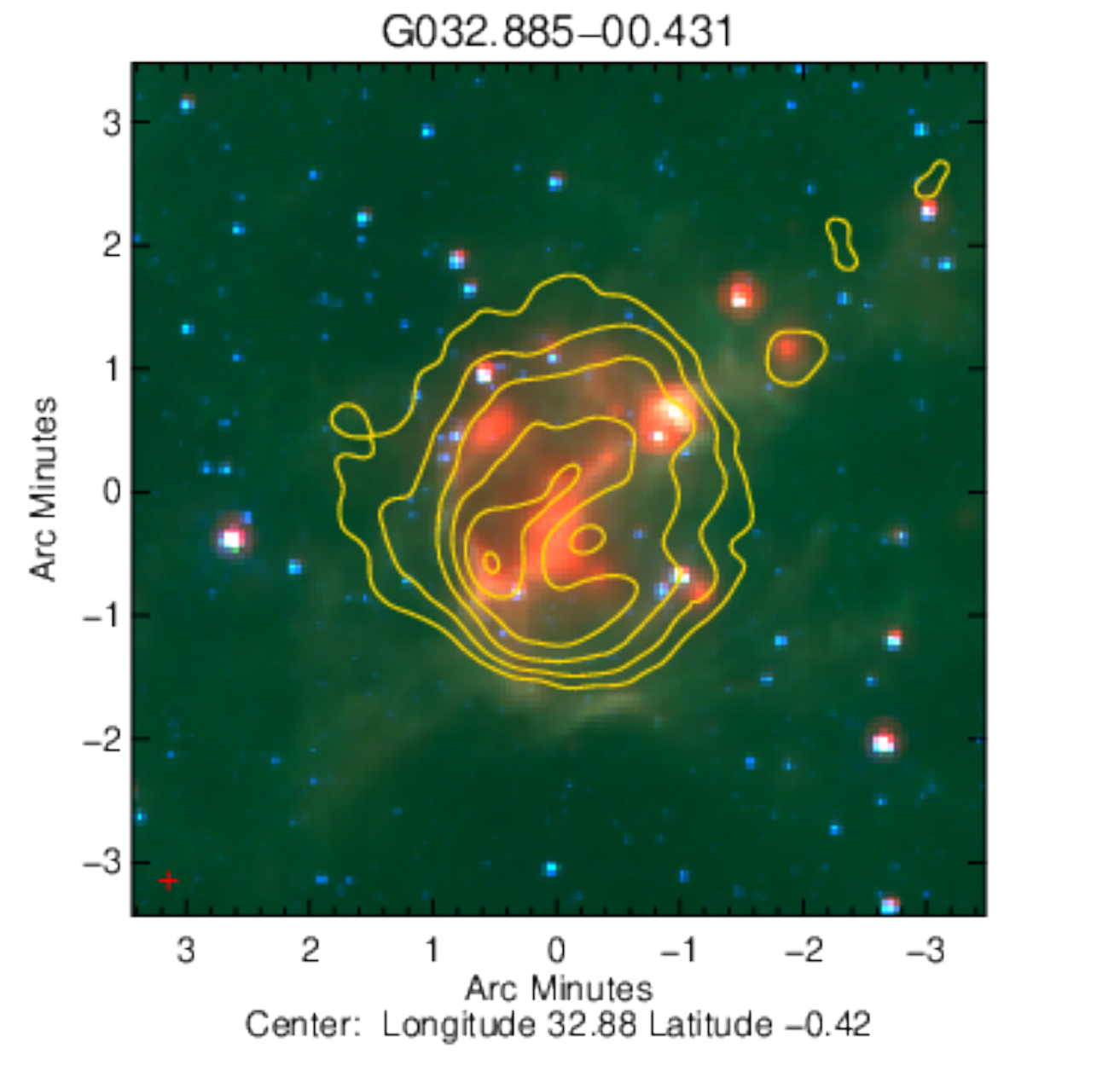}
\includegraphics[width=0.44\textwidth, trim= 0 0 0 0, angle=0]{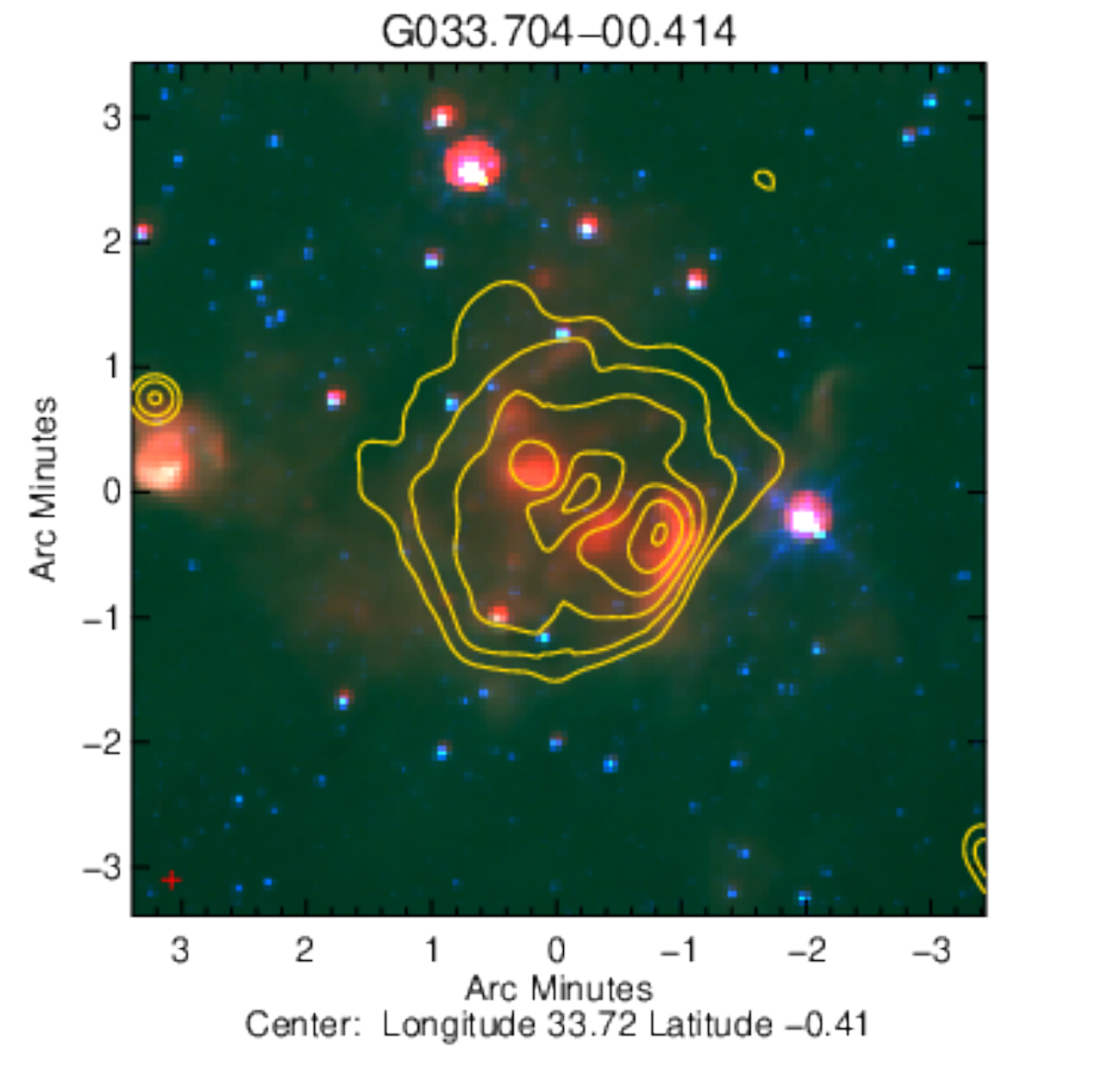}

\caption{In the upper, middle and lower panels we show examples of sources classified as extended radio sources, ionization fronts and evolved \hii\ regions, respectively. In all cases the background image is a false color composite image produced by combining the GLIMPSE 5.8 and 8.0\,$\mu$m bands and the MIPSGAL 24\,$\mu$m band; these are shown in  {blue, green, and red}, respectively. The yellow contours show the distribution of the GLOSTAR 5.8\,GHz radio emission.}
\label{fig:extended_types}
\end{figure*}

\subsubsection{Classification of extended radio sources ($Y$-factor $>$ 2)}

There are 225 radio sources in the catalog that have $Y$-factors greater than 2 that have not been previously classified (see Sect.\,\ref{sect:literature} for details). We can assume that any extended radio source is also likely to have thermal bremsstrahlung emission associated with evolved \hii\ regions. In some cases, although the extended radio emission is positionally coincident with the mid-infrared emission, the morphological correlation is poor, and in these cases, we classify the emission more loosely as being associated with larger scale extended structures, the nature of which is unclear. In other cases, the radio emission appears to be associated with a section of an ionized rim of an evolved \hii\ region and these are classified as ionization fronts. Although in these cases the radio emission is likely to be associated with an \hii\ region, we are only capturing part of the emission and want to make the distinction between compact \hii\ regions where we can be confident we are capturing all of the emission and more extended sources where we are only capturing a less well determined fraction of the total integrated flux density. 

We have classified 57 radio sources as being associated with large-scale emission and a further 31 sources have been associated with ionization fronts. We have classified 71 of the extended radio sources as \hii\ regions. In Fig.\,\ref{fig:extended_types} we present three-color composite images of two examples of each of these three morphological types of structures. We classify another two sources as potential PNe due to the compact nature of the mid-infrared emission and relative isolation and lack of any nebulosity typically observed towards star forming regions. The nature of the remaining 68 extended radio sources that are not classified as either extended emission, ionization front or \hii\ region is still unclear.

\subsubsection{Classification of compact radio sources ($Y$-factor $<$ 2)}


There are 1284 compact radio source in the catalog, of which 133 have been previously classified. In the previous section, we have seen that the spectral index is useful for separating the compact radio sources into thermal and non-thermal bremsstrahlung emission. However, PNe and
{some extragalactic radio sources, known as High Frequency Peakers (\citealt{dallacasa2000})}, can have positive spectral indices  and so this alone does not definitively identify \hii\ regions. From an inspection of the mid-infrared and submillimetre dust emission towards all of the remaining compact sources (1151) we have identified another 20 \hii\ region candidates. The mid-infrared emission towards these sources generally reveals a bubble-like or complex morphology and is associated with compact dust emission.

We have identified a further 16 sources as PNe candidates due to the correlation between the compact radio and mid-infrared emission and the lack of any evidence of dust emission from the ATLASGAL maps. This will be further explored via the information provided by the GLOSTAR RRL data. We have also classified 6 more sources as being associated with an ionization front and 4 others as being extended emission; although in these cases the radio emission is compact, it is coincident with a strong and/or extended 8\,$\mu$m emission. The nature of the remaining 1105 not classified is uncertain. The mean value of the spectral index, for those 523 with this parameter, is $\sim-0.5$ {with} a standard deviation of $0.3$. Suggesting that the vast majority of these unclassified sources are likely to be extragalactic background objects.

{To calculate the number of extragalactic background objects expected in the surveyed area (57600 arcmin$^{2}$), we have followed \citet{Fomalont1991}. These authors observed
an area free of stars, and showed that the number of background objects at 5\,GHz above a flux $S$, is described by}

\[
\left(\frac{N}{{\rm arcmin}^2}\right)=\left(0.42\pm 0.05\right)\left(\frac{S}{30\,\mu{\rm Jy}}\right)^{-1.18 \pm 0.19}.
\]

\noindent{Even though our observed frequency is {slightly} different,
this study is the nearest in frequency of this kind to ours.
Thus we may use their results for an approximation. Using $S=450\,\mu$Jy, i. e., three times the mean noise level, the number of expected background sources in our map 
is $990\pm640$. These numbers also suggest that most of the 1151 unclassified sources are extragalactic radio sources.}



\begin{figure}
\centering
\includegraphics[width=0.49\textwidth, trim= 0 0 0 0, angle=0]{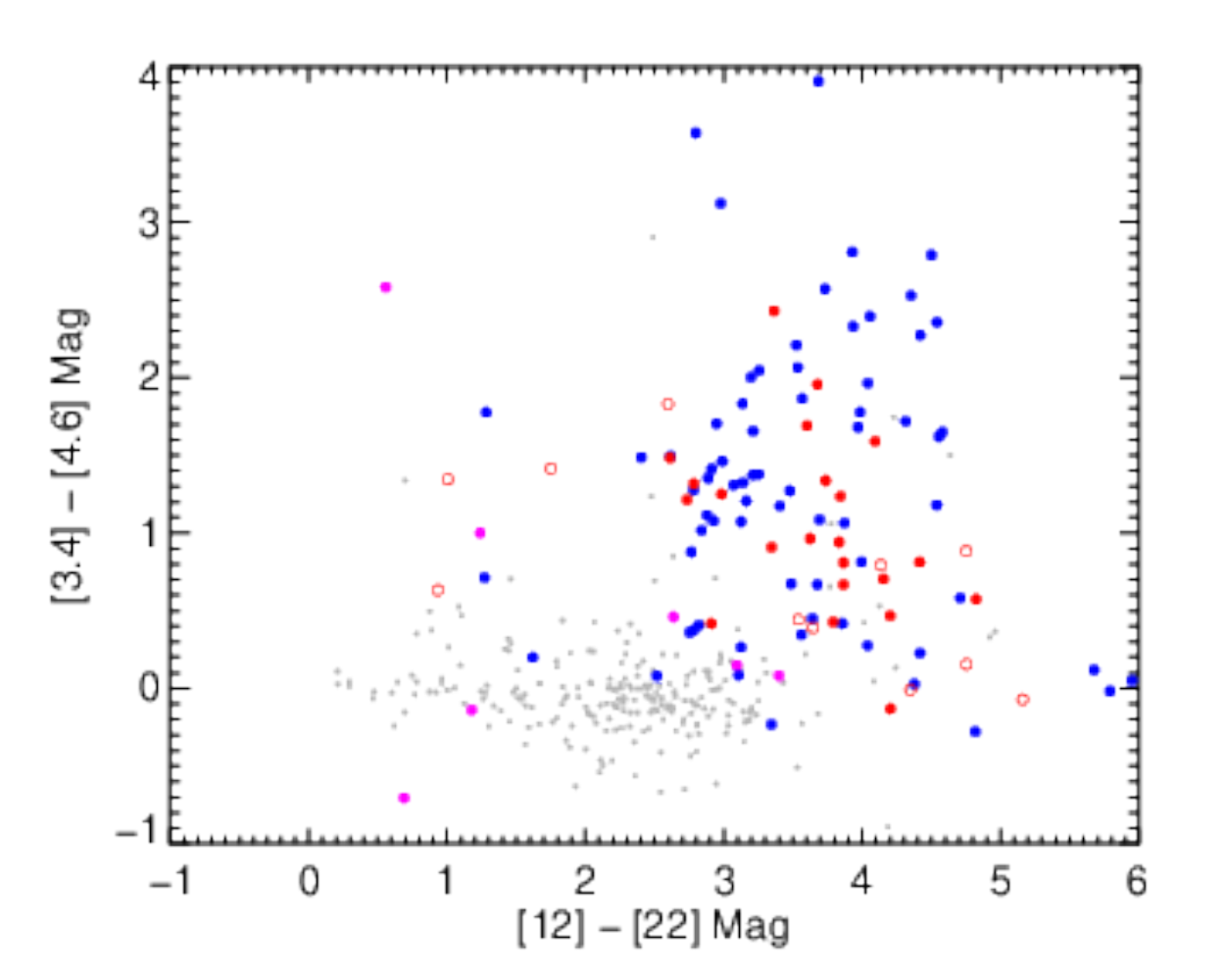}
\includegraphics[width=0.49\textwidth, trim= 0 0 0 0, angle=0]{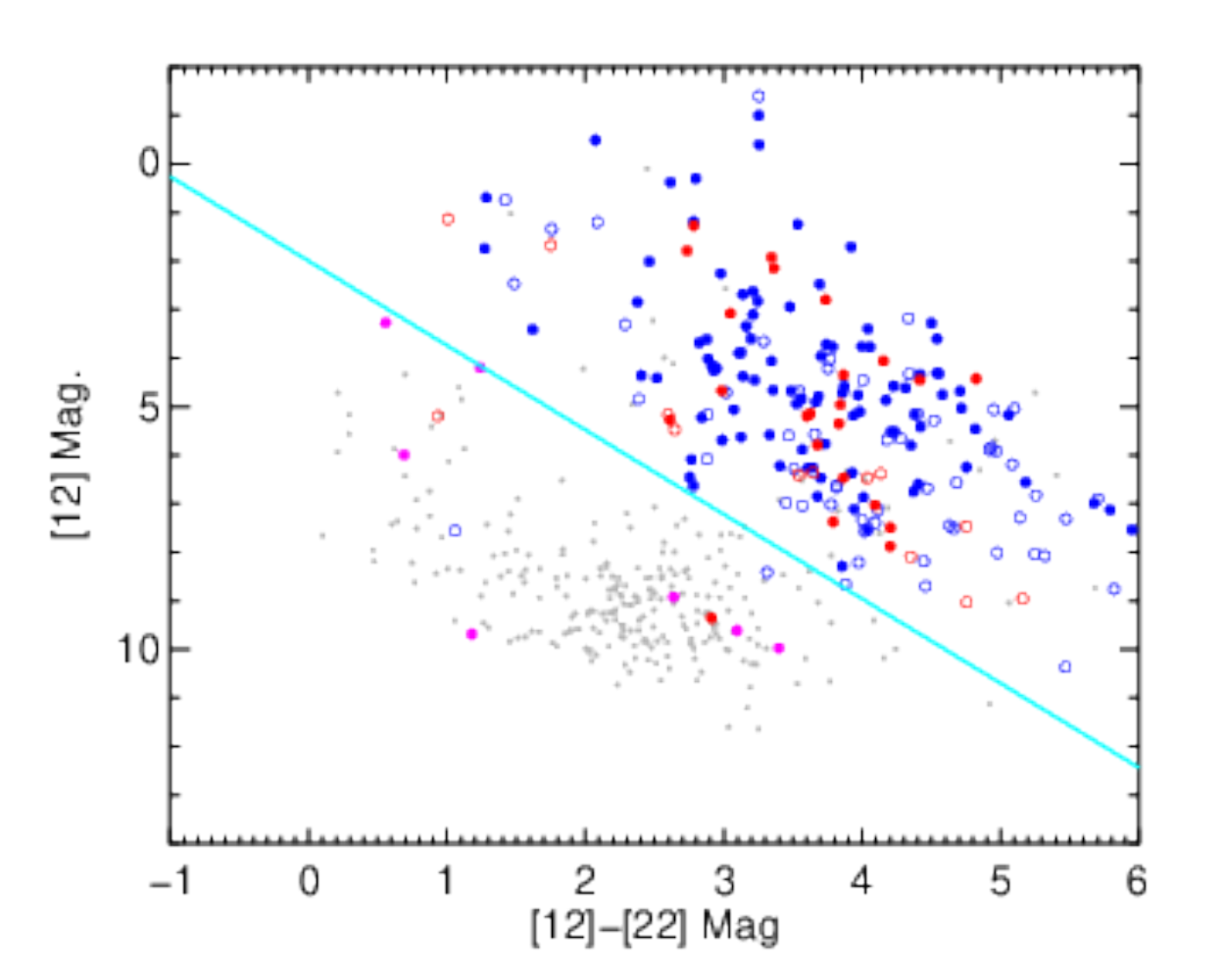}
\caption{Mid-infrared colors of compact GLOSTAR sources associated with a WISE counterpart. Sources classified as \hii\ regions from the literature are shown as filled blue circles while the \hii\ classified in this study are shown as open blue circles. Similarly, PNe identified by \citet{purcell2013} are shown as red filled circles while PNe identified in this paper are shown as open circles. Radio stars identified by \citet{purcell2013} are indicated by filled magenta circles. The smaller grey circles show the distribution of compact sources with a WISE counterpart we have been unable to classify.}
\label{fig:wise_mir_colours}
\end{figure}

\subsection{Mid-infrared colors of radio sources}

We have identified WISE counterparts for 988 radio sources from the WISE all sky point source catalog and will use these to investigate their mid-infrared color. Of these only 470 have reliable fluxes in all four bands. This increases to 475 when considering the three longest wavelength bands. We will use these to examine the colors of previously identified \hii\ regions in an effort to confirm the classifications made and to look for possible trends that can be used to define a set of criteria to identify new \hii\ region candidates. In Fig.\,\ref{fig:wise_mir_colours} we present color-color and color-magnitude diagrams showing the distribution of all \hii\ regions and PNe identified in the literature (filled blue and red circles, respectively) and from our visual inspection of the mid-infrared images (open blue and red circles, respectively). We also show the distributions of the radio stars (filled magenta circles).

Inspection of these plots shows that the locations in color-color and color-magnitude diagrams of  \hii\ regions and PNe identified in our work agree very well with those published in the literature, providing {confidence} that our classifications are robust. The unclassified radio sources (shown as grey filled circles) and radio stars almost exclusively occupy the same region of the parameter space as field stars and background objects (cf. figures B2 and B3 in \citealt{csengeri2014}) while the embedded and dust enshrouded objects (\hii\ regions and PNe) have redder colors and are clearly separated from the field stars. This is particularly clear in the color-magnitude plot shown in the lower panel of Fig.\,\ref{fig:wise_mir_colours} where all of the \hii\ regions and the vast majority of the PNe are located in an area of parameter space defined by: 

\begin{equation}
[12]\,{\rm Mag.} < 1.74 \times ([12]-[22]) +2.
\end{equation}

This is also indicated on the lower panel of Fig.\,\ref{fig:wise_mir_colours} with a cyan line.  Only two PNe (G029.493+00.931 and G033.120$-$00.894)  and 2 \hii\ regions (G030.246-00.91 and G033.704$-$00.414) are found to the left of this color criterion. One of the PNe was identified by the CORNISH team, while the other PN and both \hii\ regions have been classified by us. We note the latter of the two \hii\ regions is very evolved with a number of WISE sources being located within its boundary. 
For the other three sources it is possible that these are the result of chance alignments between radio and mid-infrared sources, but it is also possible that the colors of these objects cover a larger range or that the classifications themselves are wrong. The sample is too small to test any of these possibilities statistically but we note that the one of the PNe identified by the CORNISH team that falls on the wrong side of this criterion (G033.120$-$00.894) is positionally coincident with the WISE source J185457.96-001424.3. The positional offset is $\sim$3\arcsec\ and the photometric flux qualities are classified as AABB {and so the correlation and data are reliable}.\footnote{ The quality of the profile-fit photometry measurement is indicated by a four letter code (one for each band [W1/W2/W3/W4]; see Sect.\,\ref{sect:wise}). The letters `A' and `B'  correspond to the sources detected in a particular band with a flux signal to noise ratio $>10$ and between 3 and 10, respectively. Detections with a quality flag of `A' and `B' are considered to be reliable.}

\subsection{New compact \hii\ regions}

Our analysis has identified a total of {231} \hii\ regions, of which 46 were previously known (\citealt{kalcheva2018,ortega2010,giveon2005b}) and 90 were proposed as candidates from analysis of the WISE data by {\cite{anderson2014}}. We have therefore identified 96 new \hii\ region and confirmed the nature of the those proposed by {\cite{anderson2014}}. The majority  of the new \hii\ regions (84) have $Y$-factors $>$ 1.2 and are therefore resolved in GLOSTAR and are unlikely to have been picked up in earlier high-resolution surveys where the extended emission is filtered out (CORNISH and MAGPIS). The remaining 12 \hii\ region candidates are unresolved in GLOSTAR and so could have been picked up in earlier surveys. However, their peak fluxes are lower than the sensitivity limit of CORNISH ($7\sigma \sim 2.7\,{\rm mJy\,beam}^{-1}$) except in the case of two sources. We checked the CORNISH maps of these two sources but did not find any emission present coincident with the GLOSTAR position (the noise values are 0.34\,mJy and 0.32\,mJy for G030.371+00.483 and G030.678+00.767, respectively). Both of these sources have mid-infrared counterparts and are hence real. These sources are therefore either both extended and completely filtered out in the CORNISH data or are variable radio sources. 


\setlength{\tabcolsep}{6pt}

\begin{table*}


\begin{center}
\caption{Catalog of new  compact \hii\ region candidates. Sources that are bright enough to have been detected by CORNISH are indicated by appending a $\dagger$ to the source name. The fluxes and spectral indices are drawn from the GLOSTAR catalog (i.e., Table\,\ref{tbl:glostar_cat}). In the last two column we indicate the presence of compact (\cmark), extended (\cmark\cmark) or absence (\xmark) of any corresponding ATLASGAL or WISE emission. In the last column we indicate which are still \uchii\ region candidates after analysis (see text for details).}
\label{tbl:new_candidates}
\begin{minipage}{\linewidth}
\small
\begin{tabular}{lccccccc}
\hline \hline
\multicolumn{1}{c}{GLOSTAR}&  \multicolumn{1}{c}{$S_\nu$} &  \multicolumn{1}{c}{$\Delta S_\nu$} &   \multicolumn{1}{c}{$\alpha$} &  \multicolumn{1}{c}{$\Delta\alpha$} & {Sub-mm} & Mid-IR & \uchii\ \\
\multicolumn{1}{c}{name}&  \multicolumn{2}{c}{(mJy\,beam$^{-1}$)} & &  & {emission} & emission  & candidate\\
\hline
G029.367$-$00.317	&	2.26	&	0.24	&	\multicolumn{1}{c}{$\cdots$}	&	\multicolumn{1}{c}{$\cdots$} 						& \cmark\cmark	& \cmark\cmark  & \xmark\\
G030.154+00.583	&	0.55	&	0.13	&	\multicolumn{1}{c}{$\cdots$}	&	\multicolumn{1}{c}{$\cdots$}							& \cmark & \cmark   & \cmark\\
G030.241$-$00.592	&	1.73	&	0.18	&	$-$0.12	&	0.16						& \xmark & \cmark  & \xmark\\
G030.371+00.483$\dagger$	&	3.47	&	0.33	&	\multicolumn{1}{c}{$\cdots$}	&	\multicolumn{1}{c}{$\cdots$}				&\cmark & \cmark  & \cmark\\
G030.678+00.767$\dagger$	&	3.95	&	0.26	&	0.07	&	0.08				& \xmark & \cmark & \cmark\\
G032.051$-$00.091	&	2.12	&	0.24	&	$-$0.34	&	0.15 						&\cmark\cmark & \cmark &	\xmark \\
G032.460+00.386	&	0.67	&	0.16	&	\multicolumn{1}{c}{$\cdots$}	&	\multicolumn{1}{c}{$\cdots$}							&\cmark\cmark & \cmark\cmark	&\xmark \\
G033.714+00.256	&	1.50	&	0.22	&	\multicolumn{1}{c}{$\cdots$}	&	\multicolumn{1}{c}{$\cdots$} 							&\cmark & \cmark	&\cmark\\
G034.274$-$00.151	&	1.01	&	0.16	&	\multicolumn{1}{c}{$\cdots$}	&	\multicolumn{1}{c}{$\cdots$}						&\cmark & \cmark	&\cmark\\
G034.338$-$00.694	&	0.49	&	0.12	&	\multicolumn{1}{c}{$\cdots$}	&	\multicolumn{1}{c}{$\cdots$}	 					&\cmark & \cmark &\cmark\\
G034.440+00.060	&	1.53	&	0.35	&	\multicolumn{1}{c}{$\cdots$}	&	\multicolumn{1}{c}{$\cdots$}							&\cmark\cmark & \cmark &\cmark\\
G035.344+00.348	&	0.46	&	0.11	&	\multicolumn{1}{c}{$\cdots$}	&	\multicolumn{1}{c}{$\cdots$}							&\cmark\cmark & \cmark &\cmark\\
\hline
\end{tabular}\\

\end{minipage}

\end{center}
\end{table*}

\setlength{\tabcolsep}{6pt}

In Table\,\ref{tbl:new_candidates} we give the peak flux densities and spectral indices of the new unresolved \hii\ regions; only three sources have a detection in THOR and consequently we have only a measurement for the spectral index for three of these \hii\ regions, all of which are consistent with what is expected for optically thin \hii\ regions (i.e. $\alpha = -0.1$). This suggests that these are more evolved than the \uchii\ region stage, which are usually characterised by optically thick emission ($\alpha \sim 1$-2; \citealt{kurtz2005}). We note that no dust emission is detected towards two of these \hii\ regions (G030.241$-$00.592 and G030.678+00.767), which is also consistent with them being more evolved having already dissipated their natal clump. We further note that the latter of these is also one of the bright sources we would have expected to have been detected by CORNISH and so the spectral index and absence of dense envelope are consistent with this source being quite evolved and perhaps explains its non-detection in CORNISH.

In Table\,\ref{tbl:new_candidates} we also list the new \hii\ regions association with mid-infrared and submillimetre emission. As can be seen from this table, all of these sources are associated with mid-infrared emission and the majority are associated with dust emission. The mid-infrared emission looks to be extended in two cases indicating these are likely to be quite evolved \hii\ regions ($>$ 12\arcsec\ in diameter).\footnote{We have made these associations with dust and mid-infrared emission from inspection of the ATLASGAL and WISE maps as not all emission is captured in the catalogs. All but one of these \hii\ regions has a counterpart in WISE and 5 have a counterpart in the ATLASGAL CSC.} We have searched SIMBAD towards all of these sources. This revealed that G035.344+00.348 had been previously identified as an \hii\ region (\citealt{urquhart2011_nh3}; MSX6C G035.3449+00.3474). In the last column of Table\,\ref{tbl:new_candidates} we identify the \hii\ regions that are likely to be good \uchii\ region candidates; this is done by eliminating sources that are not associated with dust, have spectral indices indicting optically thin emission, or are extended in the mid-infrared, all of which point to sources that are more evolved than the \uchii\ phase. 

\begin{figure}
\centering
\includegraphics[width=0.49\textwidth, trim= 0 0 0 0, angle=0]{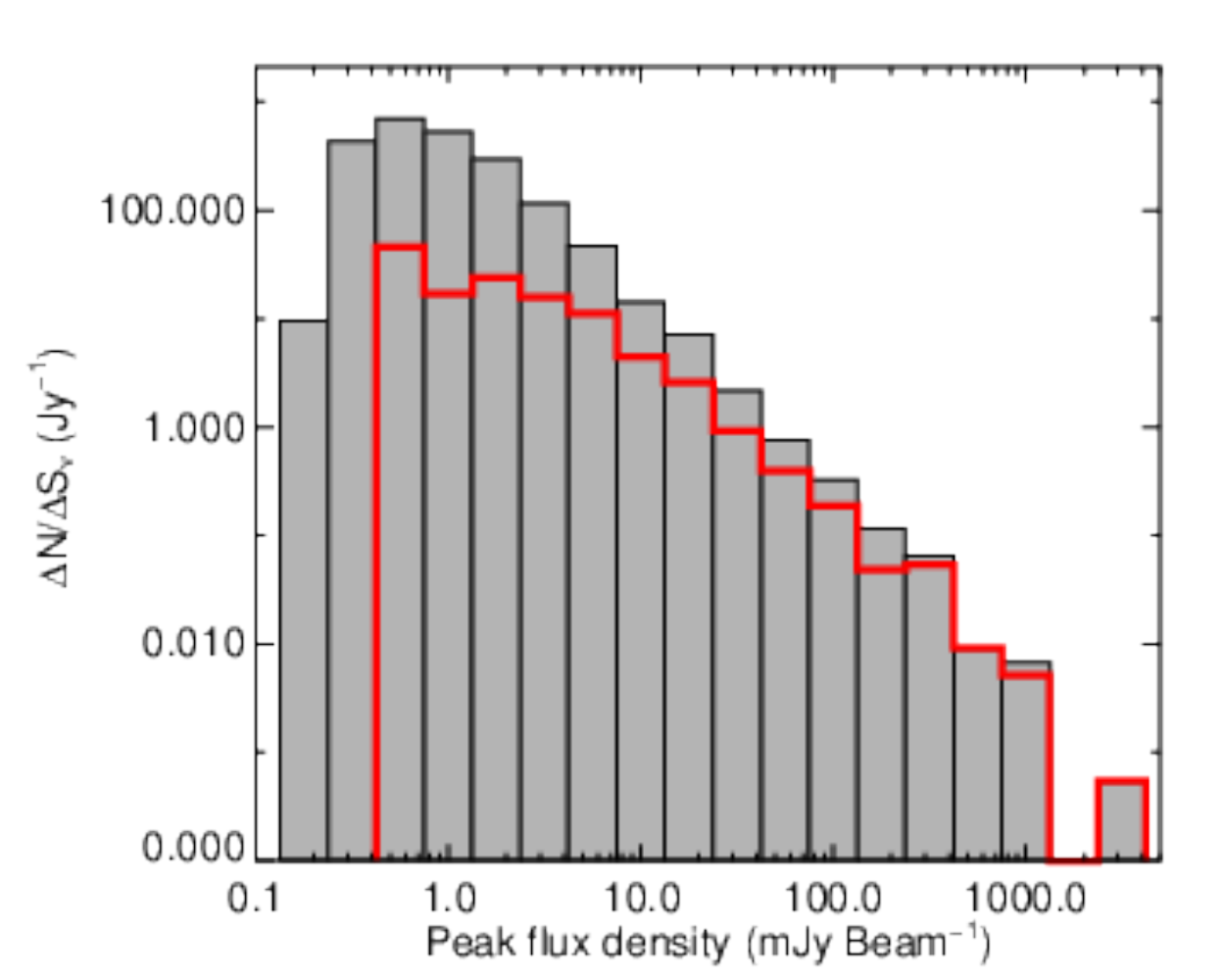}
\includegraphics[width=0.49\textwidth, trim= 0 0 0 0, angle=0]{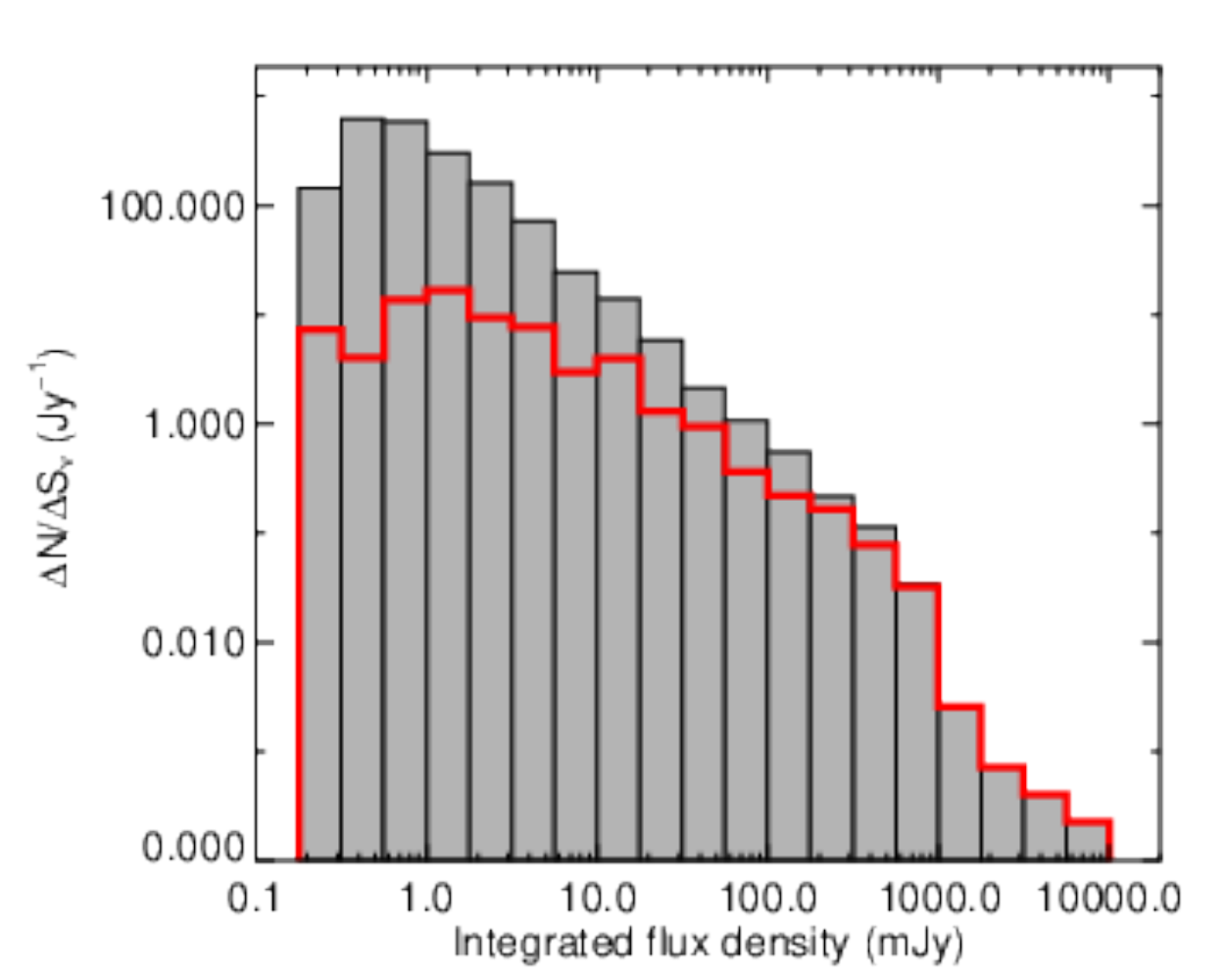}

\caption{Distribution of \hii\ region properties. In the upper and lower panels we show the peak and integrated flux density for the whole GLOSTAR catalog (grey) and those classified as \hii\ regions (red). The bin sizes in both plots is 0.25\,dex.}
\label{fig:flux_dist_hii}
\end{figure}

\section{Properties and Galactic distribution of the \hii\ regions}
\label{sect:properties_galactic_distribution_HII_regions}

From a combination of visual inspection of mid-infrared images and cross-matching with other studies we have identified a total of 231 \hii\ regions in our catalog. This is the most complete sample of \hii\ regions identified in this part of the sky to date incorporating \hii\ regions with sizes of a few arcseconds to a few arcminutes. In this section we will provide an analysis of their properties and distributions.

\subsection{Observed and measured properties}

In Fig.\,\ref{fig:flux_dist_hii} we present plots of the peak and integrated flux density for the whole GLOSTAR catalog and of the \hii\ regions identified. By comparing these two samples we see that nearly all of the very bright radio sources are associated with \hii\ regions but that the fraction of the population decreases as the flux decreases and turns over between 0.5 and 2\,mJy\,beam$^{-1}$ and between 1-2\,mJy for the peak and integrated flux densities respectively. This indicates that there are relatively fewer weak \hii\ regions in comparison to the total number of radio sources detected at low surface brightnesses. The fact that we find no \hii\ region with fluxes below the turnover in the peak flux density may indicate a minimum flux density threshold for \hii\ regions. However, this corresponds to a flux density of $\sim 0.4$\,mJy\,beam$^{-1}$, which is close to the 3-4$\sigma$ noise found towards the Galactic mid-plane where the vast majority of \hii\ regions are to be found. Radio sources detected below this threshold are primarily extragalactic background sources located away from the mid-plane where the noise level drops off significantly (see lower panel of Fig.\,\ref{fig.noise_histogram_fn_lat}).

\begin{figure}
\centering

\includegraphics[width=0.49\textwidth, trim= 0 0 0 0, angle=0]{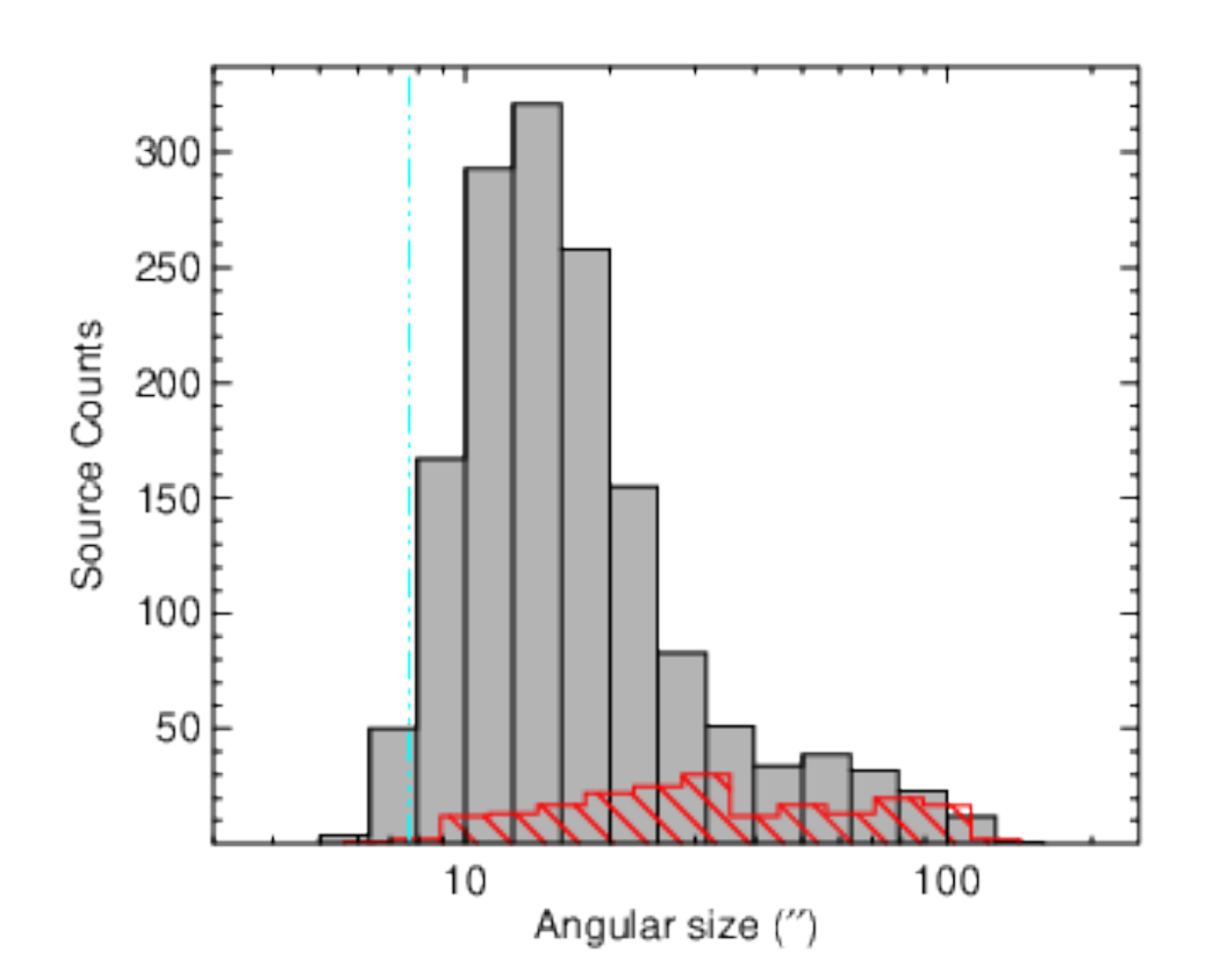}
\includegraphics[width=0.49\textwidth, trim= 0 0 0 0, angle=0]{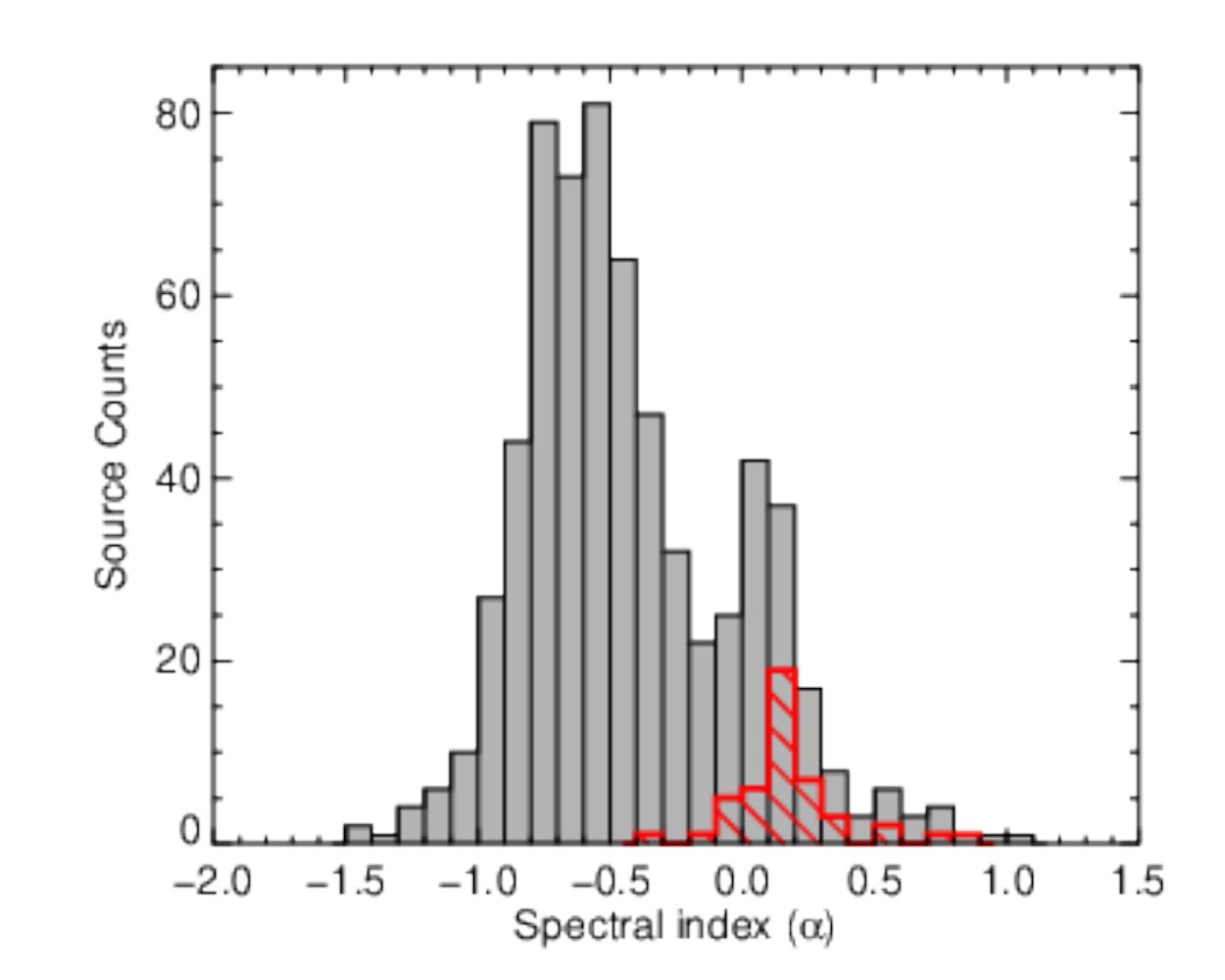}

\caption{Distribution of \hii\ region properties. In the upper and lower panels we show the angular radius and spectral index for the whole GLOSTAR catalog (grey) and those classified as \hii\ regions (red). The dashed-dotted vertical line drawn on the upper panel shows the radius of the beam. The bin sizes used in the upper and lower panels are 0.1\,dex and 0.1,  respectively.}

\label{fig:size_dist_hii}
\end{figure}

In Fig.\,\ref{fig:size_dist_hii} we compare the source sizes and spectral indices for the whole GLOSTAR sample and those identified as \hii\ regions.   It is clear from the upper panel in this figure that the vast majority of the radio sources are unresolved with the numbers dropping off steeply for sizes larger than $\sim$15\arcsec. The distribution of the \hii\ region sizes appears to be roughly uniform over the whole range of sizes, which is consistent with other studies (e.g., \citealt{kurtz1999,Masque2017}), and is to be expected given that the sample is likely to cover a larger range of distances and ages.  We also note that the \hii\ regions contribute a significant fraction of the radio sources with  sizes above $\sim$30\arcsec\ with the others being classified as ionization fronts or extended emission sources. 
We find very few radio sources with sizes larger than $\sim$100\arcsec, but this is a consequence of the interferometric snapshot observations themselves as the poor $uv$-coverage puts limits on the angular size of structures that can be reliably imaged. 

\begin{figure}
\centering
\includegraphics[height=0.80\linewidth, trim= 0 0 0 0, angle=0]{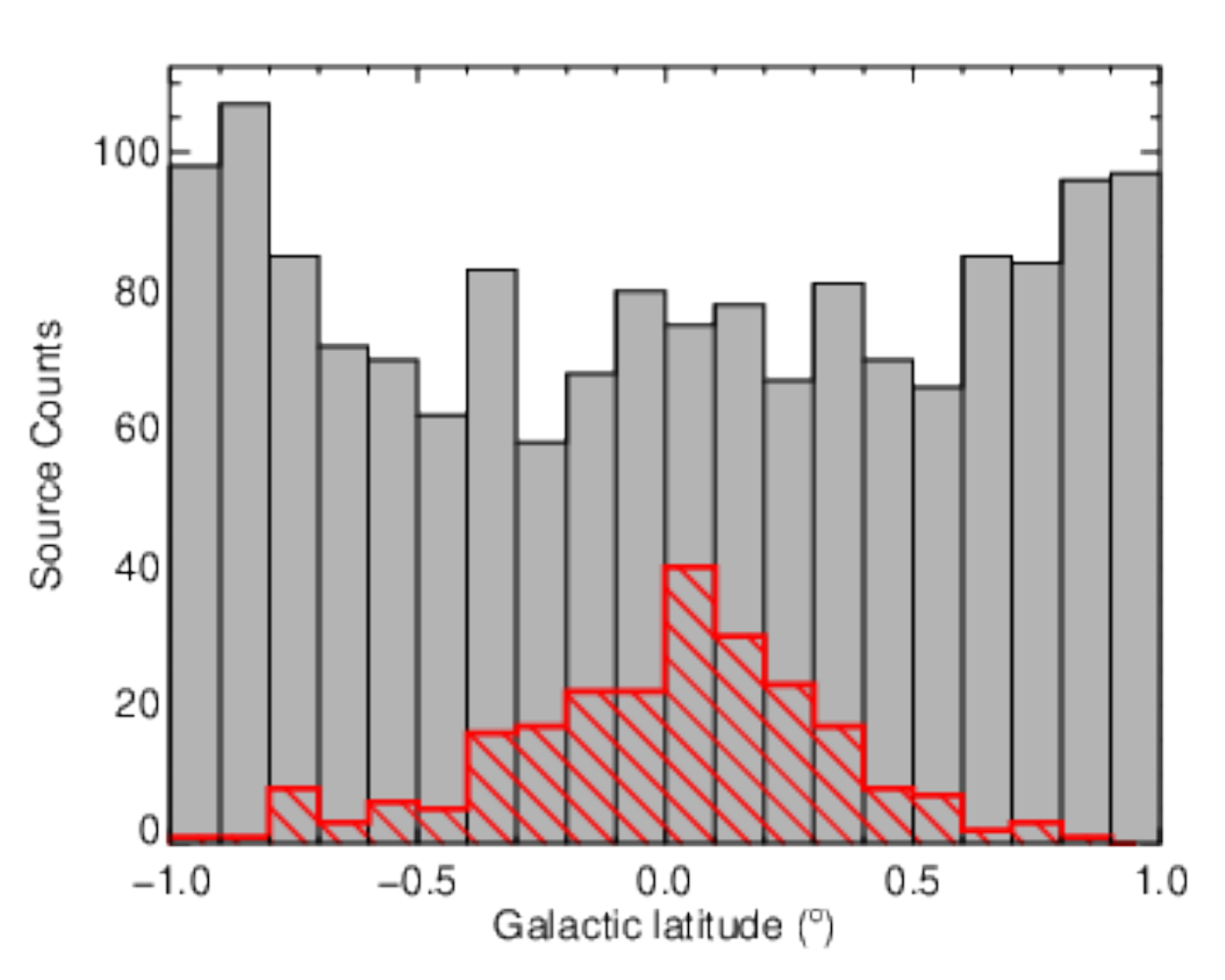}
\caption{Distribution of GLOSTAR radio sources as a function of  Galactic latitude. The whole catalog is shown as the grey histogram while the sources identified as \hii\ regions are shown as the hatched red histogram. The bin size used for both distributions is 0.1\degr.}
\label{fig:gal_lat_dist}
\end{figure}

\begin{figure*}
\centering
\includegraphics[width=0.98\textwidth, trim= 0 0 0 0, angle=0]{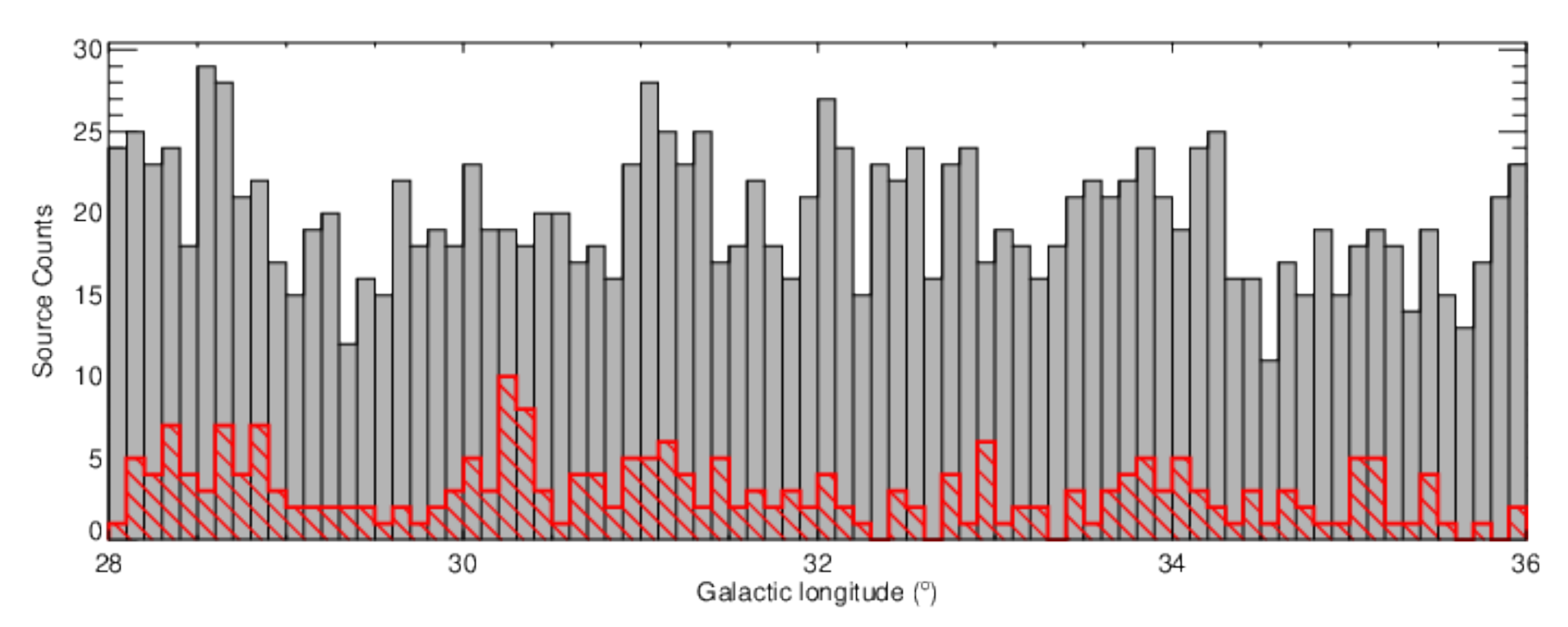}
\caption{Distribution of GLOSTAR radio sources as a function of Galactic longitude. The whole catalog is shown as the grey histogram while the sources identified as \hii\ regions are shown as the hatched red histogram. The bin size used for both distributions is 0.1\degr.}
\label{fig:gal_long_dist}
\end{figure*}

The spectral index distribution of the whole sample is bimodal and peaks at $\alpha$ = $-$0.6 and 0.1; the former is likely to be associated with synchrotron emission from the extragalactic background population while the latter is likely to be associated with thermal emission from Galactic sources such as \hii\ regions and PNe. It is therefore reassuring to see the distribution of \hii\ regions correlating with the thermal emission peak. This is comparable with the spectral index calculation of the THOR team (\citealt{wang2018}), but our sources show a higher density of positive index.\\

\subsection{Galactic distribution}

The majority of the detected radio sources are likely to be extragalactic which is expected to be smoothly distributed across the whole sky. However, the \hii\ region population should show significant variations in Galactic longitude and latitude with peaks that correlate with the most active star forming complexes and the distribution of molecular material, both of which are concentrated in the Galactic mid-plane.

In Fig.\,\ref{fig:gal_lat_dist} we show the Galactic latitude distribution of all the radio sources and the \hii\ region subsample. It is clear from this plot that the \hii\ regions are indeed tightly concentrated around the Galactic mid-plane with a peak at $b \sim 0\degr$, which is consistent with previous studies of \hii\ regions (\citealt{urquhart2013_cornish,kalcheva2018}) and dense massive star forming clumps (\citealt{urquhart2014_atlas}). The distribution of the full sample is relatively flat for the inner part of the mid-plane but rises towards at the edges of the surveyed region ($|b| \sim 1\degr$), which is probably related to the reduction in the local noise value as one move away from the Galactic mid-plane where much of the extended emission from extended \hii\ regions and supernova remnants are located (see Figs.\,\ref{fig.integrated_map} and \ref{fig.noise_histogram_fn_lat}).

In Fig.\,\ref{fig:gal_long_dist} we show the distribution of all GLOSTAR sources and the \hii\ regions as a function of the Galactic longitude. Before we discuss the longitude distribution we should bear in mind that since we have excluded emission associated with the 27 large scale structures, many of which are associated with active \hii\ regions, the peaks seen in the \hii\ region distribution are likely to be somewhat suppressed. All of these large scale structures are  Galactic in origin and are unlikely to affect the distribution of the extragalactic population, which are rarely resolved.

The distribution of the whole catalog can be described as being broadly flat. As expected, the distribution of \hii\ regions reveals some significant peaks. We note the strong peak near $\ell\sim$30\degr\, which is related with emission from the massive star forming regions W43 main and W43 south. This region is located where the Scutum-Centaurus arm is interacting with the  end of the Galactic long bar (\citealt{RodriguezCombes2008,nguyen2011}), which is expected to lead to a rise in the star formation activity in this area.

The latitude and longitude distributions are consistent with the distribution of the high-reliability ($\sigma \geq 7$) CORNISH catalog (\citealt{purcell2013}) and the lower frequency THOR catalog (\citealt{bihr2016, wang2018}), which provides confidence that the catalog we have produced is reliable.

\section{\label{conclusions} Summary and conclusions}

The GLOSTAR survey is a VLA survey 
of the Galactic mid-plane 1st quadrant ($-10\degr < \ell < 80\degr$  and $|b| <  1\degr$). In this paper we present the radio continuum data for a 16 square degree part of the survey covering 8 degrees in longitude and 2 degrees in latitude and centered on $\ell = 32\degr$, $b=0\degr$ (i.e., $28\degr < \ell < 36\degr$  and $|b| <  1\degr$). We use the \blobcat\ source extraction algorithm to produce a catalog of radio sources and identify a number of different type of morphologies, from unresolved to compact and extended. 

We have also identified 27 large scale structures that are over-resolved at the GLOSTAR resolution ($\sim$18\arcsec) and where the emission has {been} broken up into multiple radio sources. In these cases we have regrouped the various fragments and give the parameters for these objects separately. The individual fragments are removed from the final catalog, which consists of \catsize distinct radio sources. The initial catalog has been cross-matched with other radio and mid-infrared catalogs to verify that the extracted sources are real. We have visually examined mid-infrared and submillimetre images for radio sources without a counterpart to identify and remove artifacts. We discuss the observed properties of this new catalog of sources and compare them to previous studies. 

We have used the multi-wavelength analysis and searched the literature to identify the nature of the radio sources. This has resulted in the identification of {231} \hii\ regions. Of these, only 46 had been previously identified with {high} resolution radio surveys, and a further 90 \hii\ region candidates identified in the mid-infrared by \citet{anderson2014} have now been confirmed. We have identified a further {96} \hii\ regions not previously known. The majority of the new \hii\ regions are well-resolved (71 with $Y$-factor $> 2$) and have been missed by previous high-resolution interferometric surveys due to their lack of short baseline data and consequent poor sensitivity to large angular scales. Twelve of the new \hii\ regions identified are considered to be unresolved (i.e., $Y$-factor $< 1.2$), of these we identify 8 \uchii\ region candidates.

We have investigated the  properties of the full sample of \hii\ regions and compared them to the previously identified samples. The improved flux sensitivity ($\sim$100-400\,$\mu$Jy\,beam$^{-1}$) and sensitivity to larger angular scales ($\sim$20-200\arcsec) of GLOSTAR compare with other surveys, has increased the number of \hii\ regions in this part of the Galaxy by a factor of 4.

In addition to the \hii\ regions we have also identified 15 radio stars, 46 PNe, 3 pulsars, 3 compact SNR-candidates  and 99 radio sources associated with more extended structures such as the ionization fronts of much more extended \hii\ regions. We are unable to classify 1186 radio sources; these can be broadly categorized as being weak and unresolved, the majority of which are likely to be extragalactic background sources. 

Here we present a catalog of \catsize radio sources and have investigated their nature and physical properties. The data presented here is only a small fraction of the total GLOSTAR coverage, which also includes spectral line and higher-resolution data obtained with the VLA in B-array. This will be complemented with observations that are being performed with the 100-m Effelsberg telescope, which will provide sensitivity to scales larger than seen in the D-configuration observations.

\begin{acknowledgements}
We would like to thank the referee for constructive comments and suggestions that have helped to improve this work. S.-N.X.M. is a member of the International Max-Planck Research School at the Universities of Bonn and Cologne (IMPRS). N.R. acknowledges support from the Infosys Foundation through the Infosys Young Investigator grant. H.B. acknowledges support from the European Research Council under the Horizon 2020 Framework Program via the ERC Consolidator Grant CSF-648505. This research was partially funded by the ERC Advanced Investigator Grant GLOSTAR (247078). It made use of information from the ATLASGAL database at \url{http://atlasgal.mpifr-bonn.mpg.de/cgi-bin/ATLASGAL_DATABASE.cgi} supported by the MPIfR, Bonn, as well as information from the CORNISH database at \url{http://cornish.leeds.ac.uk/public/index.php} which
was constructed with support from the Science and Technology Facilities Council of the UK. This publication also makes use of data products from the Wide-field Infrared Survey Explorer, which is a joint project of the University of California, Los Angeles, and the Jet Propulsion Laboratory/California Institute of Technology, funded by the National Aeronautics and Space Administration. This research has made use of the SIMBAD database, operated at CDS, Strasbourg, France. We have used the collaborative tool Overleaf available at: https://www.overleaf.com/. 
\end{acknowledgements}

\bibliographystyle{aa}

\clearpage
\onecolumn

\end{document}